%% file: 00-main.tex
\documentclass[acmtog,nonacm]{acmart} 
\acmJournal{TOG}

\input{98-pre}

\begin{document}

\title{Neural Kinematic Bases for Fluids}

\author{Yibo Liu}
\email{liuyibo@uvic.ca}
\orcid{0009-0007-3537-1572}
\affiliation{%
  \institution{University of Victoria}
  \country{Canada}
  \city{Victoria}
}

\author{Zhixin Fang}
\email{zhixinfang@inworld.ai}
\orcid{0009-0004-8153-9549}
\affiliation{%
  \institution{Inworld AI}
  \country{Canada}
  \city{Vancouver}
}

\author{Sune Darkner}
\email{darkner@di.ku.dk}
\orcid{0000-0001-6114-7100}
\affiliation{%
  \institution{University of Copenhagen}
  \country{Denmark}
  \city{Copenhagen}
}

\author{Noam Aigerman}
\email{noam.aigerman@umontreal.ca}
\orcid{0000-0002-9116-4662}
\affiliation{%
  \institution{University of Montreal}
  \country{Canada}
  \city{Montreal}
}

\author{Kenny Erleben}
\email{kenny@di.ku.dk}
\orcid{0000-0001-6808-4747}
\affiliation{%
  \institution{University of Copenhagen}
  \country{Denmark}
  \city{Copenhagen}
}

\author{Paul Kry}
\email{kry@cs.mcgill.ca}
\orcid{0000-0003-4176-6857}
\affiliation{%
  \institution{McGill University}
  \country{Canada}
  \city{Montreal}
}

\author{Teseo Schneider}
\email{teseo@uvic.ca}
\orcid{0000-0002-5969-636X}
\affiliation{%
  \institution{University of Victoria}
  \city{Victoria}
  \country{Canada}
}

\renewcommand{\shortauthors}{Liu et al.}

\begin{abstract}
\input{01-abstract}    
\end{abstract}



  
\input{01-teaser}

\maketitle

\input{02-intro}
\input{03-related}
\input{04-method}
\input{05-results}

\input{06-conclustions}

\input{08-ack}

\bibliographystyle{ACM-Reference-Format}
\bibliography{literature.bib}

\clearpage
\appendix
\input{09-append}

\end{document}

%% file: 98-pre.tex
\usepackage[english]{babel}
\usepackage{libertinus}
\usepackage{algpseudocodex}
\usepackage{algorithm}
\usepackage{amsfonts}
\usepackage{amsmath}
\usepackage{amsthm}
\usepackage{mathtools}
\usepackage{enumerate}
\usepackage{csquotes}
\usepackage{hyperref}
\usepackage{multirow}
\usepackage{cleveref}
\usepackage{wrapfig}
\usepackage{svg}

\usepackage{tikz}
\usetikzlibrary{shapes.geometric, arrows}

\newcommand{\vo}{\varphi}

\newcommand{\ai}{\alpha_i}

\newcommand{\RR}{\mathbb{R}}

\DeclareMathOperator*{\argmin}{arg\,min}

\definecolor{teseoCol}{rgb}{.15, .68, .38}
\definecolor{kennyCol}{rgb}{.68, .15, .38}
\definecolor{YiboCol}{rgb}{.18, .68, .94}

\definecolor{reviewcol}{rgb}{0.2, 0.6, 0.86}
\newcommand{\review}[1]{{\leavevmode #1}}

\definecolor{revisioncol}{rgb}{0.54, 0.16, 0.88}

\definecolor{revisecol}{rgb}{0.54, 0.16, 0.88}
\newcommand{\revise}[1]{{\leavevmode #1}}

%% file: 01-abstract.tex
We propose mesh-free fluid simulations that exploit a kinematic neural basis for velocity fields represented by an MLP. We design a set of losses that ensures that these neural bases \review{approximate} fundamental physical properties such as orthogonality, divergence-free, boundary alignment, and smoothness. Our neural bases can then be used to fit an input sketch of a flow, which will inherit the same fundamental properties from the bases. We then can animate such flow in real-time using standard time integrators. Our neural bases can accommodate different domains, \review{moving boundaries}, and naturally extend to three dimensions.

%% file: 01-teaser.tex
\begin{teaserfigure}
    \centering
  \includegraphics[width=.19\linewidth,trim={0.15cm 0.1cm 0.2cm 0.15cm},clip]{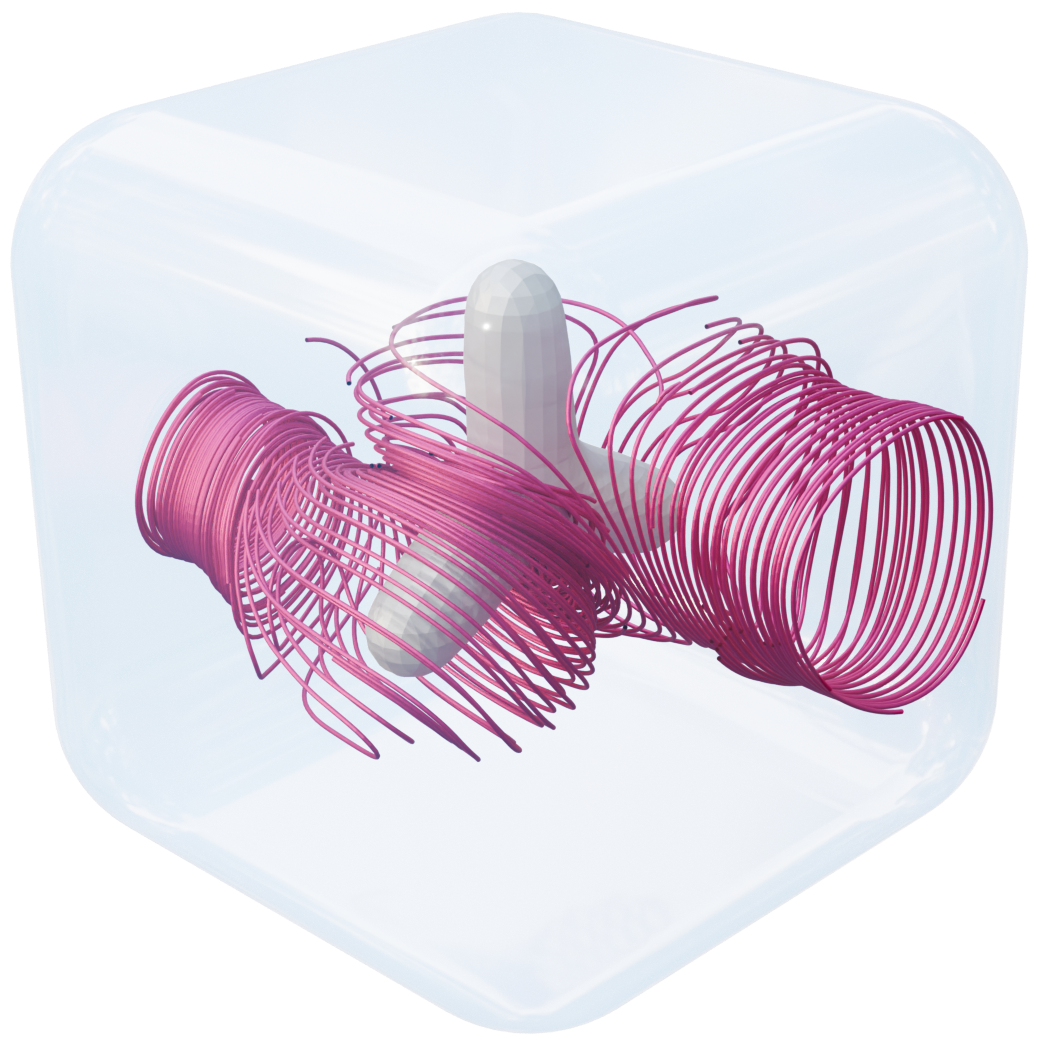}  \hfill
  \includegraphics[width=.19\linewidth,trim={0.15cm 0.1cm 0.2cm 0.15cm},clip]{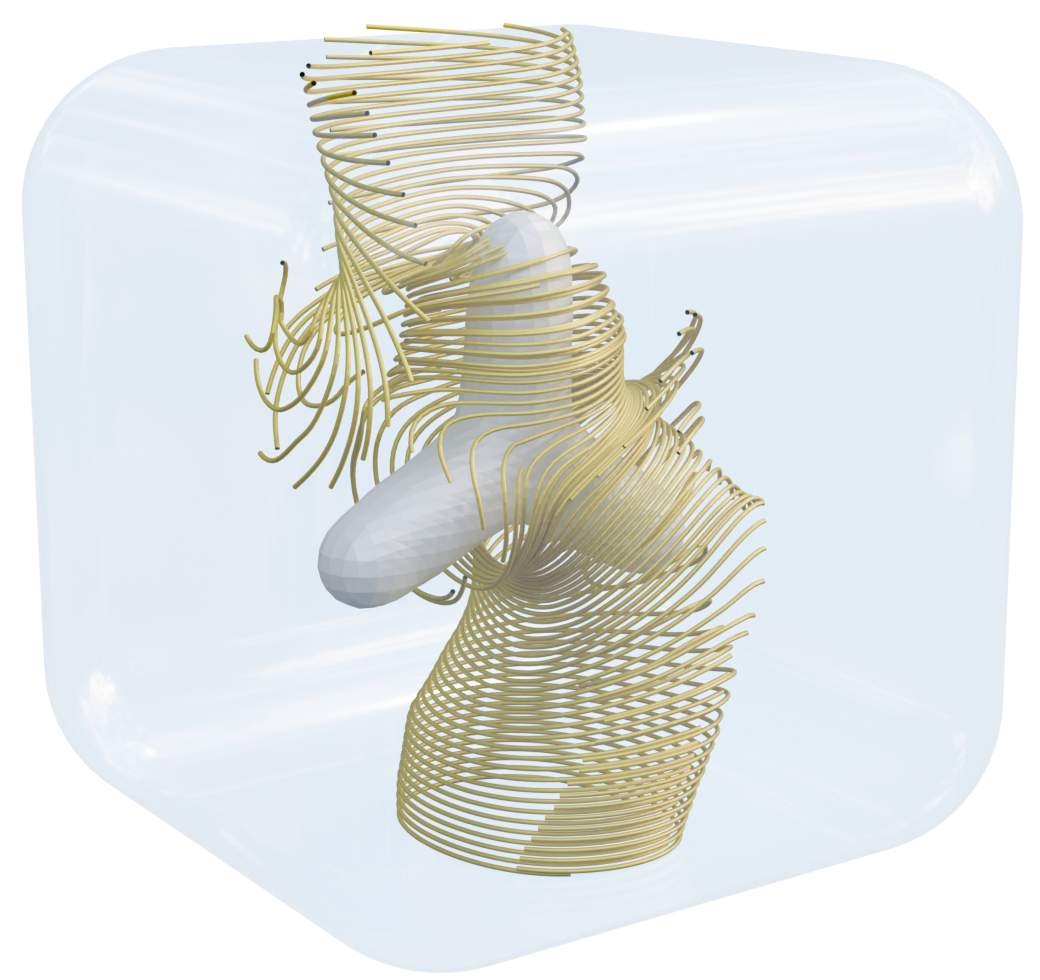}  \hfill
  \includegraphics[width=.19\linewidth,trim={0.1cm 0.1cm 0.2cm 0.15cm},clip]{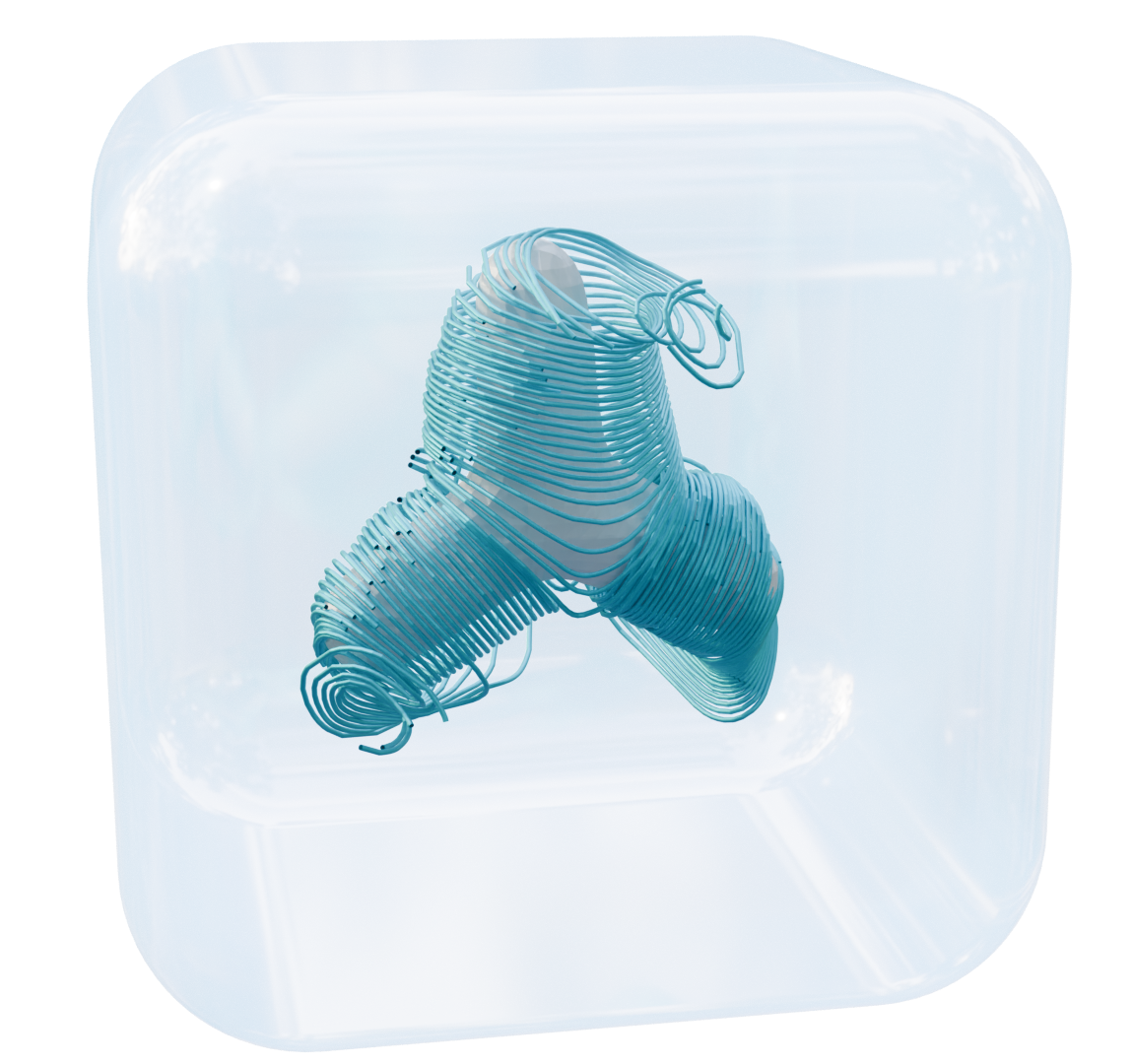}  \hfill
  \includegraphics[width=.19\linewidth,trim={0.15cm 0.1cm 0.1cm 0.15cm},clip]{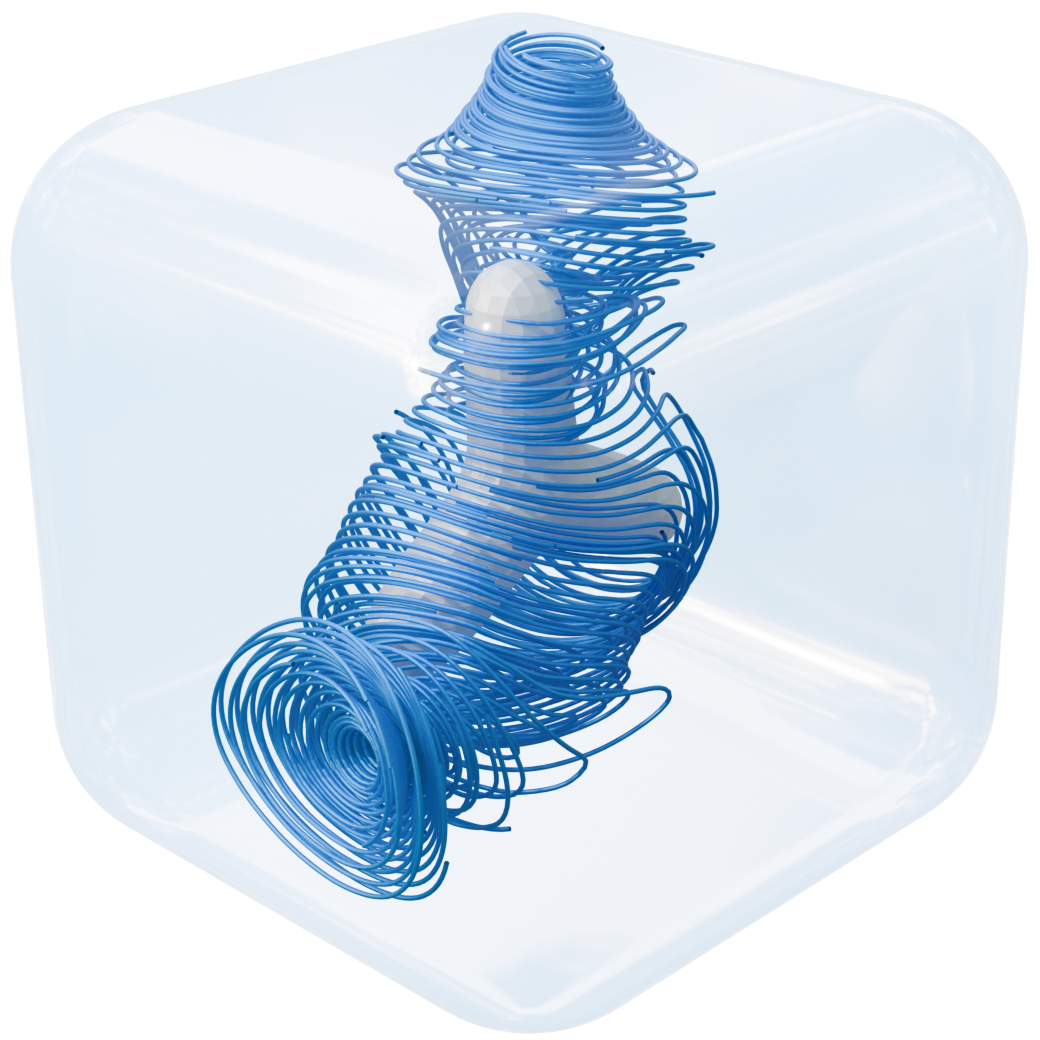}  \hfill
  \includegraphics[width=.19\linewidth,trim={0.15cm 0.1cm 0.2cm 0.15cm},clip]{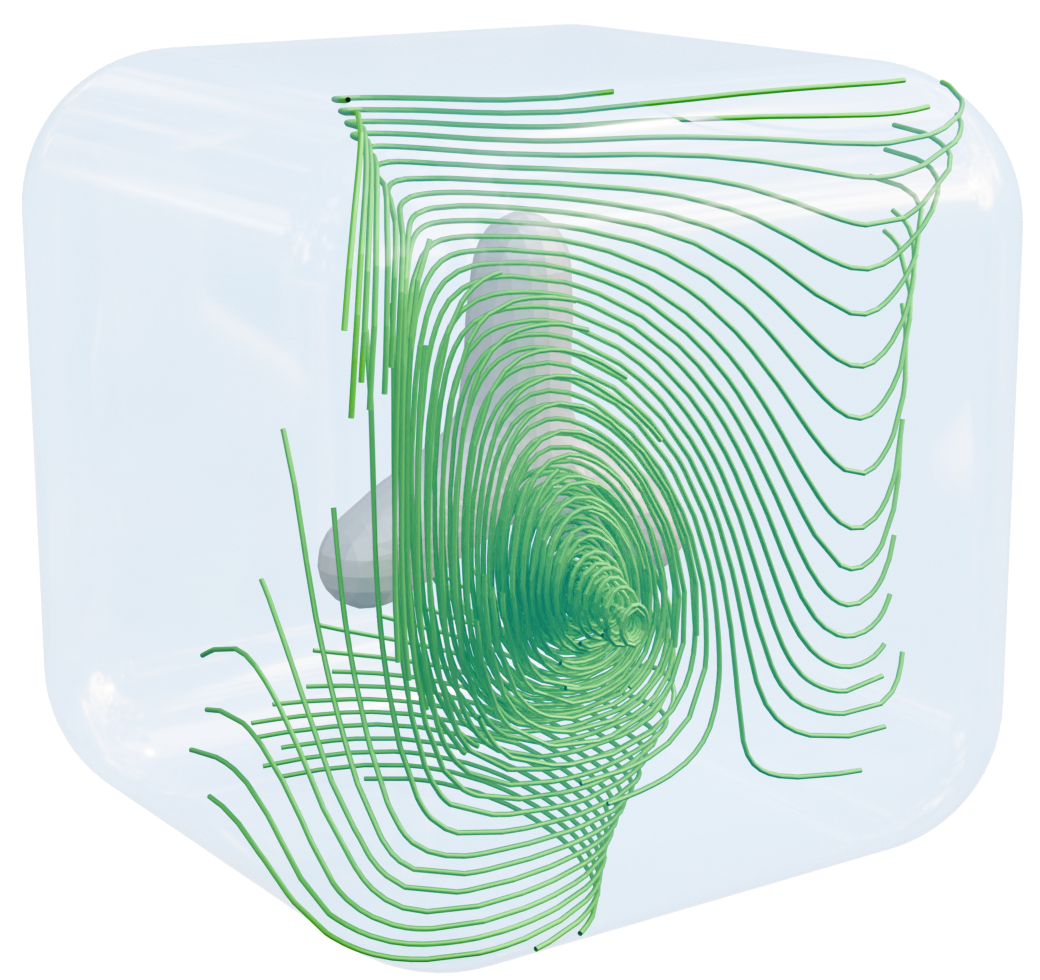} 
  \includegraphics[width=.19\linewidth,trim={0.15cm 0.1cm 0.2cm 0.15cm},clip]{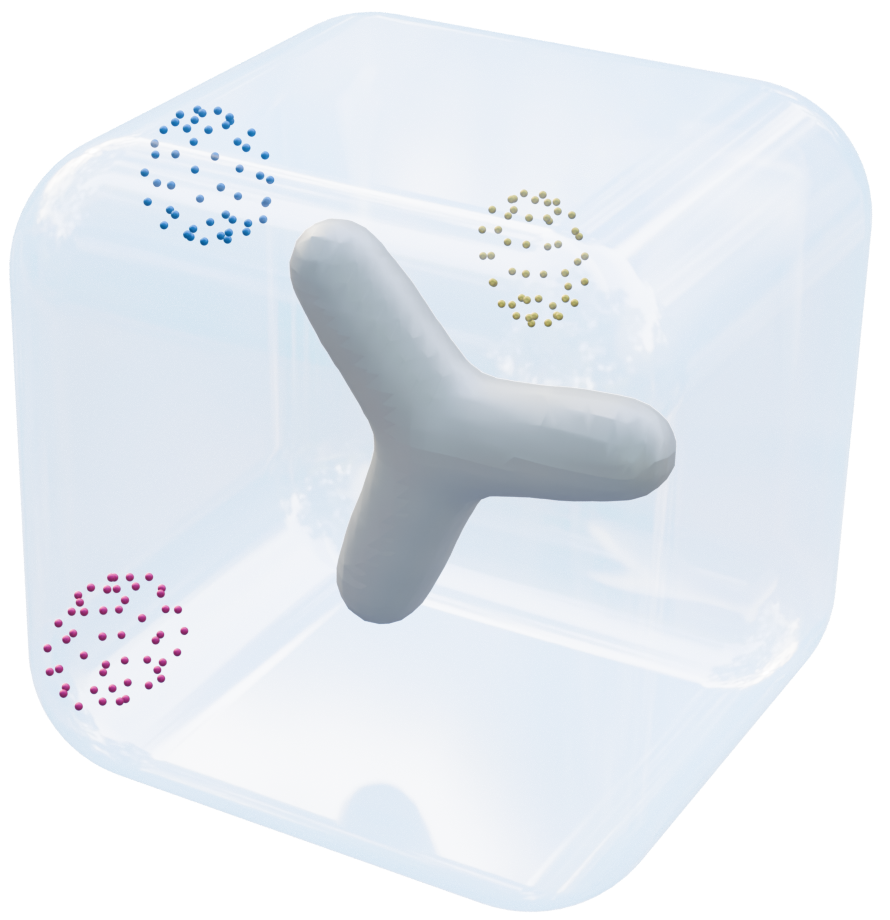}  \hfill
  \includegraphics[width=.19\linewidth,trim={0.15cm 0.1cm 0.2cm 0.15cm},clip]{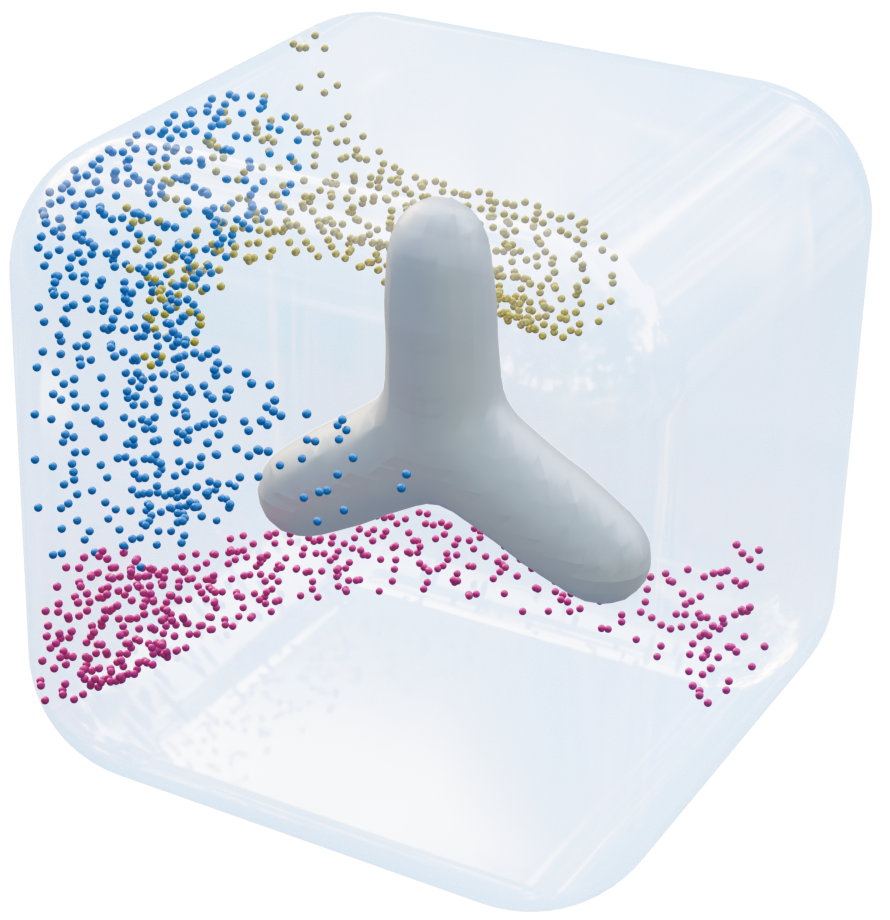}  \hfill
  \includegraphics[width=.19\linewidth,trim={0.15cm 0.1cm 0.2cm 0.15cm},clip]{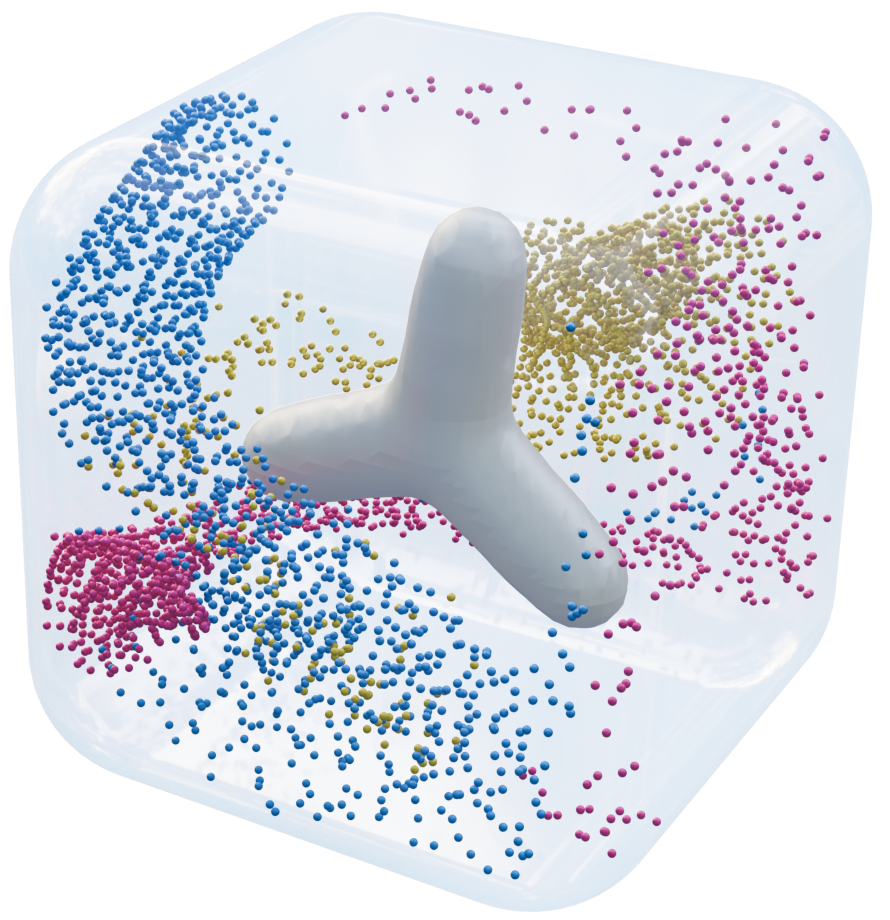}  \hfill
  \includegraphics[width=.19\linewidth,trim={0.15cm 0.1cm 0.2cm 0.15cm},clip]{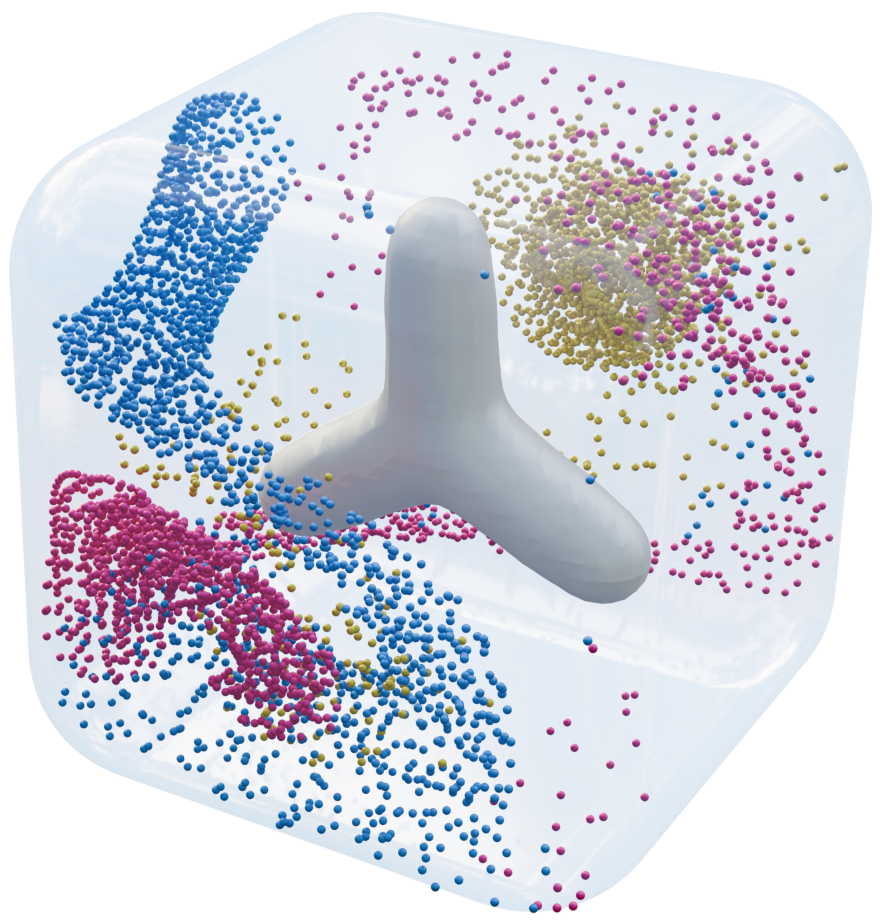}  \hfill
  \includegraphics[width=.19\linewidth,trim={0.15cm 0.1cm 0.2cm 0.15cm},clip]{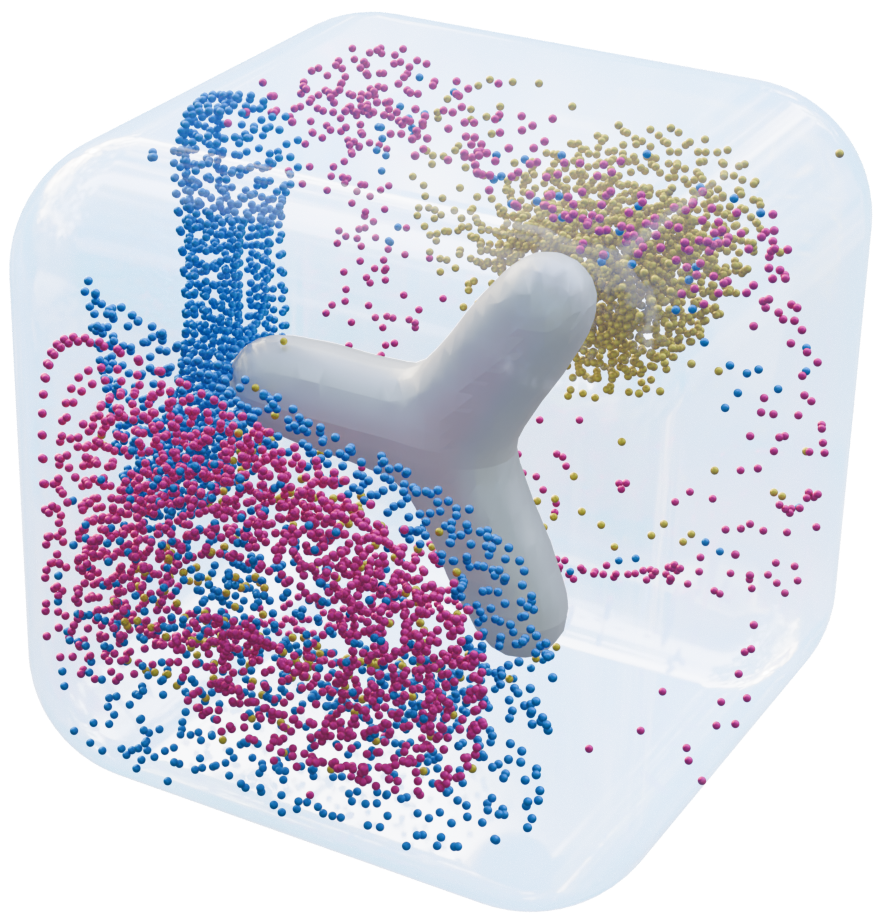} 
  \includegraphics[width=.19\linewidth,trim={0cm 0cm 0cm 0cm},clip]{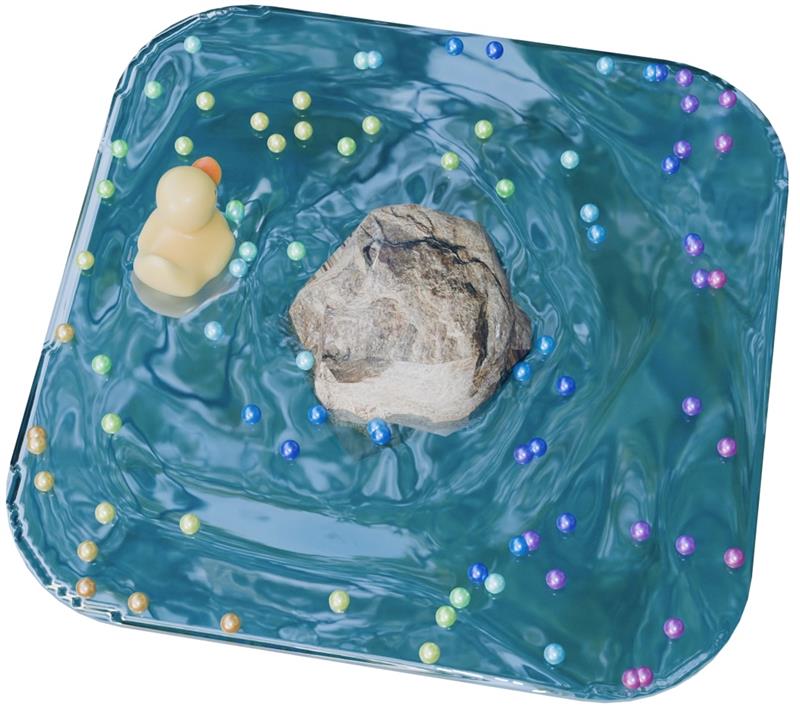}  \hfill
  \includegraphics[width=.19\linewidth]{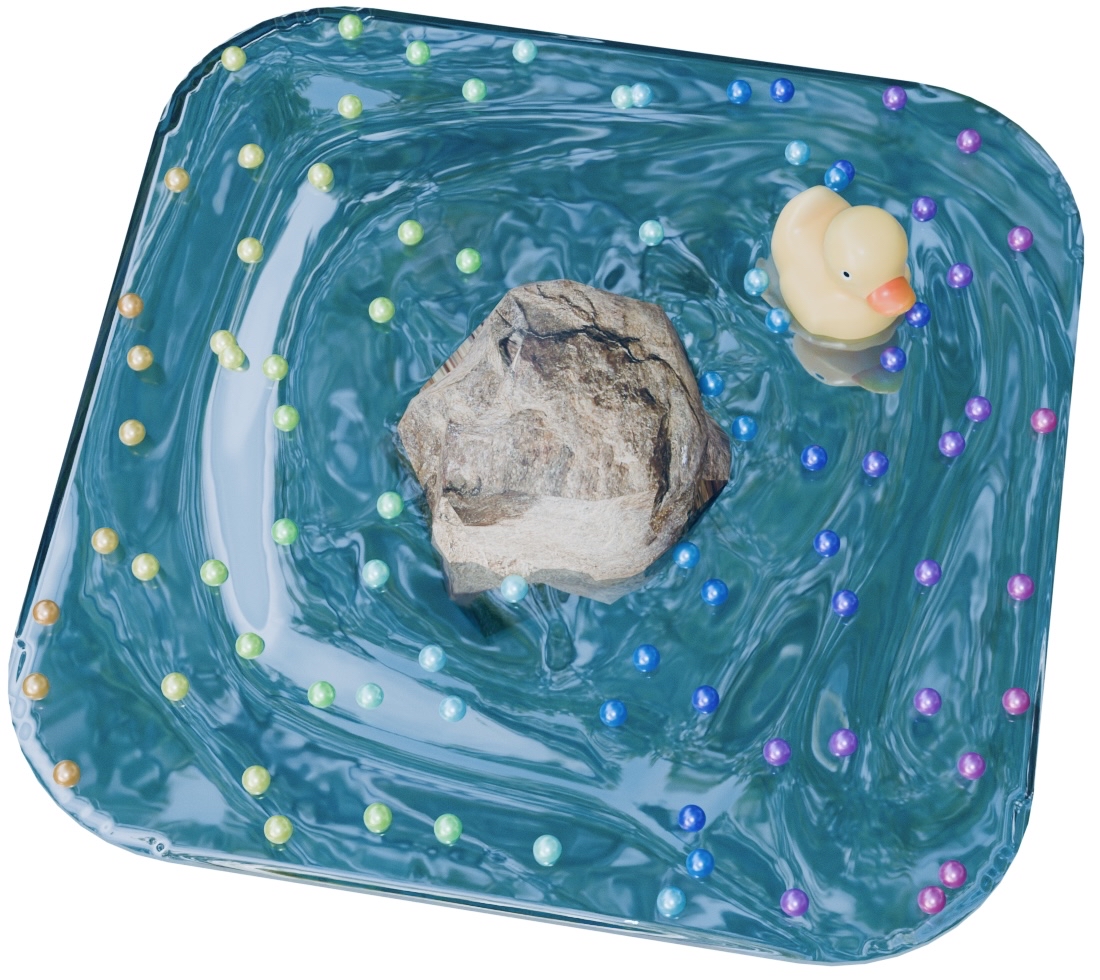}  \hfill
  \includegraphics[width=.19\linewidth]{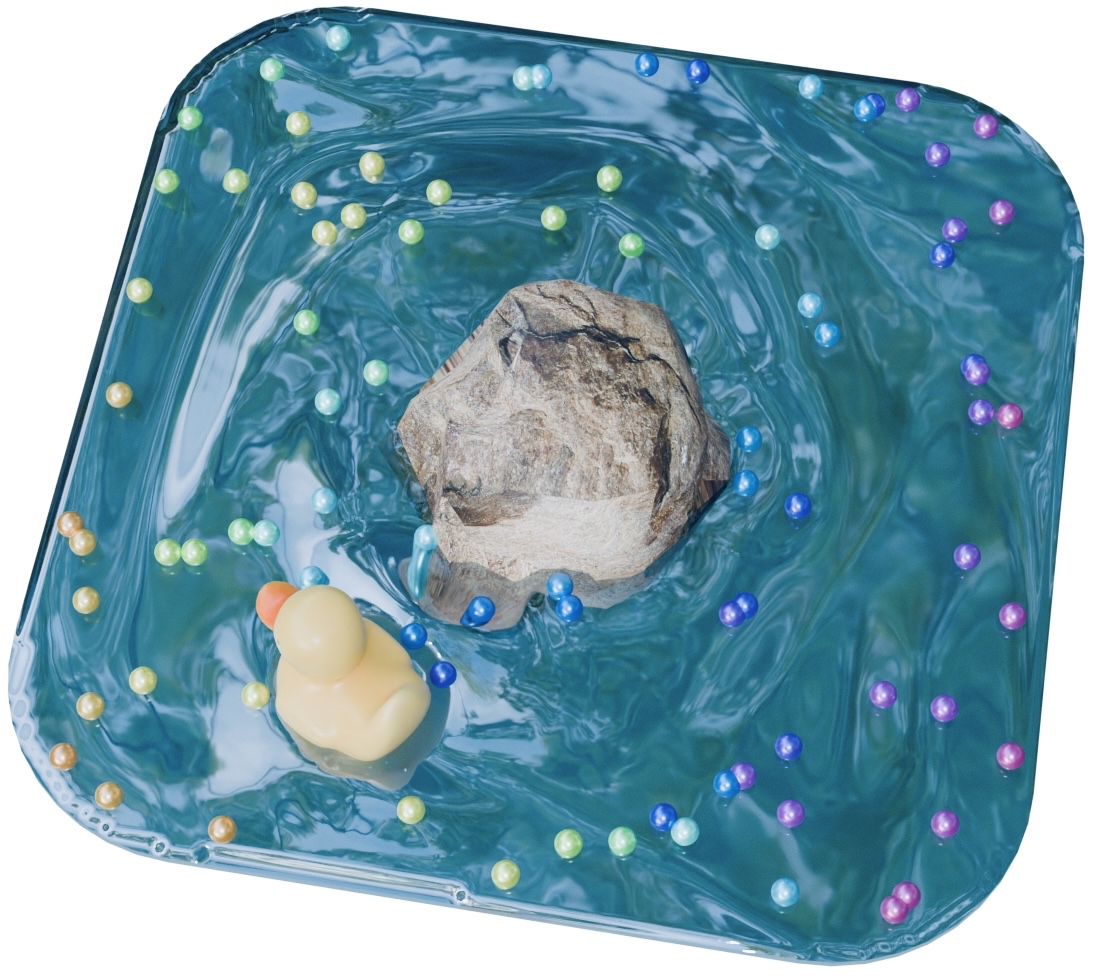}  \hfill
  \includegraphics[width=.19\linewidth]{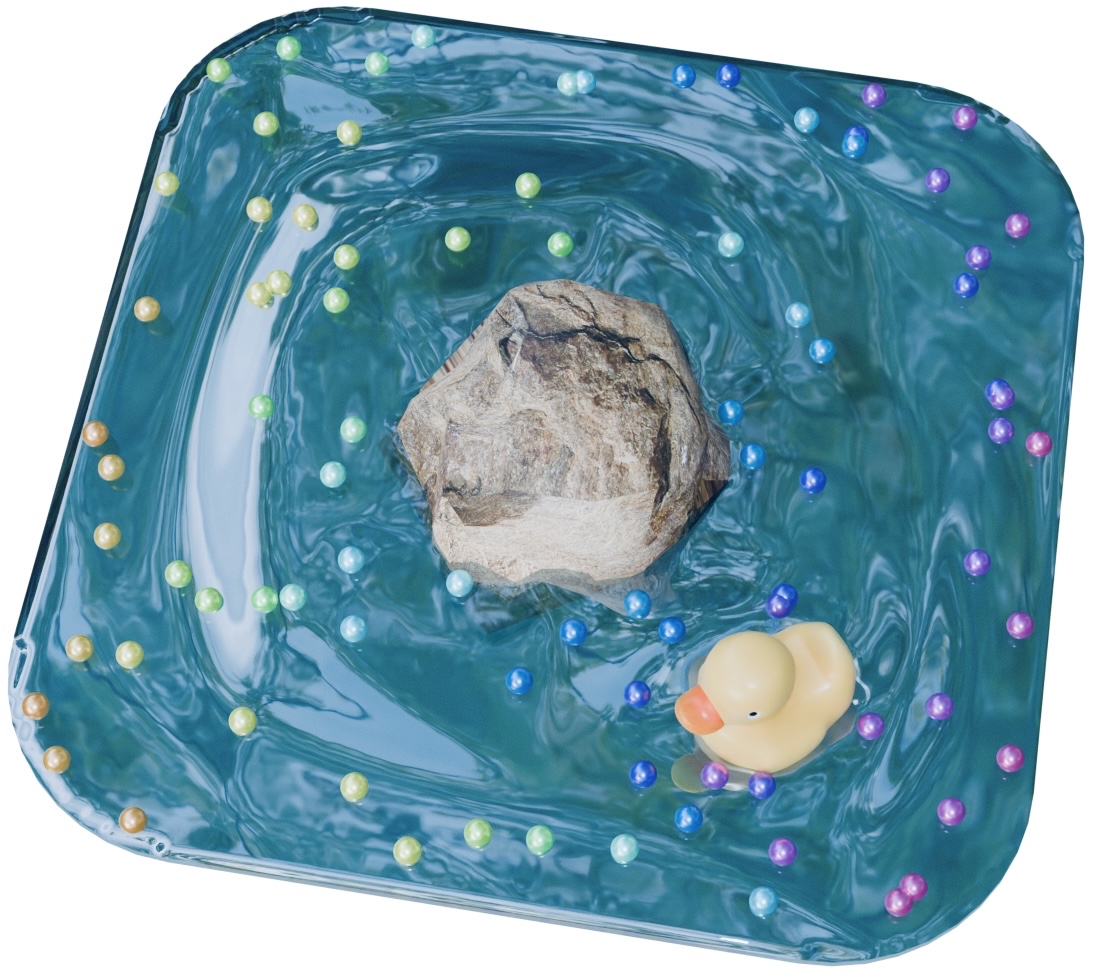}  \hfill
  \includegraphics[width=.19\linewidth]{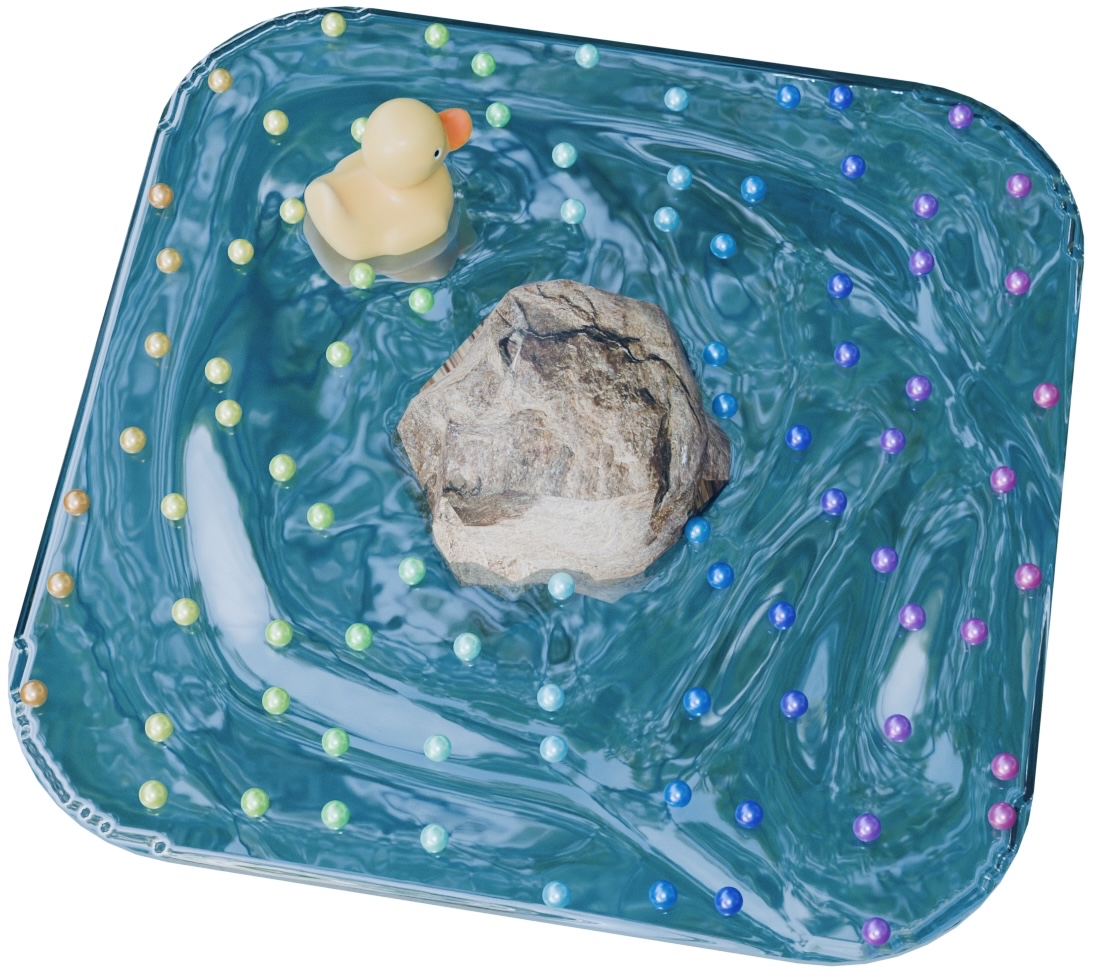}
    \caption{\review{We use neural kinematic bases (top) to fit a changing domain which we then simulate in real-time with standard semi-implicit advection (middle). \revise{The same construction also applied to two dimensions where we can simulate fluids in real time (bottom).}}}
    \label{fig:teaser}
\end{teaserfigure}

%% file: 02-intro.tex
\section{Introduction}

Animating 2D or 3D shapes from \emph{sparse} data is a complex problem at the core of computer graphics which started with pioneering work such as that of 
\citet{badler1982modelling}.
The main motivation is character animation, where the complex animation and deformation are driven by a skeleton (sparse data)~\cite{lewis2000pose}. The general problem can be abstracted as computing a sparse representation (or basis) that allows interactive animation and shape deformation. The main challenge is to ensure that any generated deformation looks physically realistic, which can be enforced by carefully designing the bases. With the advent of machine learning in character animation, the effort moved from creating bases to designing networks for such bases~\cite{li:2021:skel,Bertiche_2021_ICCV,Kavan:2024:skel}.

In parallel, machine learning enabled replacing complicated and costly accurate physical simulations by cheap networks~\cite{du:2023:dps}, in particular for fluid simulations (\Cref{sec:related}). However, these methods share the same ``problems'' as physical simulations: the input is the state of the system (initial and boundary condition), and the output is an animation. This setup lacks the fine-grained control present in animation pipelines and the ability to control it from sparse data.

We propose a novel neural kinematic basis that allows animators to quickly and interactively design and build fluid animation in two and three dimensions. Our idea is to borrow the interactivity and ease of traditional animation pipelines and combine them with the power of neural representation used in physical simulations. Similar to skeletal animation, we aim to design a basis that encodes the complex dynamics of the fluid. We represent our neural bases with an MLP, which we train purely using fundamental physical properties such as incompressibility,
smoothness, boundary alignment, and orthogonality; therefore, we do not require any ground truth. Our training data is small and is only used to allow our neural bases to generalize to the new unseen domains. Since we force the MLP network to encode fundamental laws, any animation generated from our bases will also inherit the same properties. Once we fit the initial flow, we can use standard integration techniques to develop a plausible fluid animation (\Cref{fig:teaser}).

%% file: 03-related.tex
\section{Related Work}
\label{sec:related}

Reduced-order models in fluid dynamics trace back to \citet{lumley.67}. In computer graphics, a similar concept using principal component analysis (PCA) was introduced by \citet{pentland.williams.89} as a method for reducing degrees of freedom in the deformation of solids. This foundational idea spurred extensive follow-up work, extending deformable models. For instance, \citet{barbic.james.05} proposed a fast subspace integration method for reduced-coordinate nonlinear deformable models, leveraging mass-scaled PCA to generate low-dimensional bases. They demonstrated that model reduction allows internal forces to be precomputed as cubic polynomials, enabling efficient simulations with costs independent of geometry. Building on PCA, \citet{treuille.lewis.ea06} extended the approach to fluid dynamics, constructing a reduced-dimensional velocity basis by applying PCA to velocity fields from full-dimensional simulations.

To address computational challenges in PCA-based methods, \citet{an.kim.ea08} introduced optimized cubature schemes for efficient integration of force densities associated with specific subspace deformations. Similarly, \citet{kim.james.09} developed an online, incremental reduced-order nonlinear model. Later, \citet{Kim2013resim} extended cubature schemes to efficiently perform consistent semi-Lagrangian advection within a fluid subspace, enabling re-simulation of fluid systems with modified parameters. In comparison, our work employs sparse integration techniques and semi-Lagrangian advection.

\Citet{von.tycowicz.schulz.ea13} accelerated the construction of reduced dynamical systems by approximating reduced forces; \review{\Citet{review10.1145/3197517.3201325} uses data-drive approach to quickly predict fluid behavior around 3D obstacles}. Our work also incorporates approximations, using smoothed boundary indicators and domain masks. Additionally, \citet{deWitt2012} represented fluids as linear combinations of eigenfunctions of the vector Laplacian, performing time integration via Galerkin discretization of the Navier-Stokes equations. While this method supported immersed rigid bodies by projecting out velocity components corresponding to boundary flux, \citet{Cui2018} extended the approach to handle Dirichlet and Neumann boundary conditions. They further improved scalability, following \citet{Jones2016}, by utilizing sine and cosine transforms to reduce storage demands for eigenfunctions. Inspired by Laplacian eigenfunction methods, \citet{Mercier2020} and \citet{Wicke2009} developed reduced basis functions on regular tiles, enforcing consistency constraints between adjacent tiles. Our approach adopts a basis-function perspective but employs neural representations instead of eigenfunctions. 
Comparing to \citet{Cui2018}, which is limited to generating bases in a single rectangular domain and handles obstacle boundaries via an explicit post-processing step, our neural network is trained to generalize across arbitrary domains and evaluates bases on-the-fly during simulation. While \citet{Cui2018} require at least 5.5 hours of precomputation per domain, our model is trained at a one-time cost, after which it supports unlimited simulations without domain-specific precomputation. By treating both obstacles and rectangular boundaries consistently during basis generation, our method naturally satisfies no-slip boundary conditions (Appendix \ref{sec:append}).

\citet{yang.li.ea15} identified modal matrix construction, cubature training, and dataset generation as key bottlenecks in traditional approaches. In contrast, our method trains basis functions to preserve geometric invariants in fluid simulations. Similarly, \citet{xu.barbic.16} precomputed separate reduced models for different object poses and combined them at runtime into a unified dynamic system. We build on this idea by utilizing local geometric invariants to address complex nonlinear spaces.

\citet{romero.casas.ea21} augmented linear handle-based subspace formulations with nonlinear, learning-based corrections to decouple internal and external contact-driven effects. In our work, contact handling is achieved by conditioning neural basis functions on contextual information. Likewise, \citet{aigerman.gupta.ea22} used neural networks to predict piecewise linear mappings of arbitrary meshes, incorporating smoothing techniques to handle discontinuities. We adopt a similar gradient-smoothing strategy to address these challenges. {Similar to \citet{revision10.1145/3588432.3591521}, which presents a reduced-order simulation method using unsupervised learning for neural network training, we design a set of physics-informed losses, inspired by \citet{pinn}, to enable unsupervised training of our neural network.

In the domain of Eulerian fluid configurations, \citet{Wu2023} learned latent space embeddings and applied linear time integration operators based on the sines and cosines of latent variables. Further, \citet{chen.xiang.ea23} and \citet{chang.chen.ea23} introduced discretization-independent reduced-order modeling, representing displacement fields as continuous maps encoded by implicit neural fields. Extending these ideas, \citet{chen.chiarmonte.ea23} applied neural fields to the material point method, while \citet{tao.ea24} proposed neural implicit reduced fluid simulations, using non-linear latent embeddings to capture fluid-fluid interactions.

Our work follows the neural field paradigm to represent learned basis functions, enabling robust and efficient reduced-order fluid simulations while preserving geometric and physical invariants.

%% file: 04-method.tex
\section{Method}

\begin{figure}
    \centering
      \includegraphics[width=.32\linewidth,trim={0.7cm 0.7cm 0.7cm 0.7cm},clip]{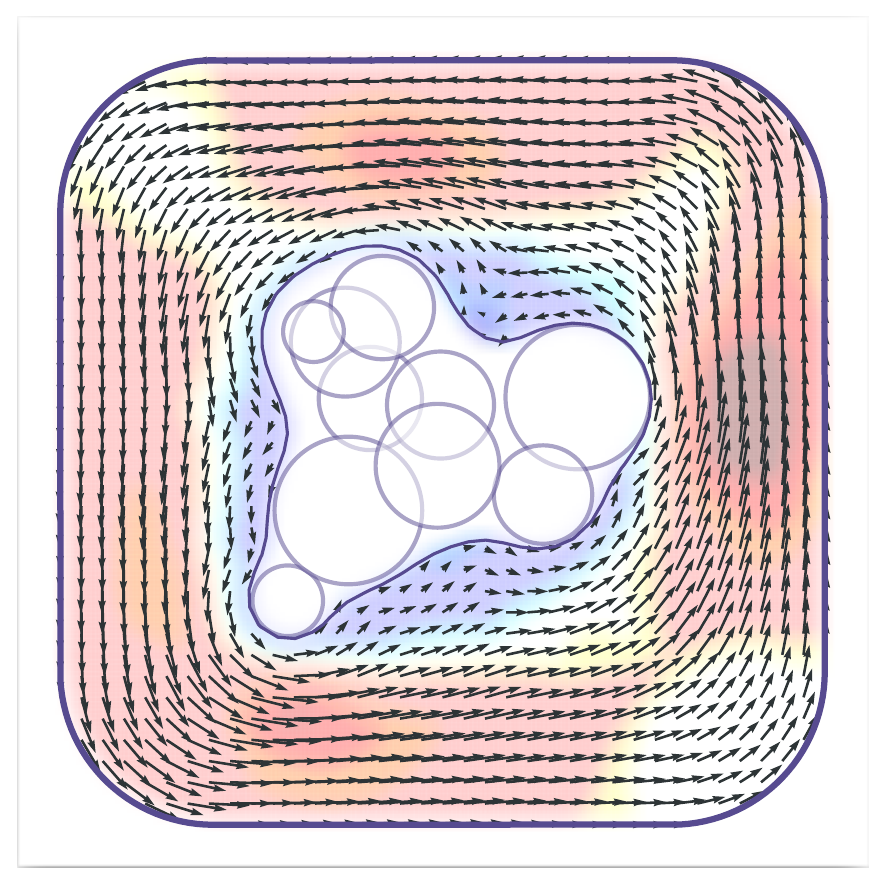} \hfill
      \includegraphics[width=.32\linewidth,trim={0.7cm 0.7cm 0.7cm 0.7cm},clip]{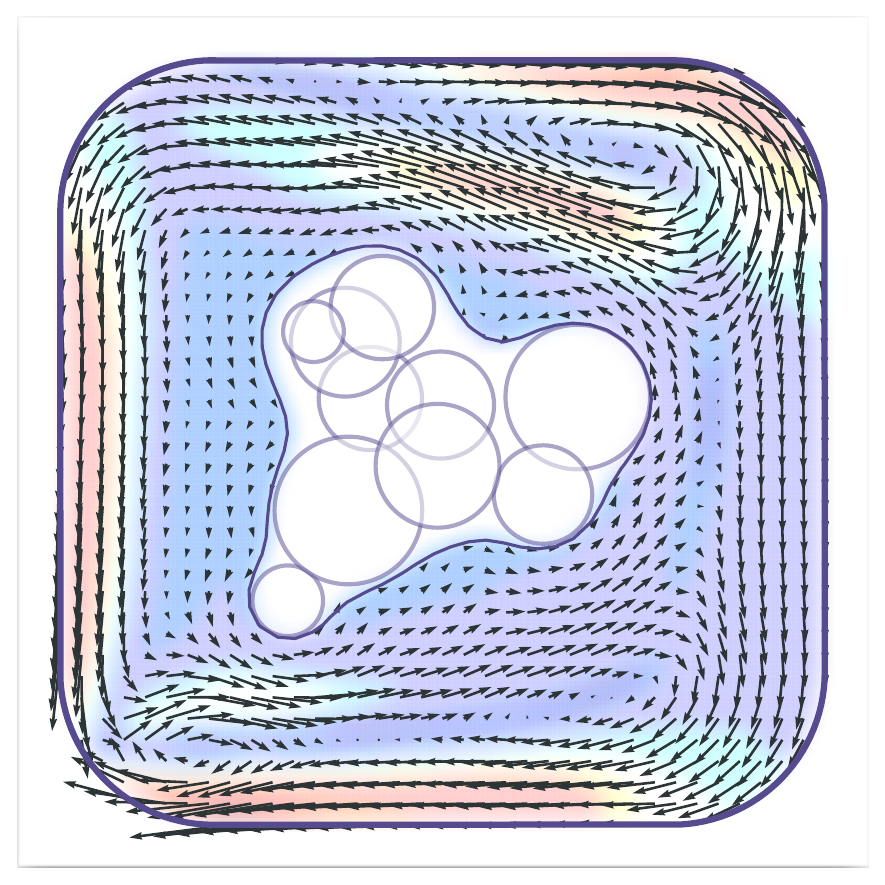} \hfill
      \includegraphics[width=.32\linewidth,trim={0.7cm 0.7cm 0.7cm 0.7cm},clip]{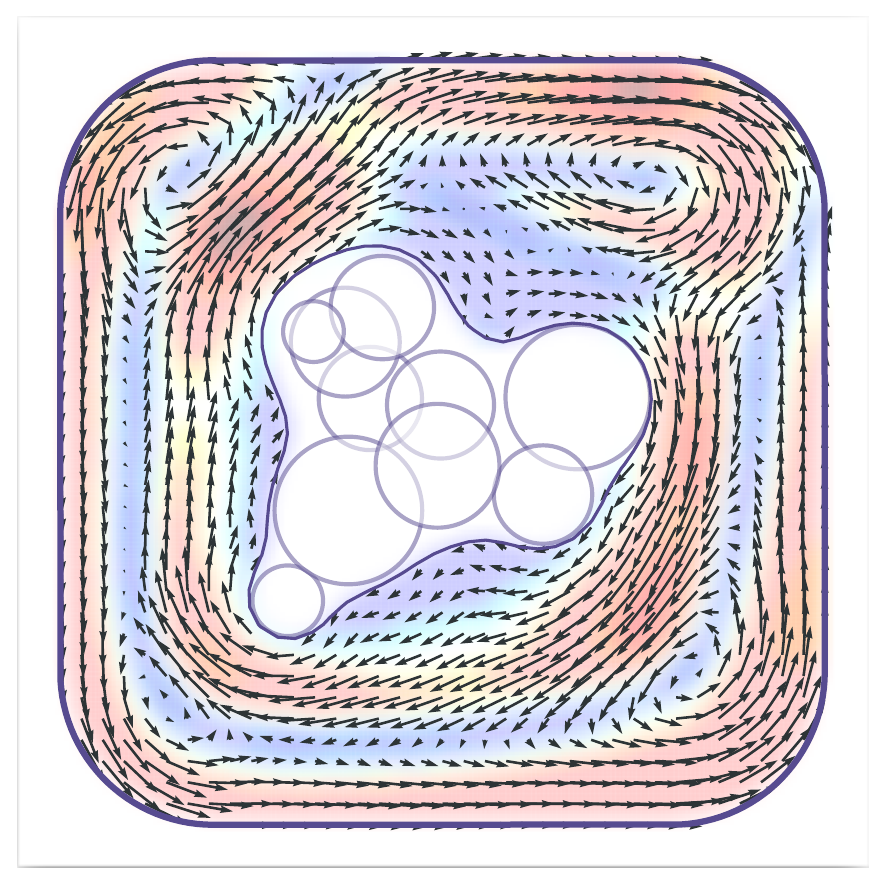} 
      \\[0.5em]
     \includegraphics[width=.32\linewidth,trim={0.7cm 0.7cm 0.7cm 0.7cm},clip]{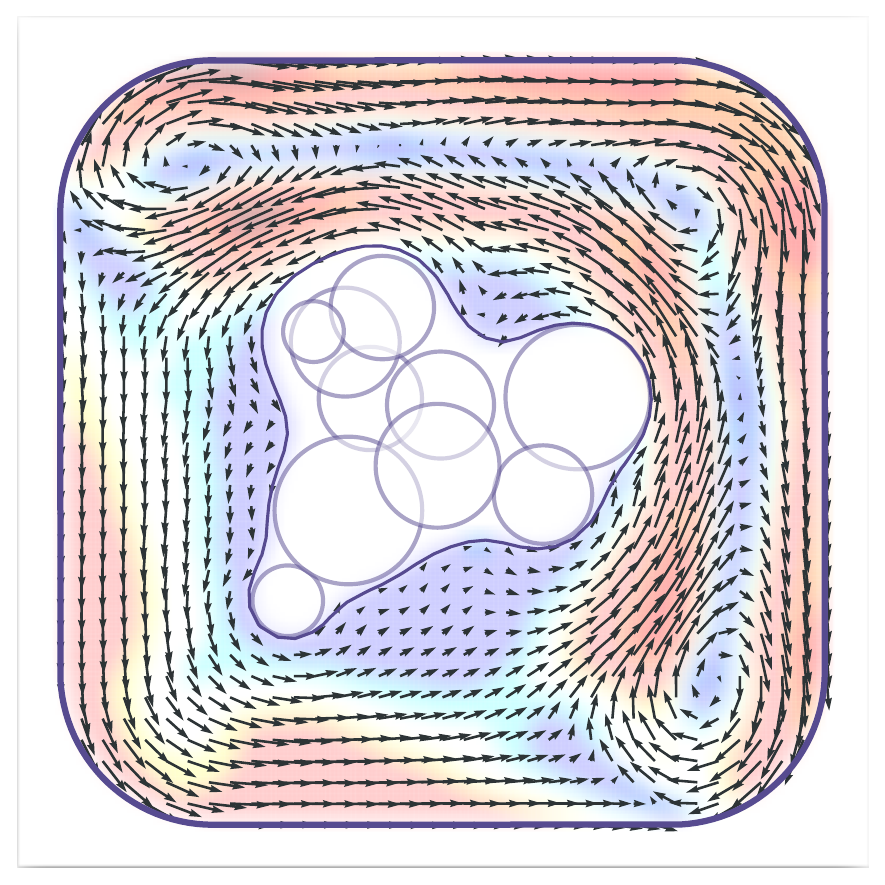} \hfill
     \includegraphics[width=.32\linewidth,trim={0.7cm 0.7cm 0.7cm 0.7cm},clip]{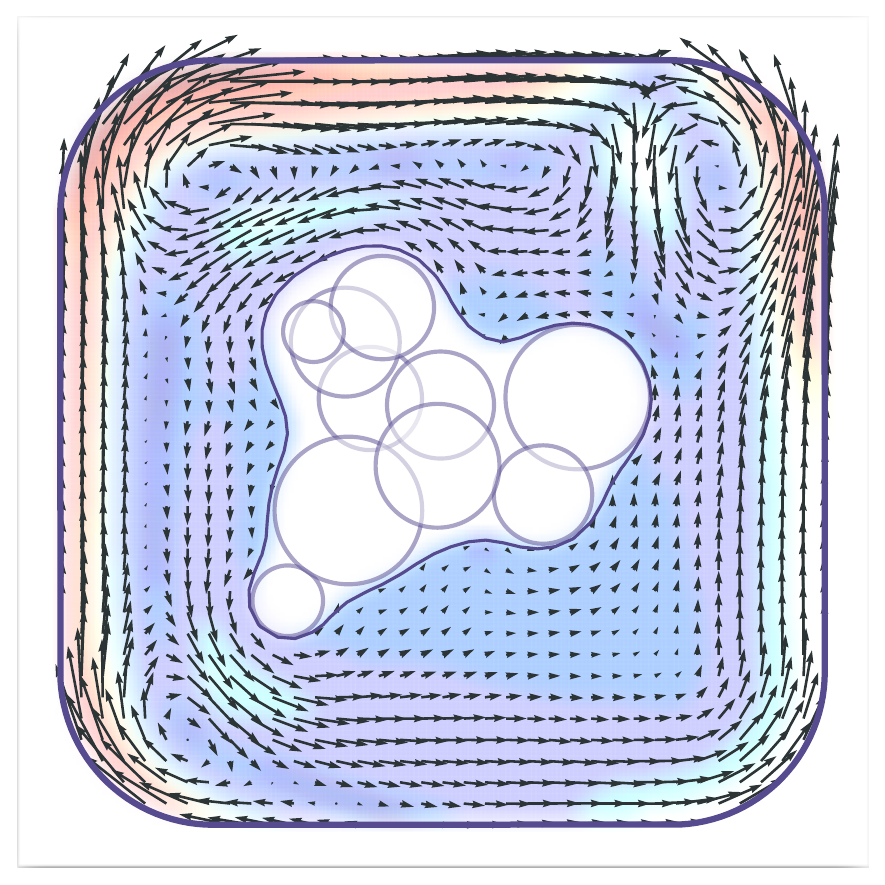} \hfill
      \includegraphics[width=.32\linewidth,trim={0.7cm 0.7cm 0.7cm 0.7cm},clip]{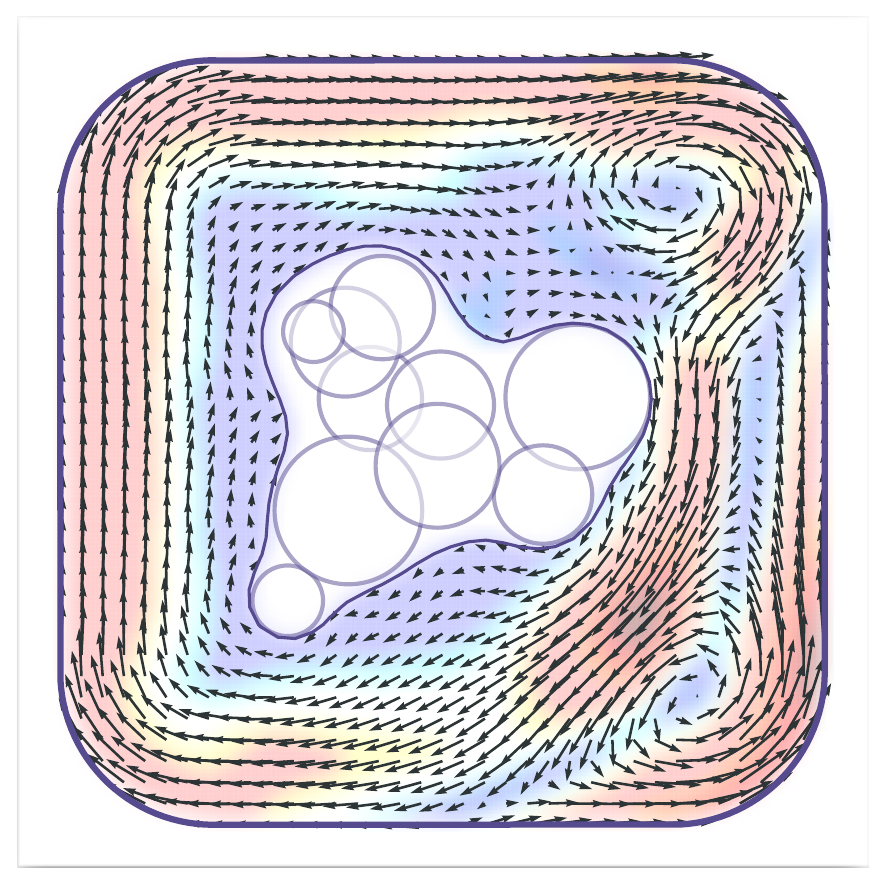} \\[0.5em]
      
     \includegraphics[width=.32\linewidth,trim={0.7cm 0.7cm 0.7cm 0.7cm},clip]{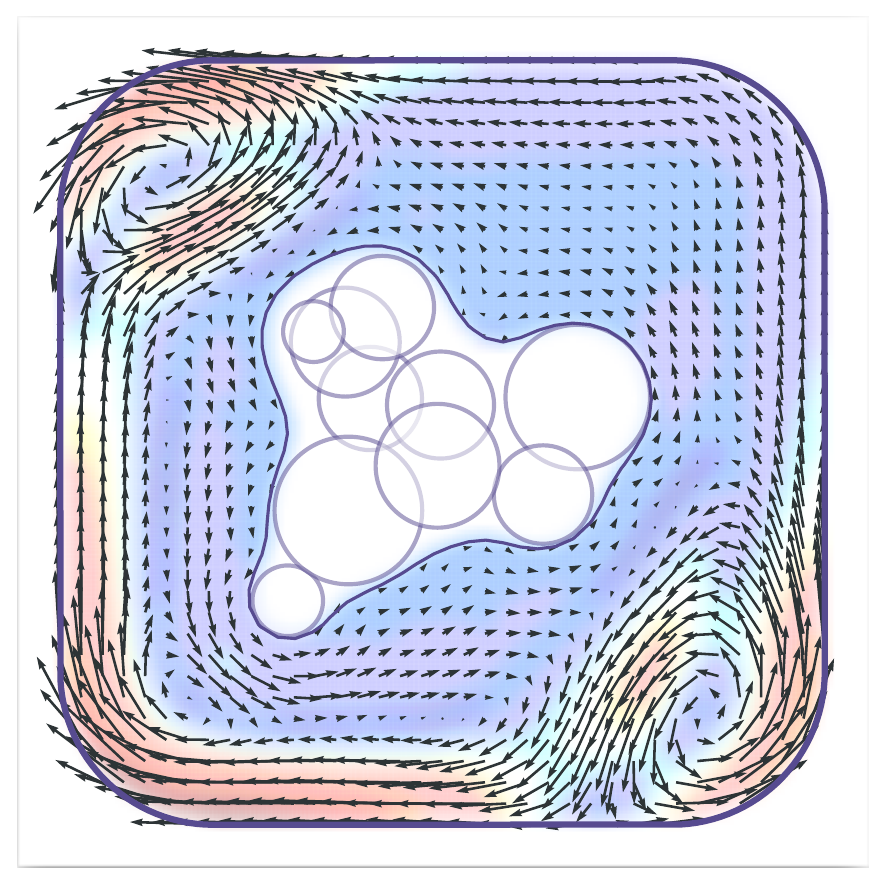} \hfill     \includegraphics[width=.32\linewidth,trim={0.7cm 0.7cm 0.7cm 0.7cm},clip]{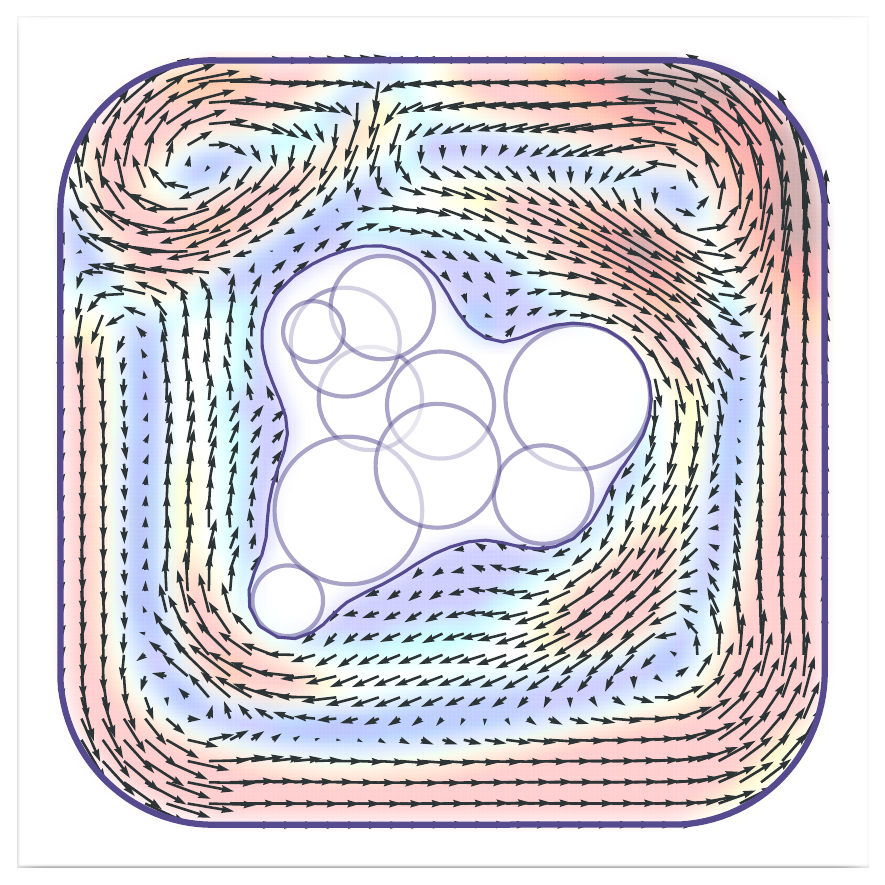} \hfill
      \includegraphics[width=.32\linewidth,trim={0.7cm 0.7cm 0.7cm 0.7cm},clip]{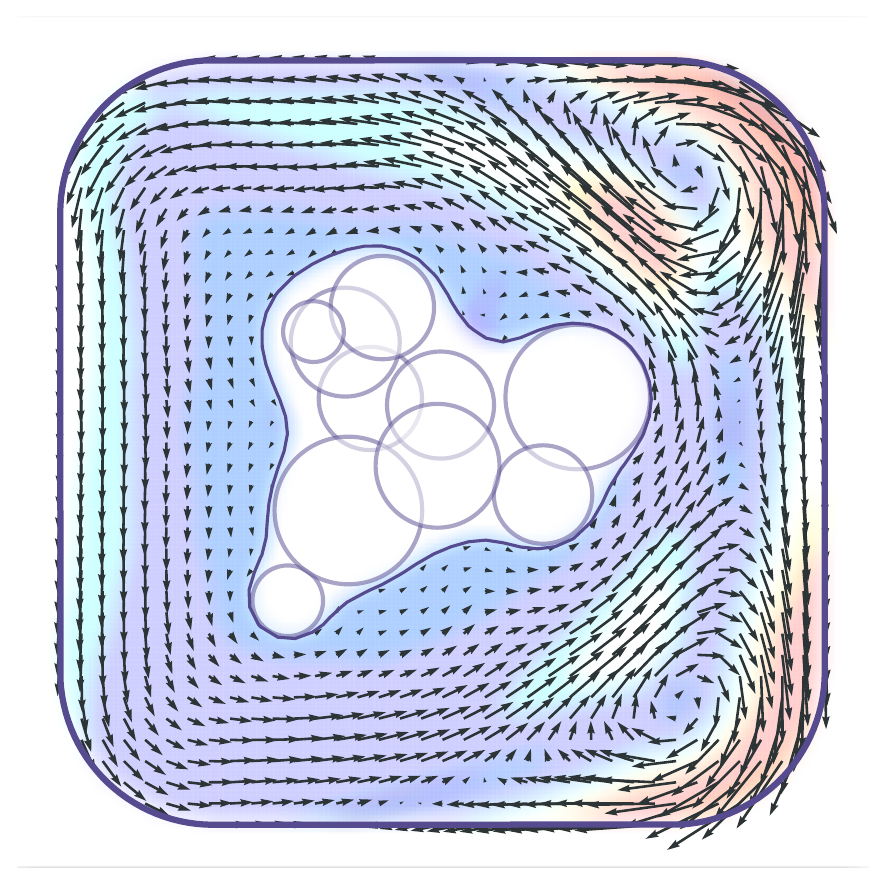}  
      
    \caption{Example of neural basis functions generated by our MLP network. We clearly see that they are parallel to the boundary, non-zero, and smooth.}
    \label{fig:bases}
\end{figure}

The core idea of our approach is to design a set of nonlinear neural basis functions (\Cref{fig:bases}) that respect common invariants in fluid flow problems. Observe that our neural bases, $\varphi_i$, are vector fields. This is a little different from finite element approaches where the shape functions would be scalar functions and the unknown coefficient would be vectors living at nodal points. Let $\Omega$ be a unit rounded square domain with $m$ potentially overlapping circular holes (\Cref{fig:bases}). On this domain, we define $b$ neural basis functions $\vo_k, k=1\dots b$ that satisfy the following common invariants.

\emph{Divergence.} Crucially, for incompressible fluid simulation (inviscid and otherwise), we require that our basis can reconstruct divergence-free velocity fields:
\[
 \text{div}(\vo_k) = 0, \qquad \forall k=1\dots b.
\]

\emph{Boundary.} To ensure that the fluid remains inside the domain, we restrict our neural bases to satisfy the slip boundary condition
\[
 \langle n , \vo_k \rangle = 0, \qquad \forall k=1\dots b,
\]
with $n$ the normal on the boundary.

\emph{Orthonormality.} Bases need to be linearly independent, additionally, we require them to be orthogonal to each other,
\[
\int_{\Omega} \langle \vo_k , \vo_l \rangle = \delta_{k,l}, \qquad \forall k,l=1\dots b.
\]

Using these bases, we can approximate any field velocity
\begin{equation}
    \label{eq:def:basis:functions}
    v(p)  = \sum_{k=1}^b \vo_k(p) \alpha_k
\end{equation}
which will obviously satisfy the above invariants and lead to a plausible fluid simulation that can be integrated with standard semi-implicit time integrators. Here, $\ai$ are the unknown coefficients we solve for during simulation. Their role will be similar to the weights of eigenmodes or coefficients of a finite element method. Note that, differently from traditional finite element bases, the velocity $v$ will satisfy common invariants independently from the choice of $\ai$. Our $\ai$ does not live at a nodal position in space but exists in an abstract global setting.  Our choice aggressively reduces the degrees of freedom.

\begin{figure}
  \centering
  \includegraphics[width=\linewidth]{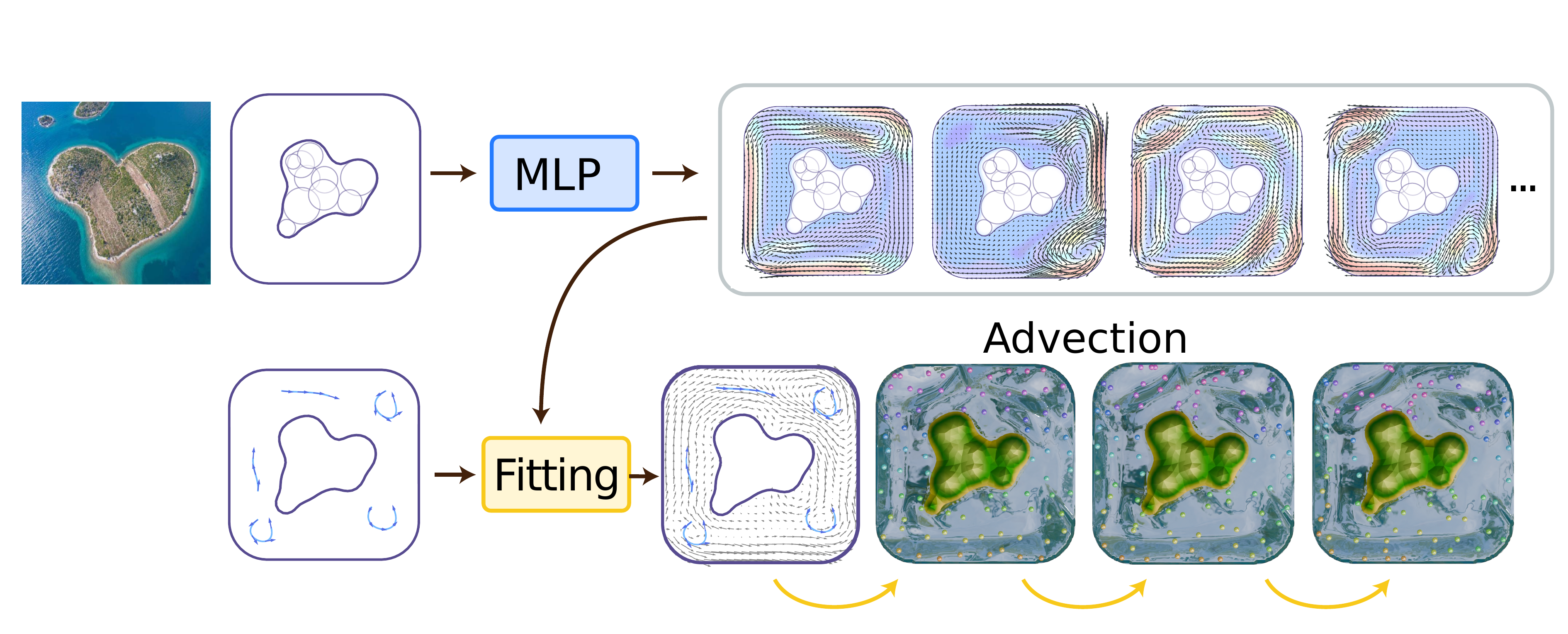}
  \caption{Overview of the pipeline of our method. We start with a sketch and our pre-trained MLP that generates fluid bases. We fit the bases to the input sketch which we then advect to generate an animation.}
  \label{fig:pipeline}
\end{figure}

Our goal is to learn a set of neural basis functions $\vo_i$ dependent on both the domain (which we encode with a set of circles) and the evaluation point, such that we can use \eqref{eq:def:basis:functions} as the \emph{kinematic neural basis} for fluid simulation.
By doing so, we can quickly evaluate specific bases (as the evaluation is done at inference time) for a given domain and thus generate fluid animation in real-time. Our design pipeline (\Cref{fig:pipeline}) starts with an input sketch of the domain $\Omega$ with a set of curves to guide the flow which we fit to our neural bases (\Cref{sec:fitting}). To obtain the neural bases, we define a set of losses (\Cref{sec:losses}) that produces bases that approximate fluid invariants in average, which we train (unsupervised) on different domains (\Cref{sec:training}). Finally, we advect the initial fluid (\Cref{sec:advection}) to generate an animation of the fluid.

\subsection{Domain}
\review{
We encode the domain with an unit square with rounded corners with radius 0.2 and an obstacle blob. To keep the representation simple, we use ten implicit circles (or spheres in 3d)
\[
f_i(p)=\|p-c_i\|^2 - r_i^2 \quad\text{with}i=1,\dots,10,
\] 
where $p$ is a point on the plane (or space), $c_i$ is the circle's center, and $r_i$ is its radius. We combine them by taking inspiration from metaballs or blobs~\cite{blinn1982generalization} to generate the implicit function
\[
f(p)=\ln\big(\sum_{i=1}^{10}e^{-k f_i(p)}\big)/k,
\]
where $k$ controls the smoothness (in our results, we always use $k=30$).}

\begin{figure}
  \centering \footnotesize
  \parbox{.03\linewidth}{\rotatebox{90}{\centering Weight}}
    \parbox{.09\linewidth}{\includegraphics[width=\linewidth]{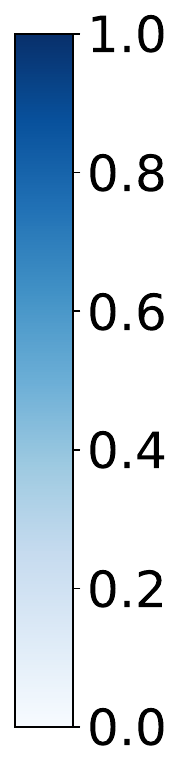}}\hfill\hfill
  \parbox{.43\linewidth}{\includegraphics[width=\linewidth,trim={0.7cm 0.7cm 0.7cm 0.7cm},clip]{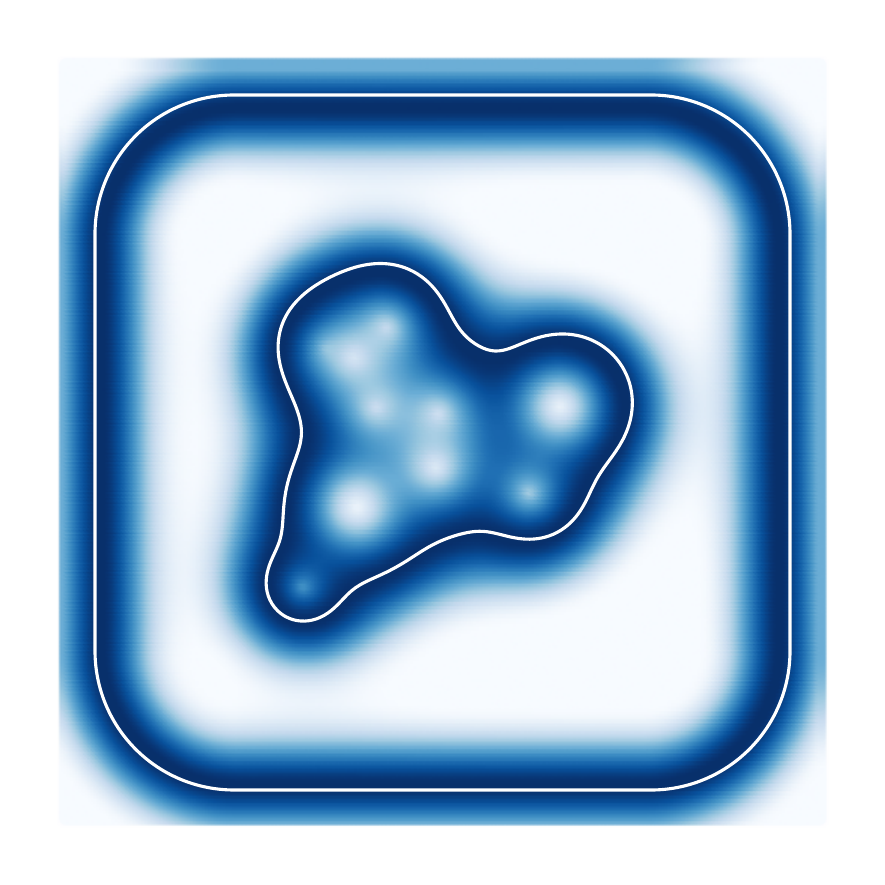}} \hfill
  \parbox{.43\linewidth}{\includegraphics[width=\linewidth,trim={0.7cm 0.7cm 0.7cm 0.7cm},clip]{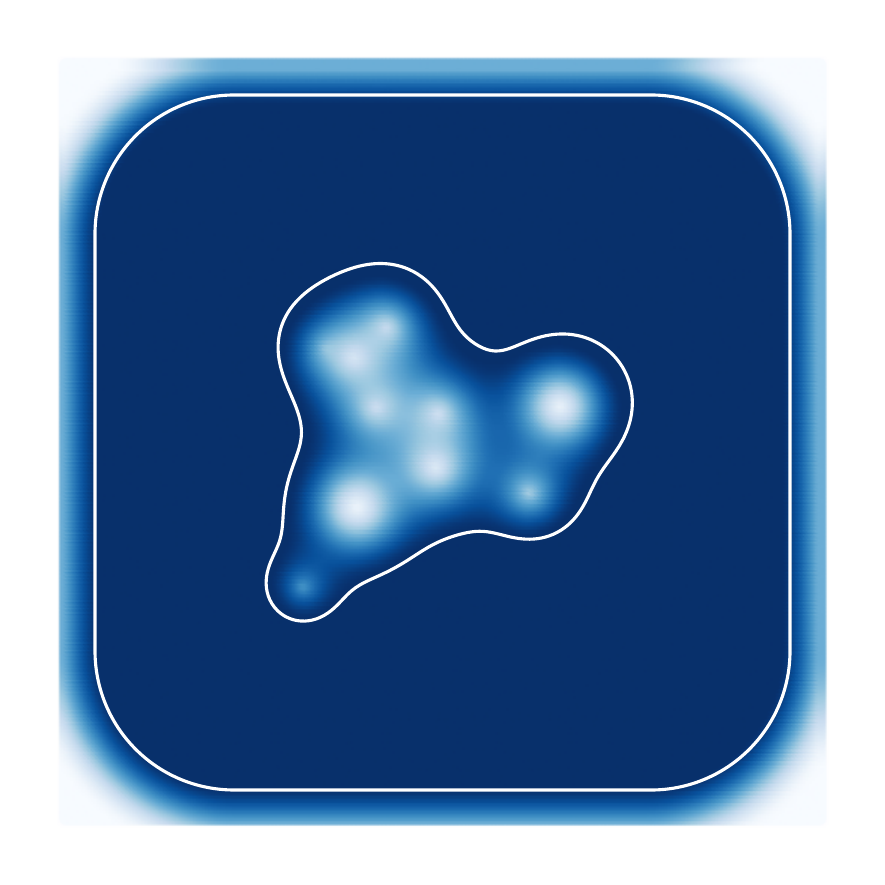}}

  \caption{\review{Boundary indicator function $w_b$ (left) and domain mask $w$ (right).}}
  \label{fig:mask}
\end{figure}

Since we sample the plane independently from the domain, we apply a soft mask to filter out the out-of-domain points. We first define a boundary indicator function (\Cref{fig:mask}, left)
\[
d(p)=\bigg(\frac{2 \lvert f(p) \rvert ^3}{\varepsilon^3}
- \frac{3 \lvert f(p)\rvert^2}{\varepsilon^2} + 1\bigg)
\]
\[
w_b(p)=
\begin{cases}
0, &\lvert f(p)\rvert > \epsilon \text{ or } d(p) < 0\\
d(p)^4 &\text{otherwise}.
\end{cases}
\] Then, we define the mask $w$ (\Cref{fig:mask}, right) by setting all values of $w_b$ out of the domain to zero and the values in the domain to 1. This function is one inside the domain and drops to zero for any point farther than $\varepsilon$ from the boundary. Note that the function is non-zero on a small region outside the domain, and therefore, we approximate our integral on the $[-\varepsilon, 1+\varepsilon]^2$ domain. To ensure that our losses are independent from the number of points and their position, we normalize them by
\[
S = Wb, \qquad\text{with}\qquad W = \sum_{i=1}^m w(p_i).
\]

\subsection{Losses}\label{sec:losses}
We will proceed by defining a set of losses that measure the physical appropriateness of our kinematic neural basis. Our basis neural fields $\vo_i$ will be computed by minimizing these losses in aggregate over a collection of $n$ randomly sampled points $p_i$. This sampling can be seen as using Monte Carlo integration. The main advantage of our approach is that we purely rely on fundamental physical properties (e.g., divergence-free or slip boundary conditions) and do not require any ground-truth data; we only require the bases to generalize over the domain.

The input of our MLP network $F$ is an evaluation point $p\in\RR^2$ and the center $c$ and radius $r$ of $m$ circles; it produces a set of $b$ basis functions $\vo_k(p)$. That is,
\[
F(p, \rho, \theta) =
    \{\vo_k(p)\}_{k=1,\dots, b},
\]
where $\rho = \{c_i, r_i\}_{i=1}^m$ is the circle set, $\theta$ the MLP parameters, and $\vo_k(p)$ our neural bases evaluated at $p$.

We can now encode the divergence invariant as an \review{average}, using the loss
\begin{equation}
    \label{eq:def:divergence:term}
    \mathcal{L}_{\text{div}}(\theta)
    =
    {\sum_{k=1}^b\sum_{i=1}^n
    \text{div}(\vo_k(p_i))^2 w(p_i)}/S.
\end{equation}
Interestingly, we do not require the individual $\vo_i$ to be divergence-free but wish for an expressive basis that can produce divergence-free fields (e.g., under different boundary conditions).

Next, we formulate a loss for the average slip boundary conditions
\begin{equation}
    \label{eq:def:boundary:term}
    \mathcal{L}_{\text{bc}}(\theta)
    =
    {\sum_{k=1}^b\sum_{i=1}^n
    \text{cossim}(\vo_k(p_i), n(p_i))^2 w_b(p_i)}/S_b
\end{equation}
where
\[
\text{cossim}(a,b)=\frac{\langle a, b\rangle}{\|a\|\|b\|}
\]
is the cosine similarly, and $n(p_i)$ is the normal of the closest point on the boundary of the domain. Since this is a discretization of a boundary integral (with a different weighting function), we normalize this loss by $S_b = b \sum_i w_p(p_i)$.

Finally, to prevent all bases from being the same, we enforce an average orthogonality using
\begin{equation}
    \label{eq:def:ortho:term}
    \mathcal{L}_{\text{orth}}(\theta)
    =
    \sum_{i=1}^n\sum_{k=1}^b\sum_{l=k+1}^b
    \langle \vo_k(p_i), \vo_l(p_i) \rangle w(p_i)/ S_o.
\end{equation}
Instead of normalizing by the number of bases, we normalize by the number of pairs $p$ ($S_o = Wp$). We note that this loss does not require that the bases have unit length $\langle \vo_k(p_i), \vo_k(p_i) \rangle = 1$ as the third sum starts at $k+1$. Instead, we explicitly require the average length of the vectors by requiring that the length is close to a target value $c$
\begin{equation}
    \label{eq:def:length:term}
    \mathcal{L}_{\text{len}}(\theta)
    =
    {\sum_{k=1}^b\bigg(
        \sum_{i=1}^n
        \|\vo_k(\theta, p_i)\|w(p_i)/W - c
        \bigg)^2}/b,
\end{equation}
and penalize small bases
\begin{equation}
    \label{eq:def:small:term}
    \mathcal{L}_{\text{small}}(\theta)
    =
    \sum_{k=1}^b\sum_{i=1}^n
    \text{ReLU}(
        \delta - \|\vo_k( p_i)\|)w(p_i)/S.
\end{equation}

Finally, to facilitate the learning process, we also encourage smoothness of the basis by adding
\begin{equation}
    \label{eq:def:smooth:term}
    \mathcal{L}_{\text{smooth}}(\theta)
    =
    {\sum_{k=1}^b\sum_{i=1}^n
    \|J_{\vo_k}( p_i)\|_F^2 w(p_i)}/S,
\end{equation}
where $J_{\vo_k}(p_i)$ is the Jacobian matrix.

We sum the aforementioned 6 losses with their own respective weight to formulate the fluid loss $\mathcal{L}_{\text{fuild}}$.

\subsection{Training}\label{sec:training}
\begin{figure}
  \includegraphics[width=\linewidth]{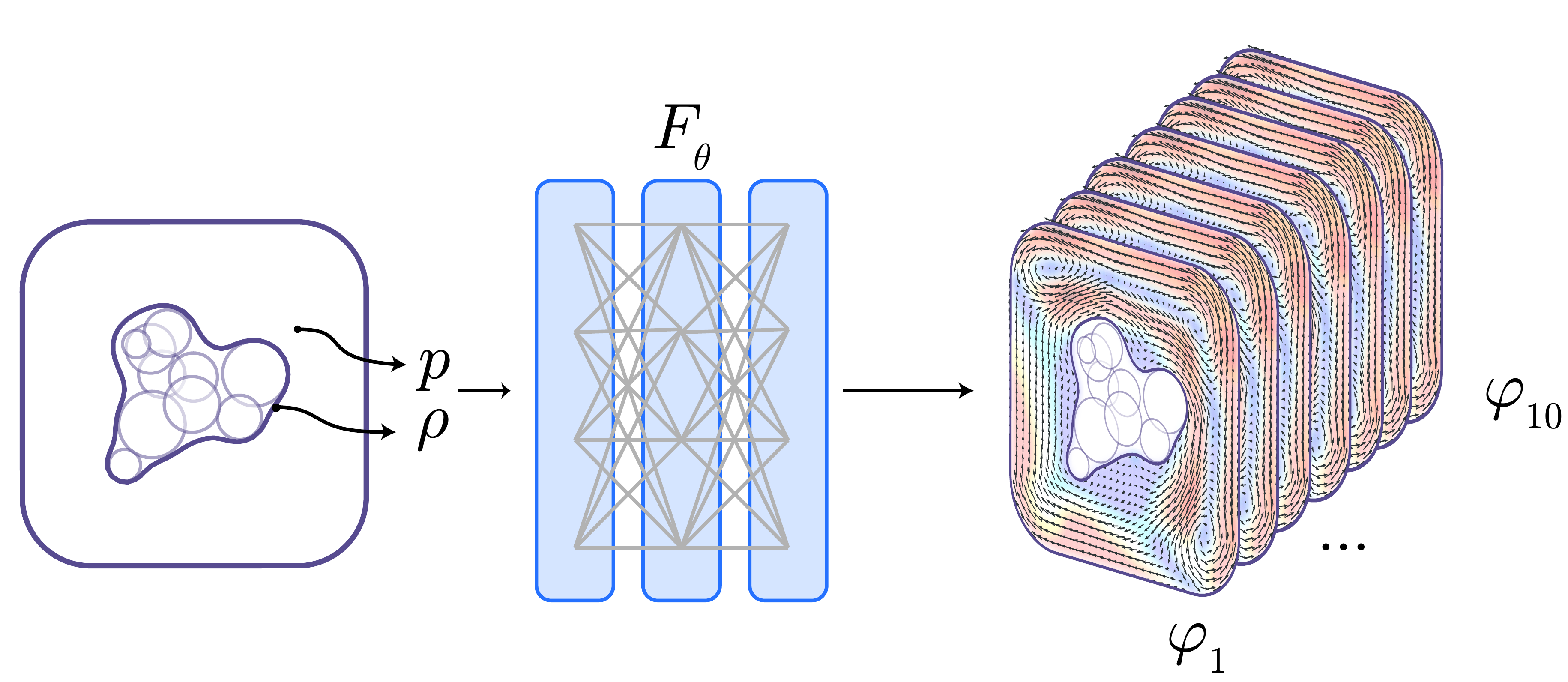}
  \caption{Overview of our MLP architecture: it takes as input a point $p$ and the set of circles' parameters $\rho$ and produces our neural bases $\varphi_i, i=1,\dots,b$.}
  \label{fig:nn}
\end{figure}

We train on a domain represented by \review{ten} circles. Our neural fields are parameterized by MLPs that have 8 fully connected layers, which use the leaky ReLU activation function (except ELU for the last layer as we want to produce negative numbers) and have 256 channels per layer (\Cref{fig:nn}). The input vector is the concatenation of the position of one sample point $p_i$ and \review{10} circle parameters. %
The output vector consists of the vector for the evaluation of the $b$ neural bases.

We assign the following weights to each term in the fluid loss function: $(w_\text{drch}, w_\text{div}, w_\text{orth}, w_\text{bc}, w_\text{len}, w_\text{small}) = \review{(0.01, 5, 100, 30, 100, 100)}$. We choose the threshold for small base penalty $\delta$ as 0.05 and the average length for length penalty as $c= 0.37$.

The MLP uses Kaiming initialization for its parameters. We train the MLP using the Adam optimizer with a learning rate initialized at 0.0005. To dynamically adjust the learning rate, we employed an ExponentialLR scheduler, applying a decay factor of 0.96, resulting in a learning rate scaled by $LR \times 0.96^{t}$ per epoch, where $t$ is the epoch.

We train on 1000 different geometric samples, containing randomly generated circles for 2D and spheres for 3D with centers in the $[0.225,0.675]^d$ square or cube and radius in the $[0.03, 0.09]$ interval, \review{randomly generated shapes with two and three components, and, in 2D, shapes representing the letters of the english alphabet}. They can be (partially) overlapping or separate. For each sample, we randomly sample $n=10^{6}$ points over the domain. The training was conducted for a total of 10 epochs with a batch size of $10^{4}$.

The implementation is based on PyTorch, with a training time of 23.6 hours for the 3D model and 22.9 hours for the 2D model, on a single NVIDIA GeForce RTX 3090 GPU.

\subsection{Fitting to a Sketch}\label{sec:fitting}

\begin{figure}
  \centering \footnotesize
  \includegraphics[width=.32\linewidth,trim={0.7cm 0.7cm 0.7cm 0.7cm},clip]{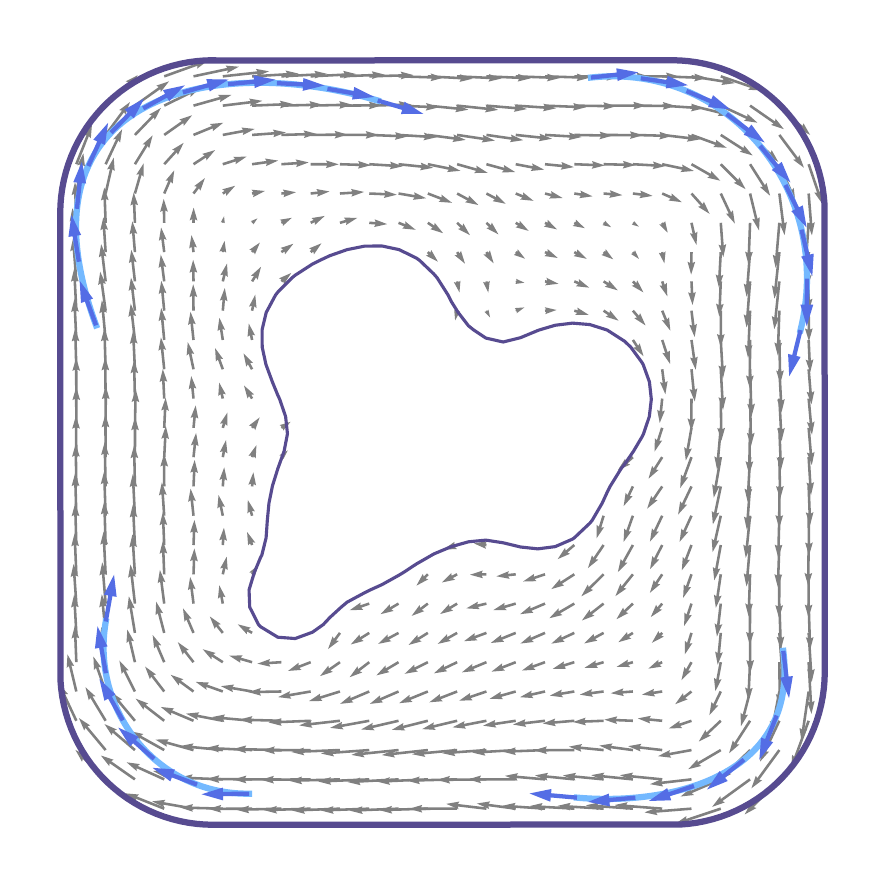} \hfill
  \includegraphics[width=.32\linewidth,trim={0.7cm 0.7cm 0.7cm 0.7cm},clip]{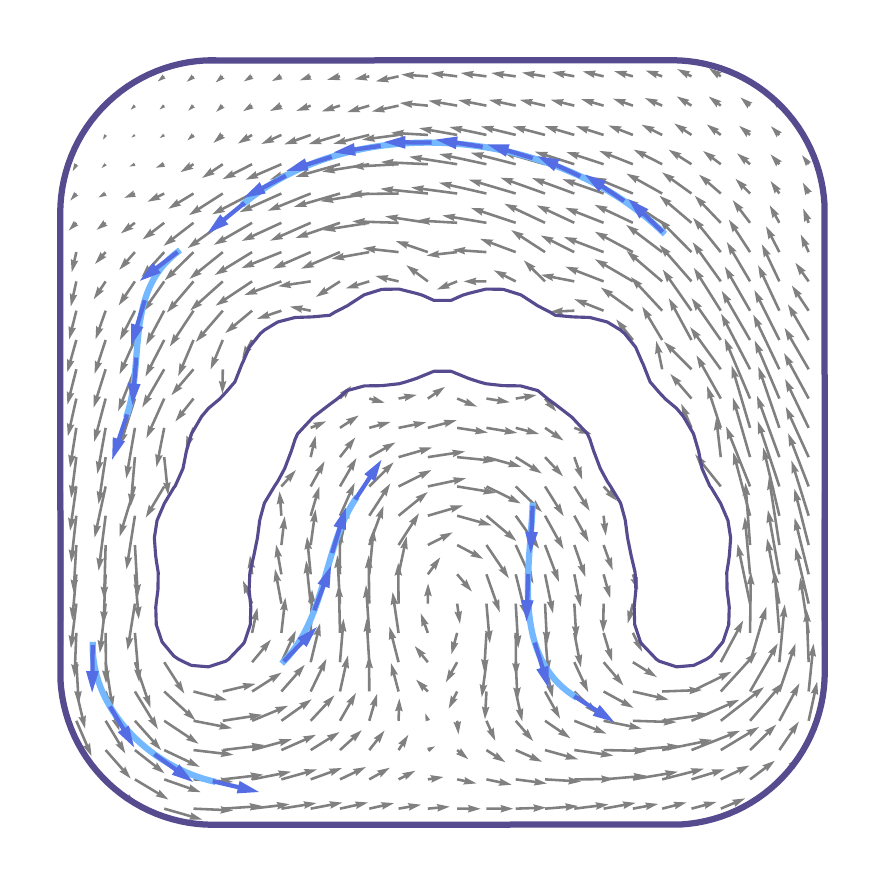} \hfill
  \includegraphics[width=.32\linewidth,trim={0.7cm 0.7cm 0.7cm 0.7cm},clip]{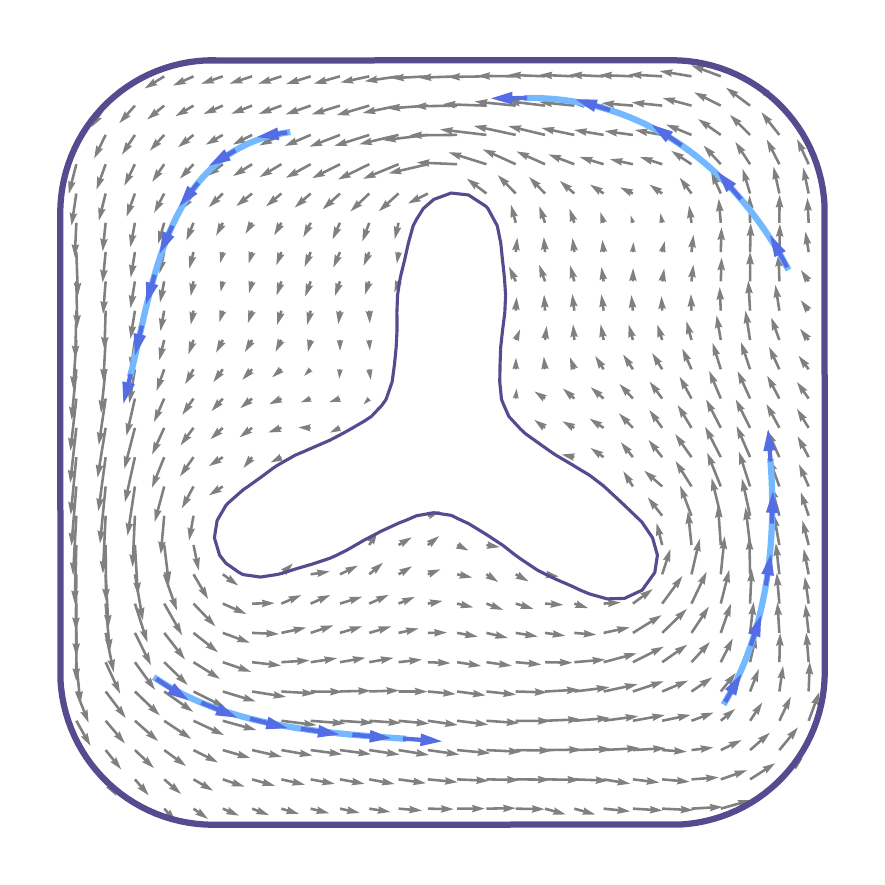} \par
  \parbox{.32\linewidth}{\centering Simple circular flow.}  \hfill
  \parbox{.32\linewidth}{\centering Counter-clockwise flow.} \hfill
  \parbox{.32\linewidth}{\centering Slow flow.}\\[1em]
 \includegraphics[width=.32\linewidth,trim={0.7cm 0.7cm 0.7cm 0.7cm},clip]{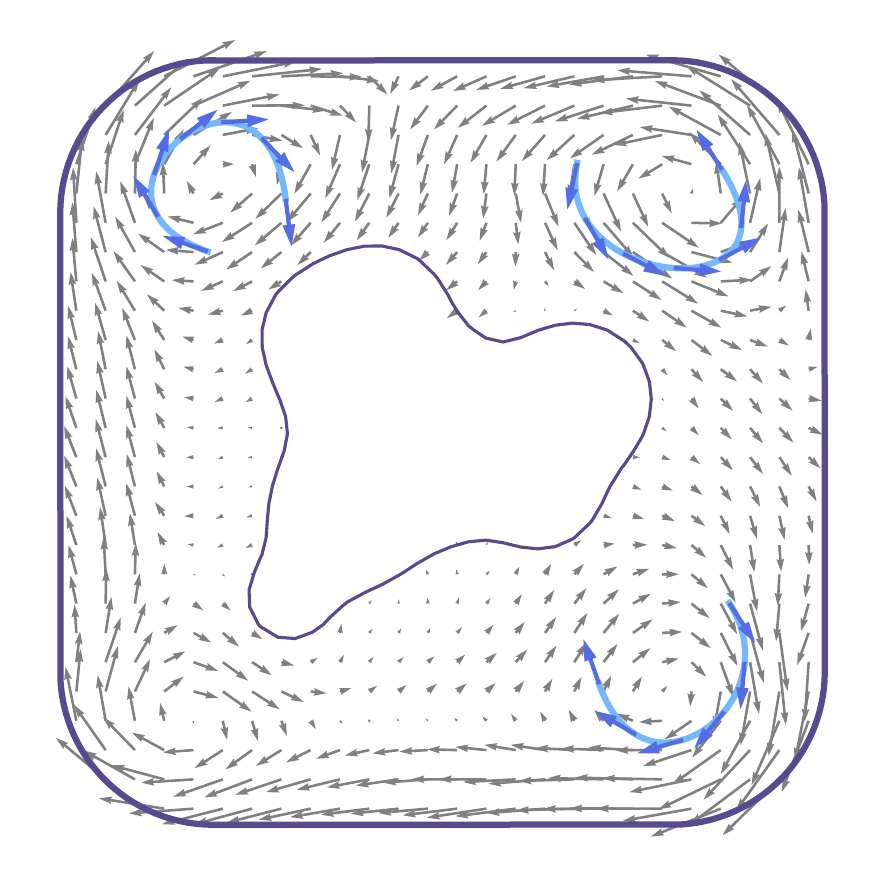} \hfill
 \includegraphics[width=.32\linewidth,trim={0.7cm 0.7cm 0.7cm 0.7cm},clip]{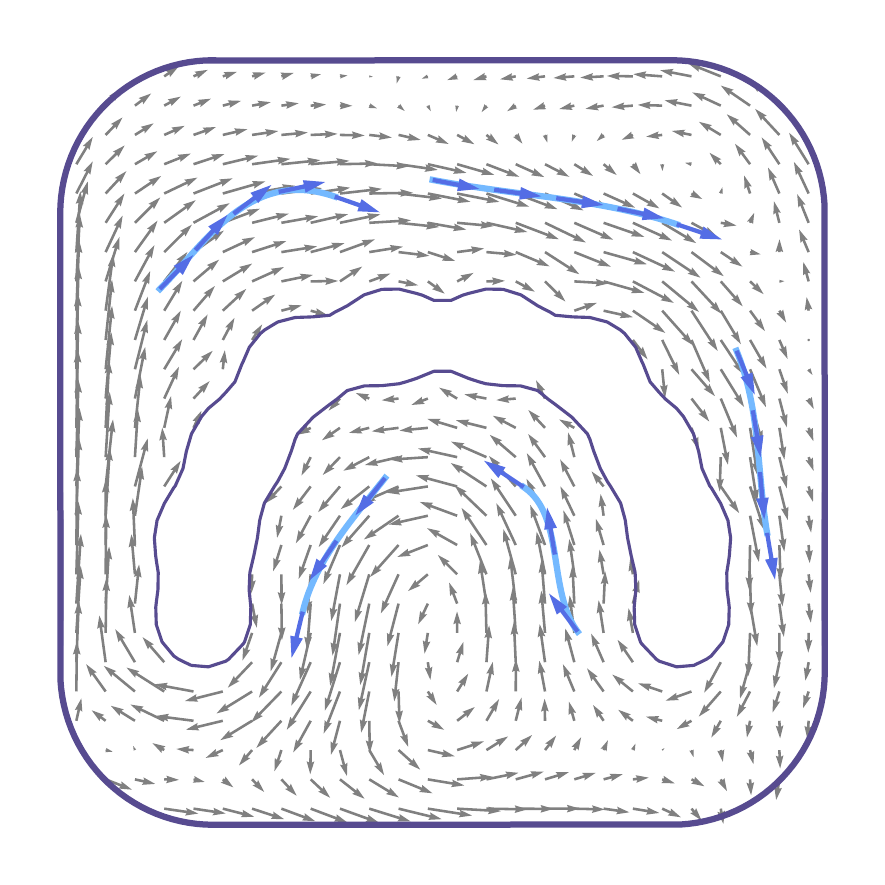} \hfill
  \includegraphics[width=.32\linewidth,trim={0.7cm 0.7cm 0.7cm 0.7cm},clip]{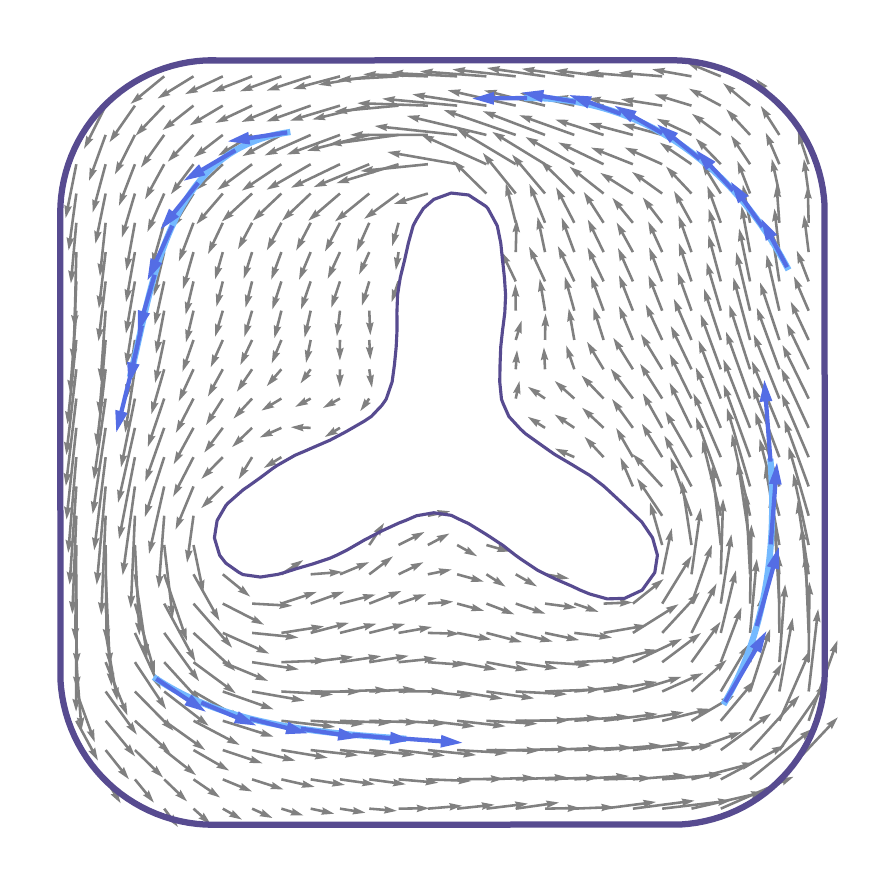} \par
     \parbox{.32\linewidth}{\centering Corner vortices.}
    \parbox{.32\linewidth}{\centering Clockwise flow.}
    \parbox{.32\linewidth}{\centering Fast flow.}\par
  \caption{Example of the different interactions possible to design a fluid flow. From the initial design to different types of editing.}
  \label{fig:sketch}
\end{figure}

We now aim to use our neural bases to animate an input image; the image contains the position of the circles and several streamlines represented by parametric curves $\gamma(t)$. We start by drawing the circles and a few curves, and this quickly generates a velocity field; by adding additional curves, we can interactively refine the flow by, for instance, adding vortices or changing the flow direction. \Cref{fig:sketch} demonstrates how the interactive sketches control the initial flow. Our bases can fit a variety of curves in a smooth manner; for instance, they fit a simple circular flow, or they allow the creation of many vortices and can control the direction or intensity of the flow.

To generate the input flow, we sample every curve $\gamma$ at $c_i$ points uniformly spaced in parametric space and compute the target velocity $t_i= \nabla \gamma(c_i) / \| \nabla \gamma(c_i)\|$ as the normalized tangent at $c_i$. We use this set of points and velocities to least-square fit an initial set of parameter $\alpha$;
\[
\alpha^0=\argmin_\alpha \sum_i \left\| \sum_{k=1}^b \vo_k(c_i) \alpha_k - t_i \right\|^2.
\]

\subsection{Advection}\label{sec:advection}
We then use $\alpha^0$ to compute the initial velocity $v^0$, which we advect using a semi-implicit integrator~\cite{stam:1999,stam:2023}. For every time step $t$, for every point, we compute the origin position
\[
 p_o = p - v^t(p) dt,
\]
where $dt$ is the time-step, note that if $p_o$ lands outside the domain, we project it to the closest point. Following the integration scheme, the new velocity is copied from the velocity at the origin position
\[
\bar v^{t+1}(p) = \sum_{k=1}^b \vo_k(p_o) \alpha_k^t.
\]
This new velocity might not be representable by our bases; therefore we compute the new $\alpha$ by least-square fit $\bar v^{t+1}(p)$ into our bases
\begin{equation} \label{eq:alpha_proj}
\alpha^{t+1}=\argmin_\alpha \sum_i \left\| \sum_{k=1}^b \vo_k(p_i) \alpha_k - \bar v^{t+1}(p_i) \right\|^2,
\end{equation}
and use it to compute
 \[
 v^{t+1} =  \sum_{k=1}^b \vo_k(p_i) \alpha_k^{t+1}.
 \]

\paragraph{Moving Domain}
\review{
Since our neural bases can be evaluated on arbitrary domains, they can naturally capture fluid flows with moving boundaries. In the time integrator, after computing the new velocity, we instead of projecting $\bar v^ {t+1}(p)$ on the same basis \eqref{eq:alpha_proj}, we project it on new set of bases defined on a different domain. %
}

%% file: 05-results.tex
\section{Results}

We will show how our neural bases generalize to different domains (\Cref{sec:gen}), how the different losses converge during the training process (\Cref{sec:conv}), and how the different losses contribute to how the bases behave (\Cref{sec:ablation}). Finally, we present two- and three-dimensional animations by fitting our kinematic neural bases and integrating them with a semi-implicit integrator (\Cref{sec:anim}).

\subsection{Generalization}\label{sec:gen}
\begin{figure*}
  \centering\footnotesize
    \parbox{.012\linewidth}{~}\hfill
  \parbox{.158\linewidth}{\centering $\mathcal{L}_{\text{len}}$}\hfill
\parbox{.158\linewidth}{\centering $\mathcal{L}_{\text{small}}$}\hfill
\parbox{.158\linewidth}{\centering $\mathcal{L}_{\text{orth}}$}\hfill
\parbox{.158\linewidth}{\centering $\mathcal{L}_{\text{div}}$}\hfill
\parbox{.158\linewidth}{\centering $\mathcal{L}_{\text{smooth}}$}\hfill
\parbox{.158\linewidth}{\centering $\mathcal{L}_{\text{bc}}$}\par
  \parbox{.012\linewidth}{\rotatebox{90}{\centering 2D}}\hfill
  \parbox{.158\linewidth}{\includegraphics[width=\linewidth]{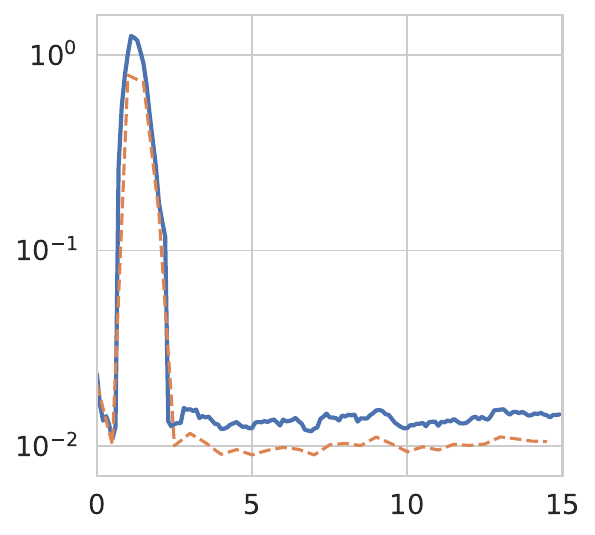}}\hfill
  \parbox{.158\linewidth}{\includegraphics[width=\linewidth]{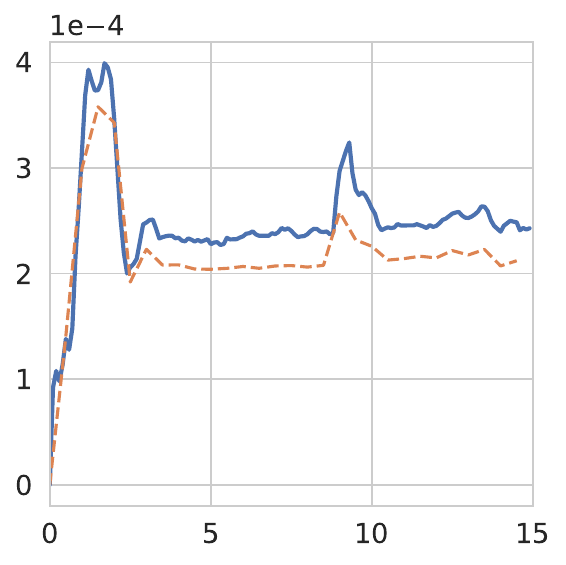}}\hfill
  \parbox{.158\linewidth}{\includegraphics[width=\linewidth]{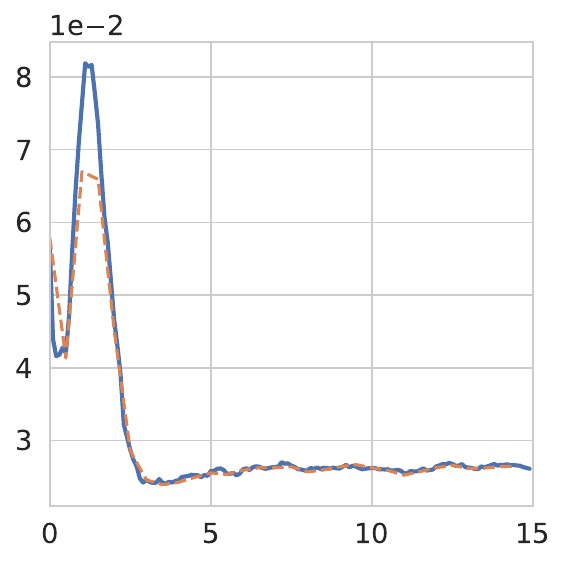}}\hfill
  \parbox{.158\linewidth}{\includegraphics[width=\linewidth]{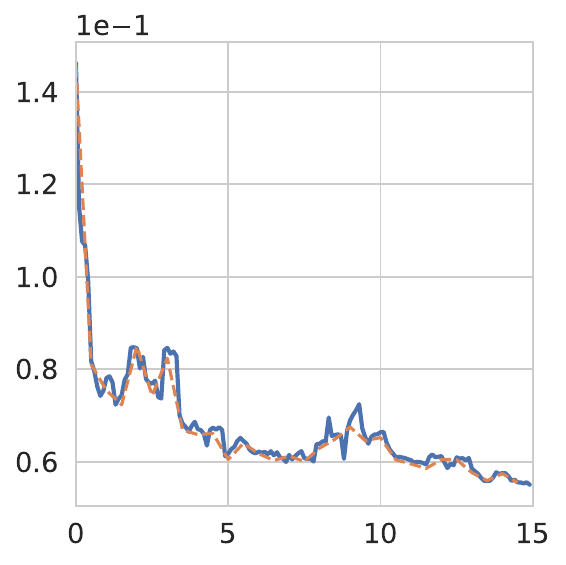}}\hfill
  \parbox{.158\linewidth}{\includegraphics[width=\linewidth]{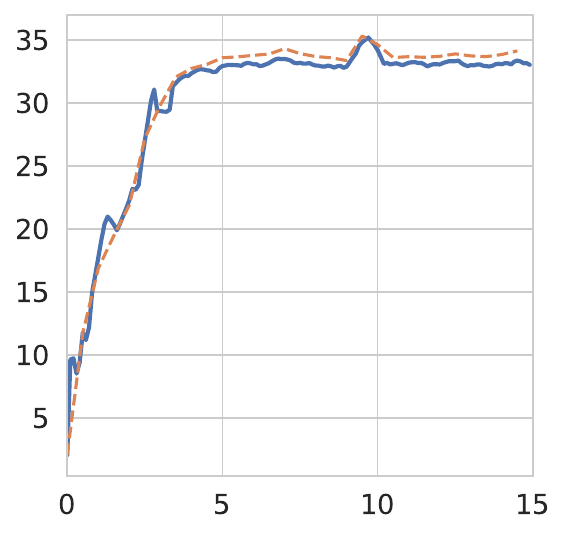}}\hfill
  \parbox{.158\linewidth}{\includegraphics[width=\linewidth]{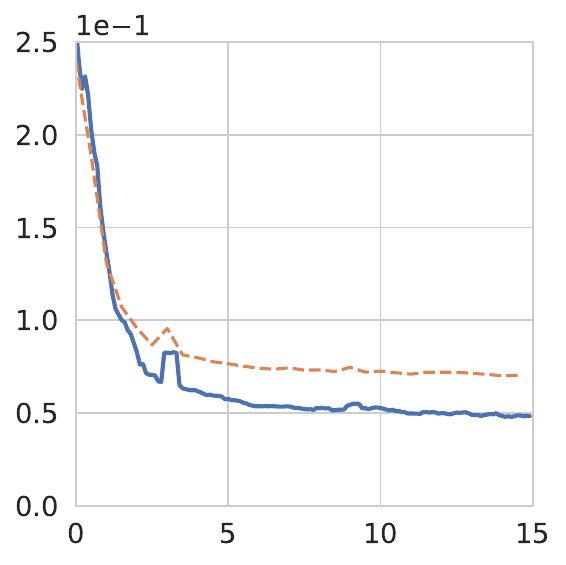} }\par
    \parbox{.012\linewidth}{\rotatebox{90}{\centering 3D}}\hfill
  \parbox{.158\linewidth}{\includegraphics[width=\linewidth]{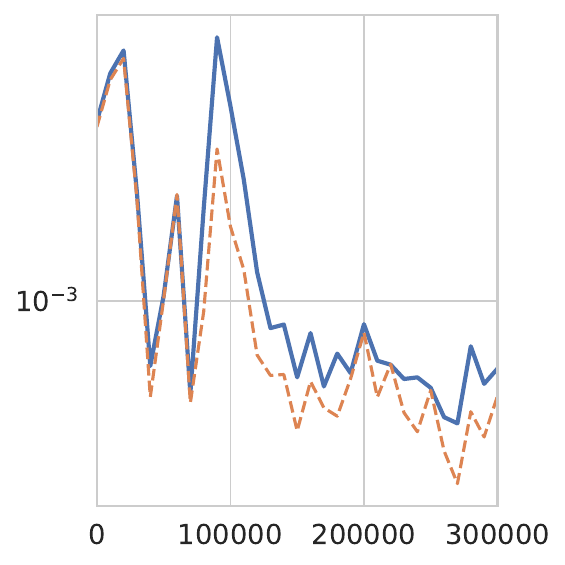}}\hfill
  \parbox{.158\linewidth}{\includegraphics[width=\linewidth]{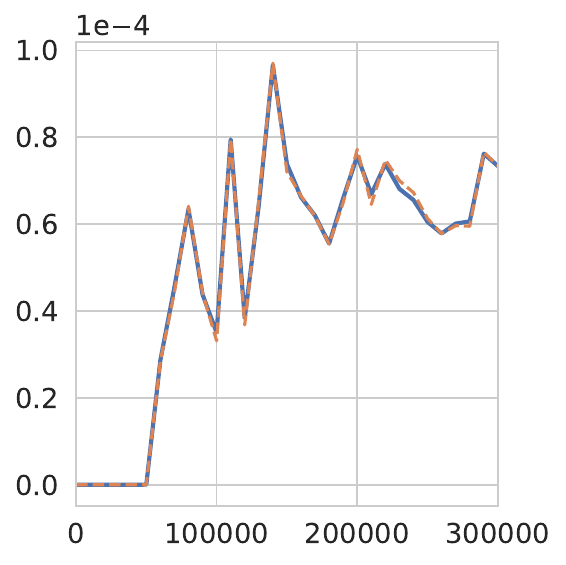}}\hfill
  \parbox{.158\linewidth}{\includegraphics[width=\linewidth]{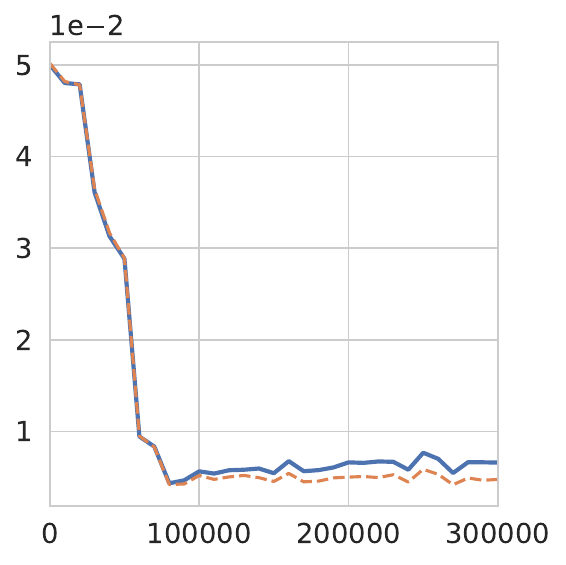}}\hfill
  \parbox{.158\linewidth}{\includegraphics[width=\linewidth]{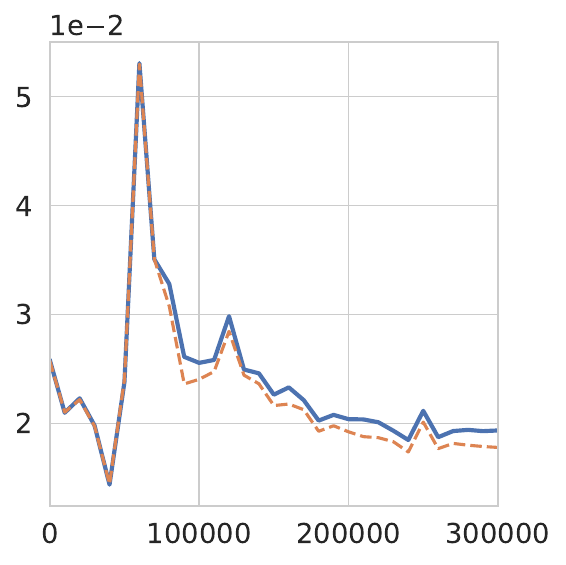}}\hfill
  \parbox{.158\linewidth}{\includegraphics[width=\linewidth]{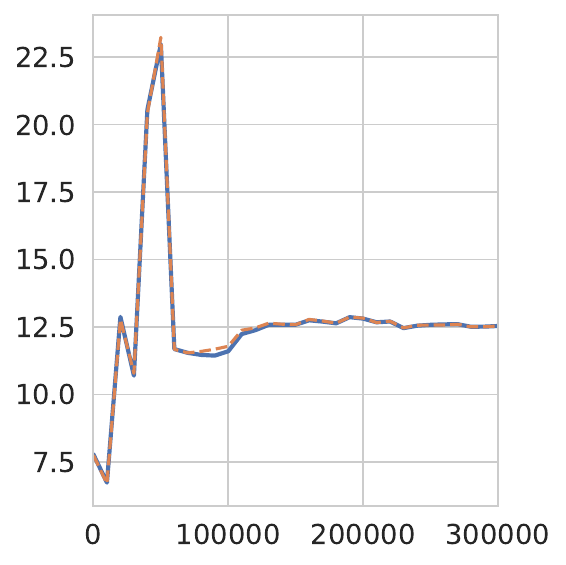}}\hfill
  \parbox{.158\linewidth}{\includegraphics[width=\linewidth]{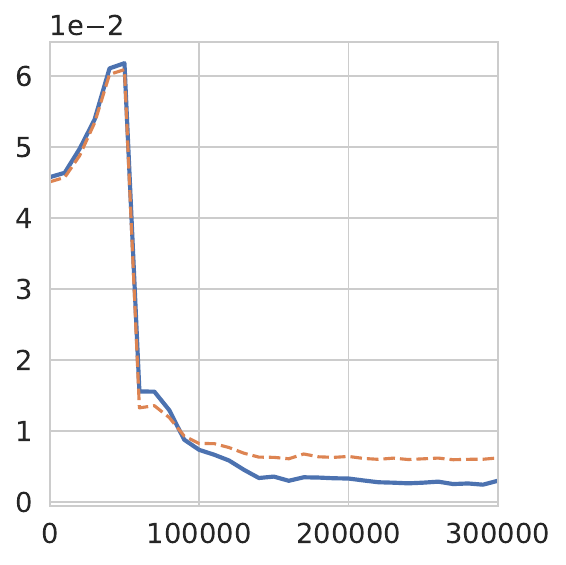} }
  \caption{Convergence of the losses over the number of epochs across the training (blue) and test (orange) set.}
  \label{fig:loss}
\end{figure*}

We evaluate our neural bases on a test dataset comprised of 100 random unseen circles/\review{spheres} with centers and radii sampled from the same distribution (Figure~\ref{fig:loss}, orange dashed, show the value of the different losses during training). At the end of the training, our neural bases have similar losses: the length losses and orthogonality are practically the same, while the boundary loss is just around 20\% larger.

\begin{figure}
  \centering\footnotesize
  \parbox{.03\linewidth}{~}\hfill
    \parbox{.46\linewidth}{\centering Divergence}\hfill
  \parbox{.46\linewidth}{\centering Boundary alignment}\par
    \parbox{.03\linewidth}{\rotatebox{90}{\centering 2D}}\hfill
  \parbox{.46\linewidth}{\includegraphics[width=\linewidth,trim={15pt 15pt 15pt 15pt},clip]{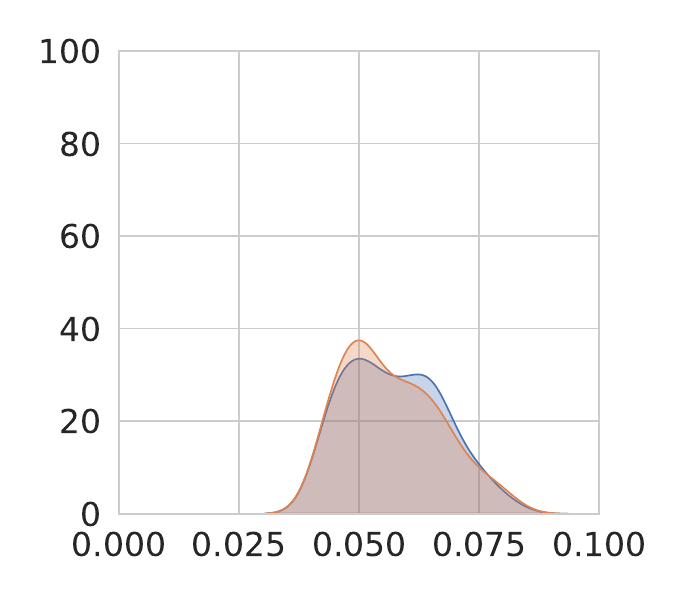}}  \hfill
  \parbox{.46\linewidth}{\includegraphics[width=\linewidth,trim={15pt 15pt 15pt 15pt},clip]{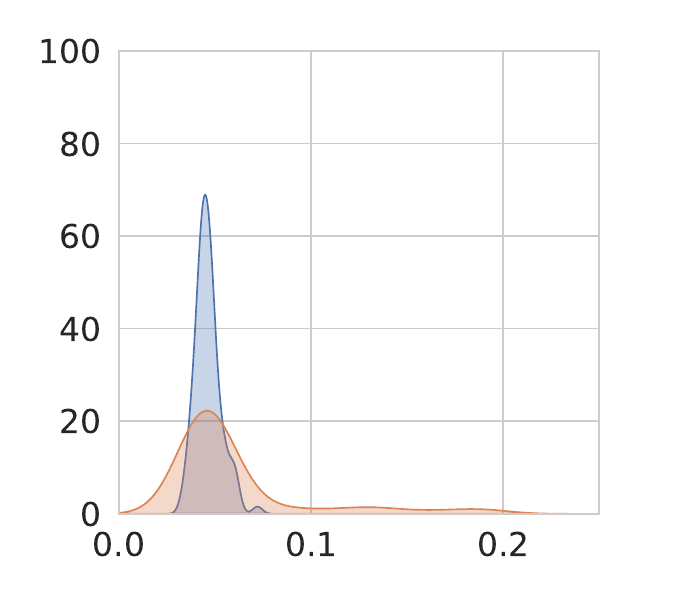}}\par
  \parbox{.03\linewidth}{\rotatebox{90}{\centering 3D}}\hfill
  \parbox{.46\linewidth}{\includegraphics[width=\linewidth,trim={15pt 0pt 10pt 0pt},clip]{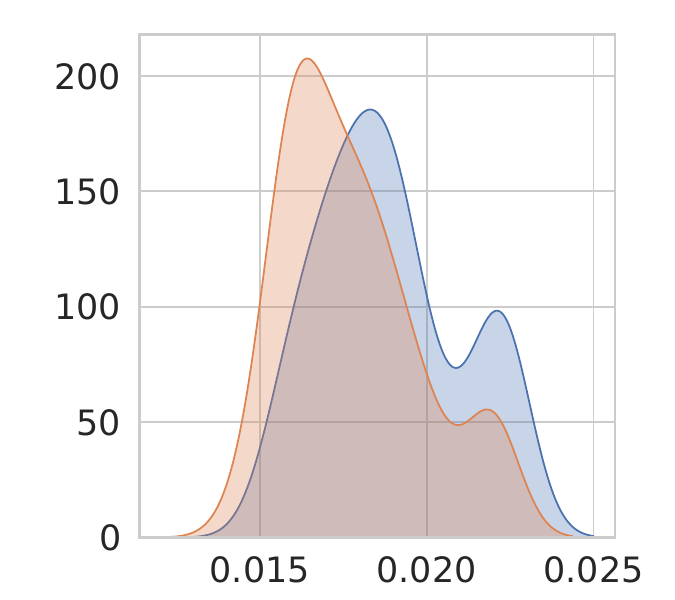}}  \hfill
  \parbox{.46\linewidth}{\includegraphics[width=\linewidth, trim={15pt 0pt 10pt 0pt},clip]{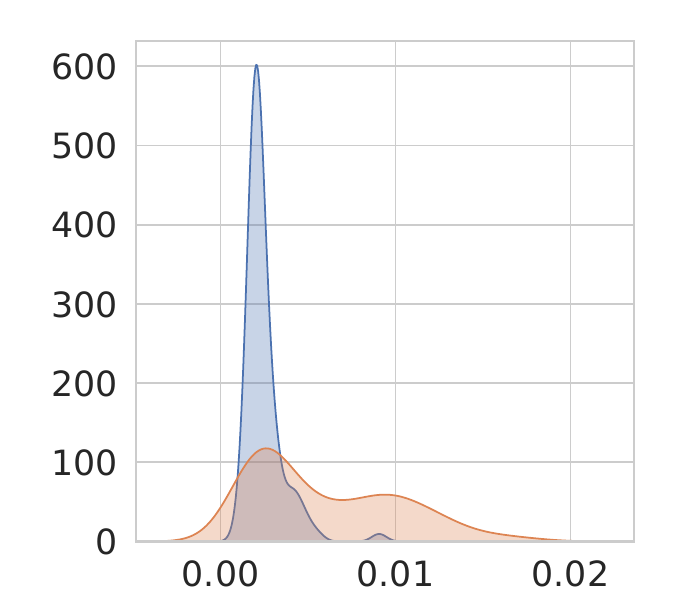}}
    \caption{Density plot of divergence loss and boundary loss on the training (blue) and test (red) sets. The $x$-axis shows loss values, and the $y$-axis indicates the probability density.}
  \label{fig:hist}
\end{figure}

\subsection{Convergence}\label{sec:conv}
Overall, our fluid loss steadily decreased over the training process. 
\review{In 2D and 3D, after the initial few epochs the small  and orthogonality loss become stationary, indicating that the bases are individually not zero, and are different (\Cref{fig:loss}). In 2D the bases also quickly reach the target length, while in 3D it slowly converges over the whole training process. It is interesting to note that in 2D subsequent epochs do not change their value, indicating that the training mostly rotates the vectors. The divergence and boundary loss  require more epochs (10-15) to reach a stationary value, since these losses are the only once affected by the shape of the domain (\Cref{fig:loss}).} We note that, as expected, the smoothness loss increases in the beginning as the bases become larger. 
\review{To evaluate the plausibility of our bases, we measure the distribution of the divergence and boundary alignment on training (blue) and test (red) set (\Cref{fig:hist}). The divergence is consistently small \review{(but not zero)} across the two sets (in average 0.5), while the boundary distribution is slightly larger for the test set. }

\subsection{Ablation study}\label{sec:ablation}

\begin{figure}
    \centering\footnotesize
    \parbox{.05\linewidth}{\rotatebox{90}{\centering Ours}}\hfill
    \parbox{.315\linewidth}{\includegraphics[width=\linewidth,trim={10pt 10pt 10pt 10pt},clip]{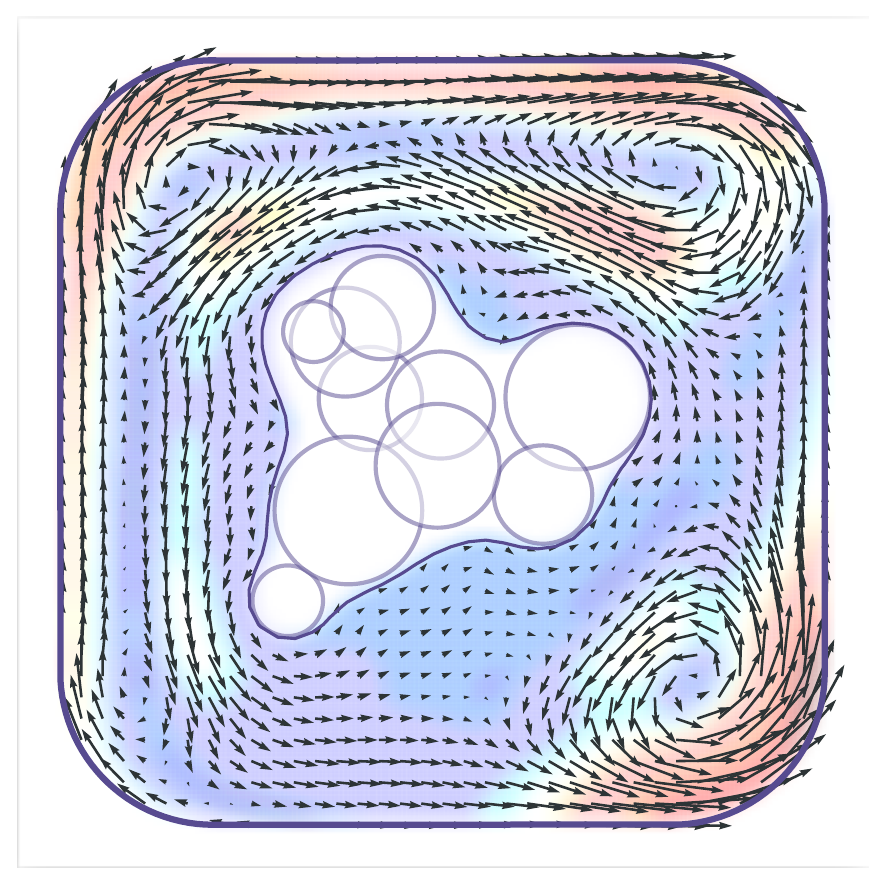}}\hfill
    \parbox{.315\linewidth}{\includegraphics[width=\linewidth,trim={10pt 10pt 10pt 10pt},clip]{figures2/bases/524/0/524_2.pdf}}\hfill
    \parbox{.315\linewidth}{\includegraphics[width=\linewidth,trim={10pt 10pt 10pt 10pt},clip]{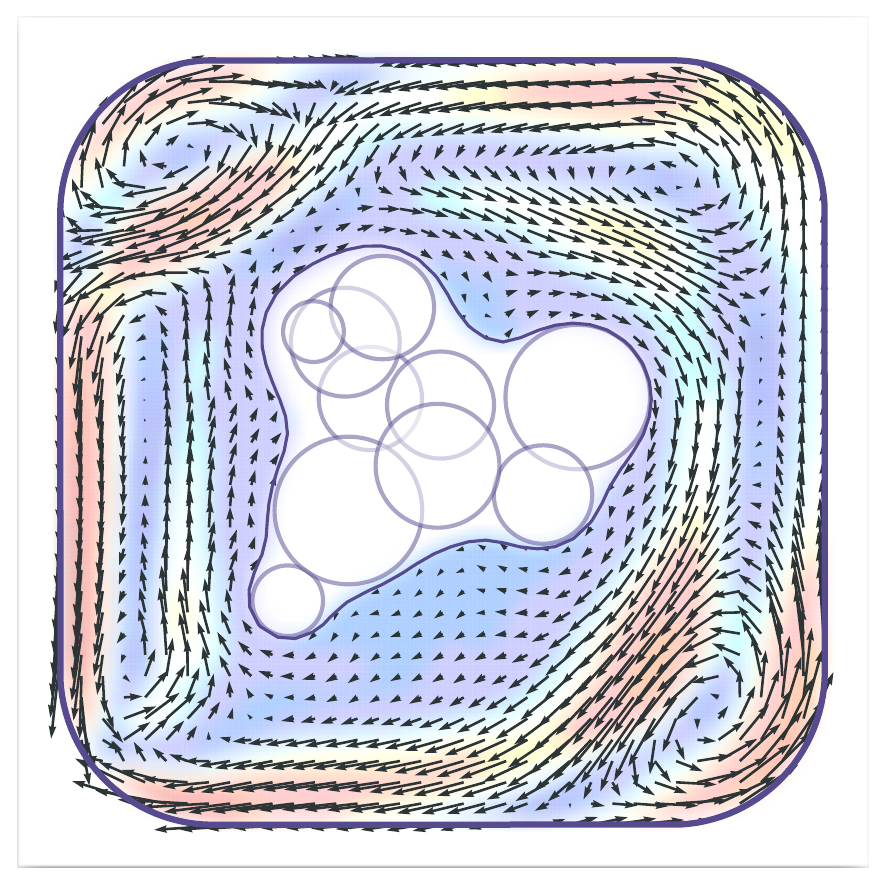}}\par
    \parbox{.05\linewidth}{\rotatebox{90}{\centering Smoothness loss}}\hfill
    \parbox{.315\linewidth}{\includegraphics[width=\linewidth,trim={10pt 10pt 10pt 10pt},clip]{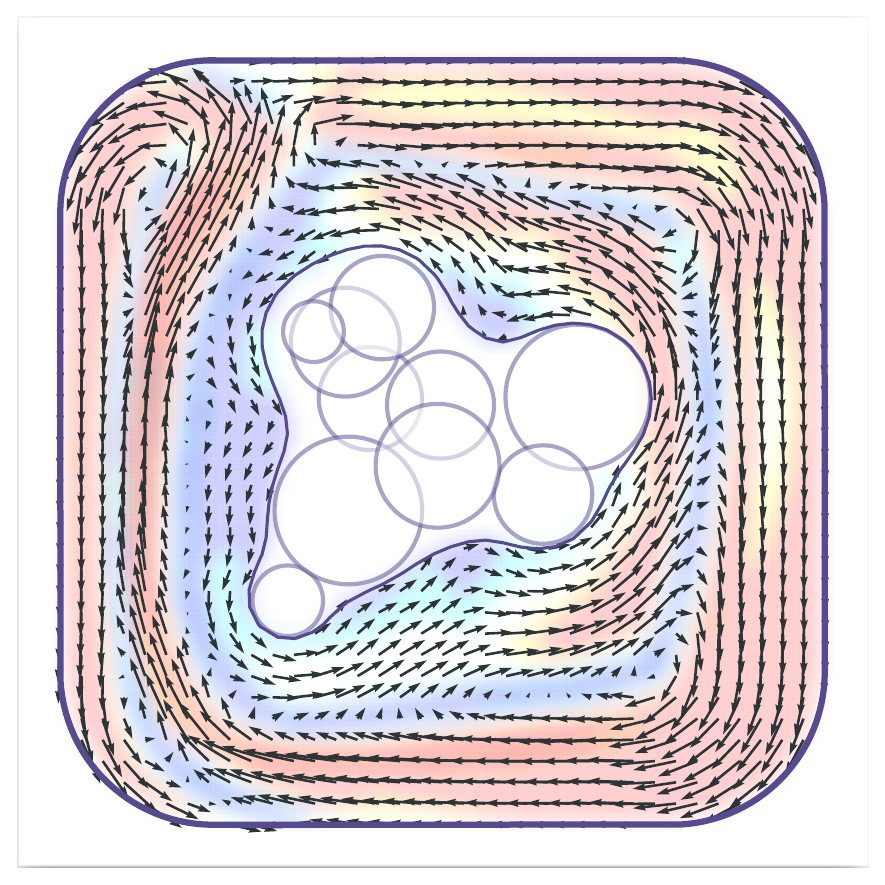}}\hfill
    \parbox{.315\linewidth}{\includegraphics[width=\linewidth,trim={10pt 10pt 10pt 10pt},clip]{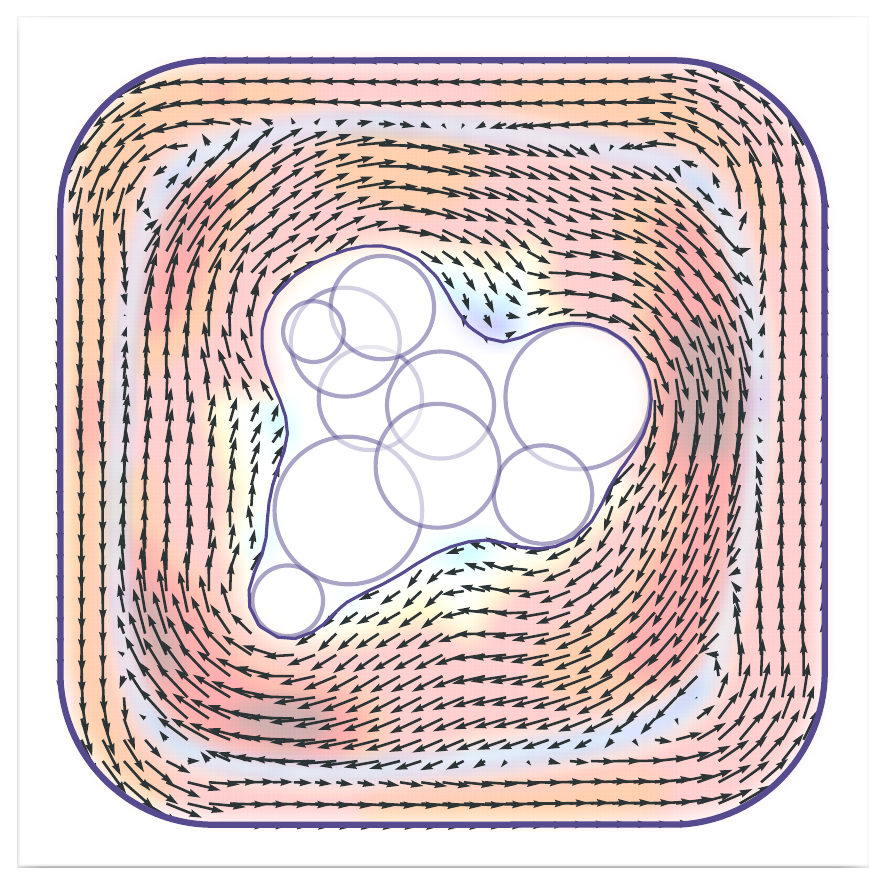}}\hfill
    \parbox{.315\linewidth}{\includegraphics[width=\linewidth,trim={10pt 10pt 10pt 10pt},clip]{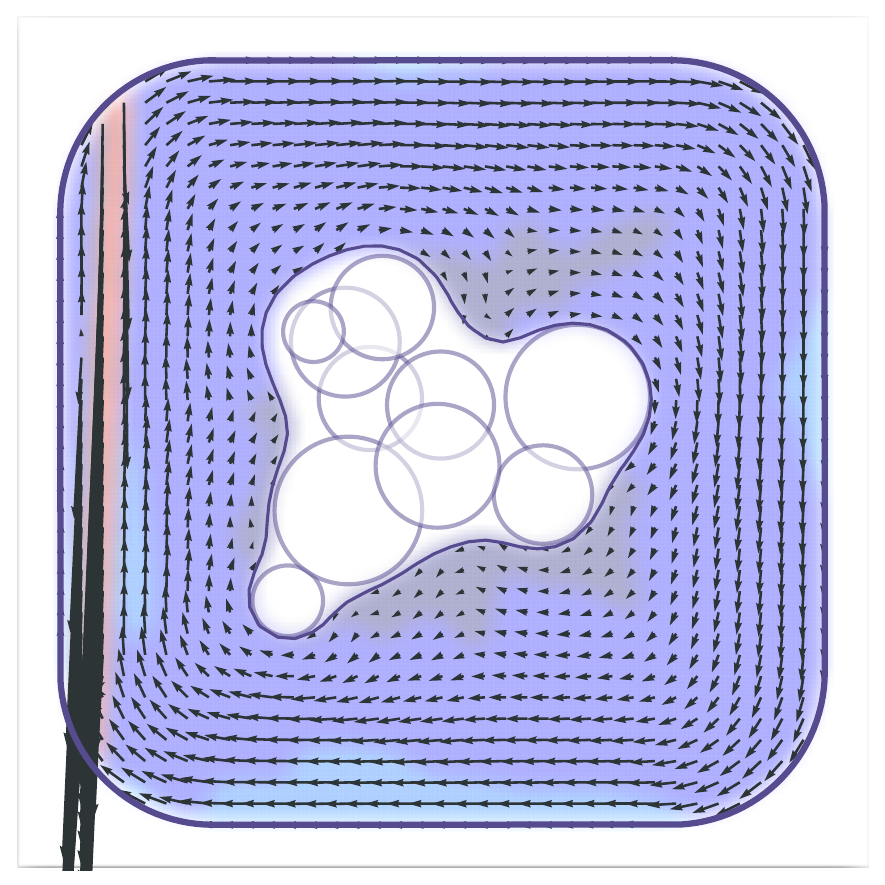}}\par
    \parbox{.05\linewidth}{\rotatebox{90}{\centering Rounded Corners}}\hfill
    \parbox{.315\linewidth}{\includegraphics[width=\linewidth,trim={10pt 10pt 10pt 10pt},clip]{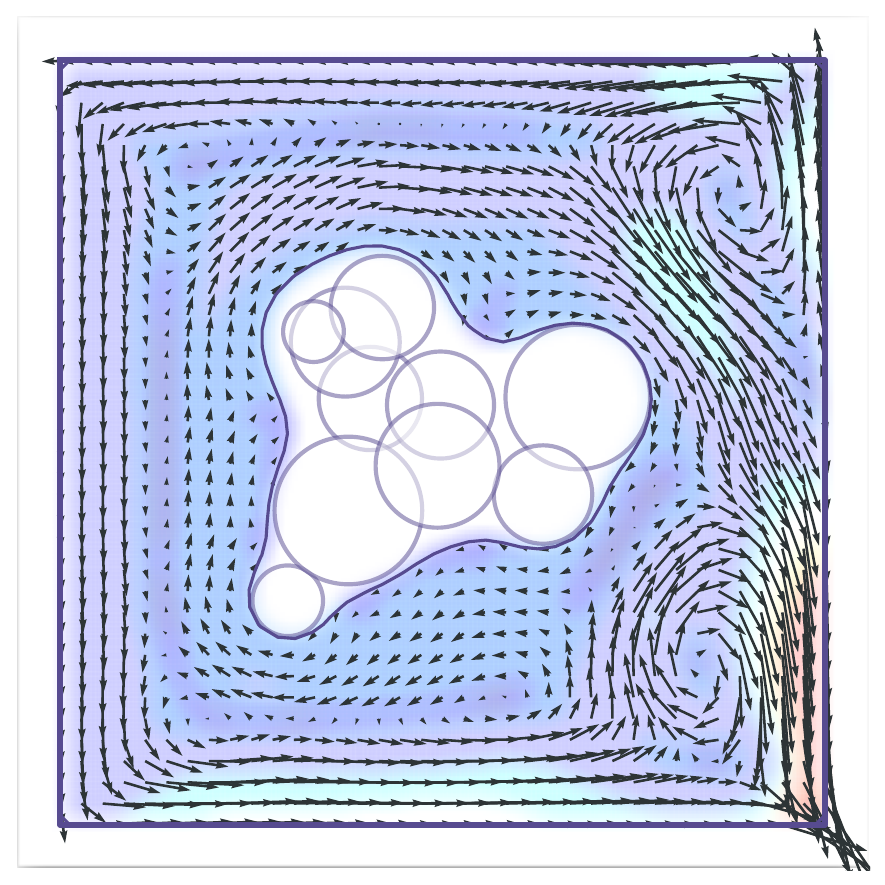}}\hfill
    \parbox{.315\linewidth}{\includegraphics[width=\linewidth,trim={10pt 10pt 10pt 10pt},clip]{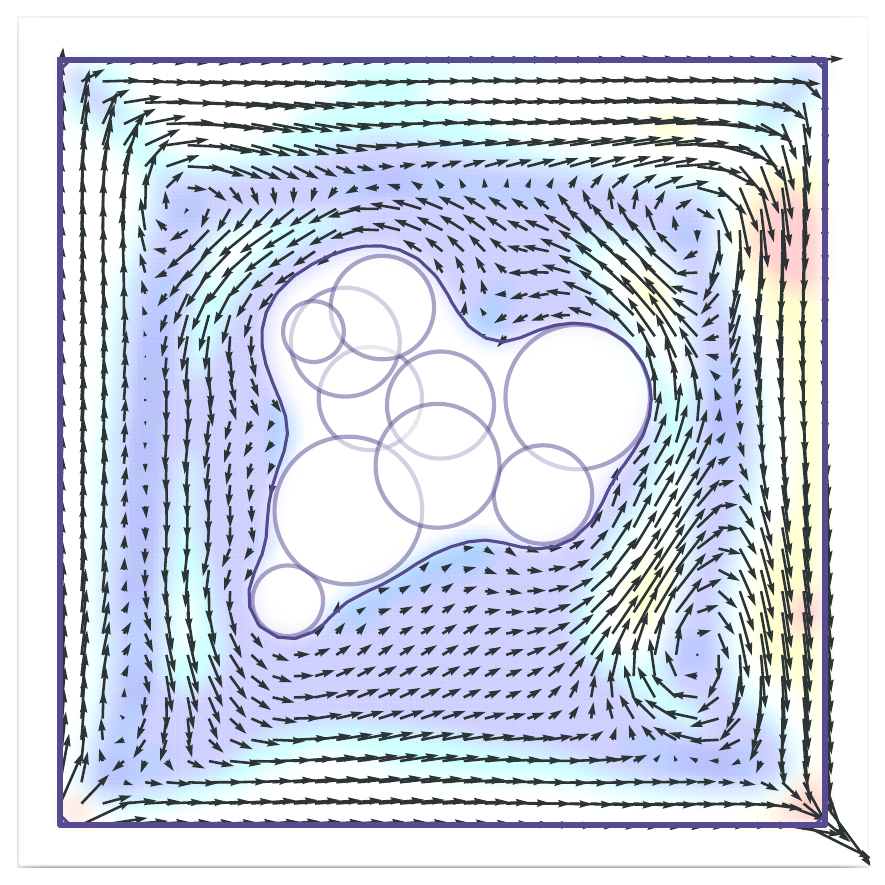}}\hfill
    \parbox{.315\linewidth}{\includegraphics[width=\linewidth,trim={10pt 10pt 10pt 10pt},clip]{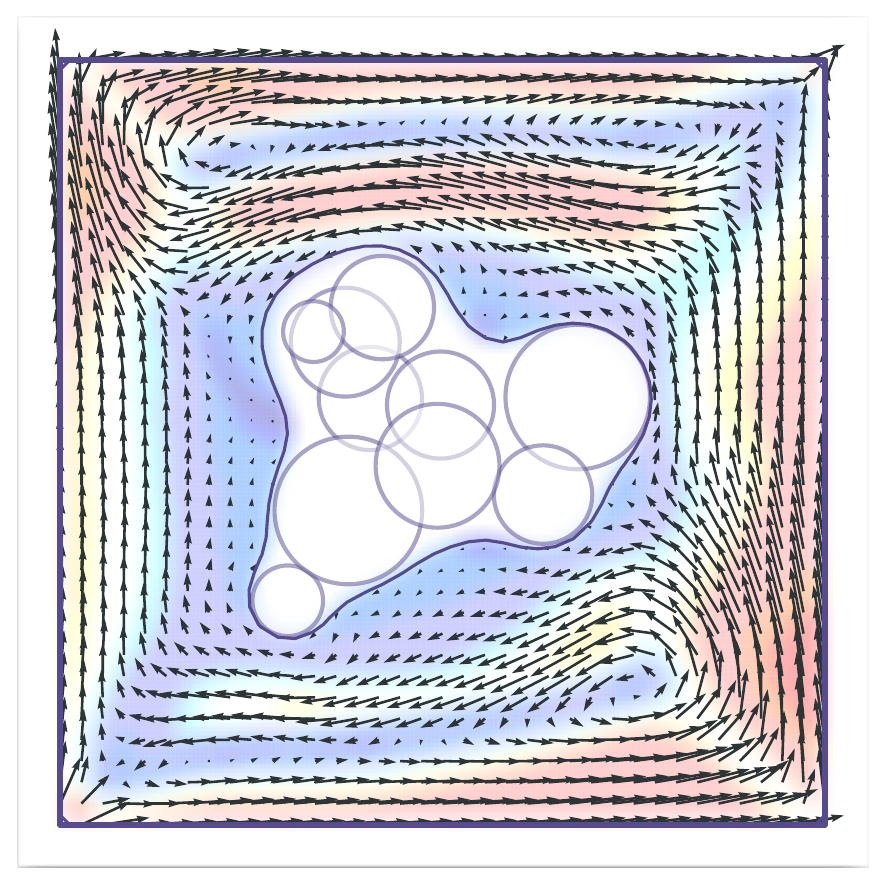}}\par
    \parbox{.05\linewidth}{\rotatebox{90}{\centering \# Sample Points}}\hfill
    \parbox{.315\linewidth}{\includegraphics[width=\linewidth,trim={10pt 10pt 10pt 10pt},clip]{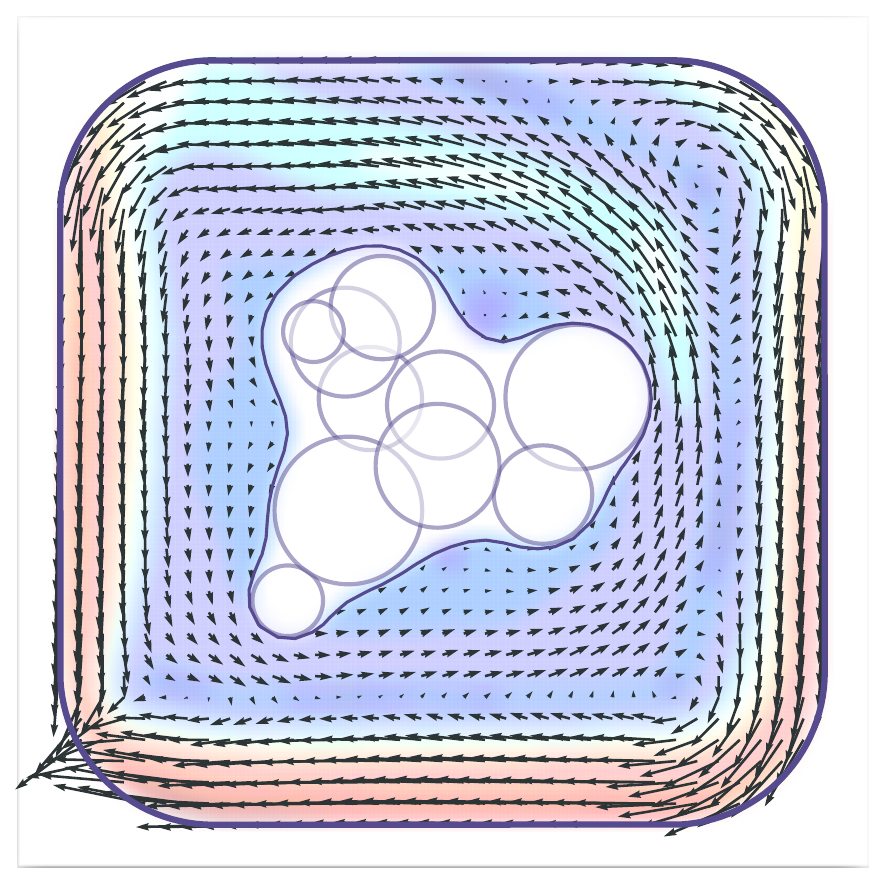}}\hfill
    \parbox{.315\linewidth}{\includegraphics[width=\linewidth,trim={10pt 10pt 10pt 10pt},clip]{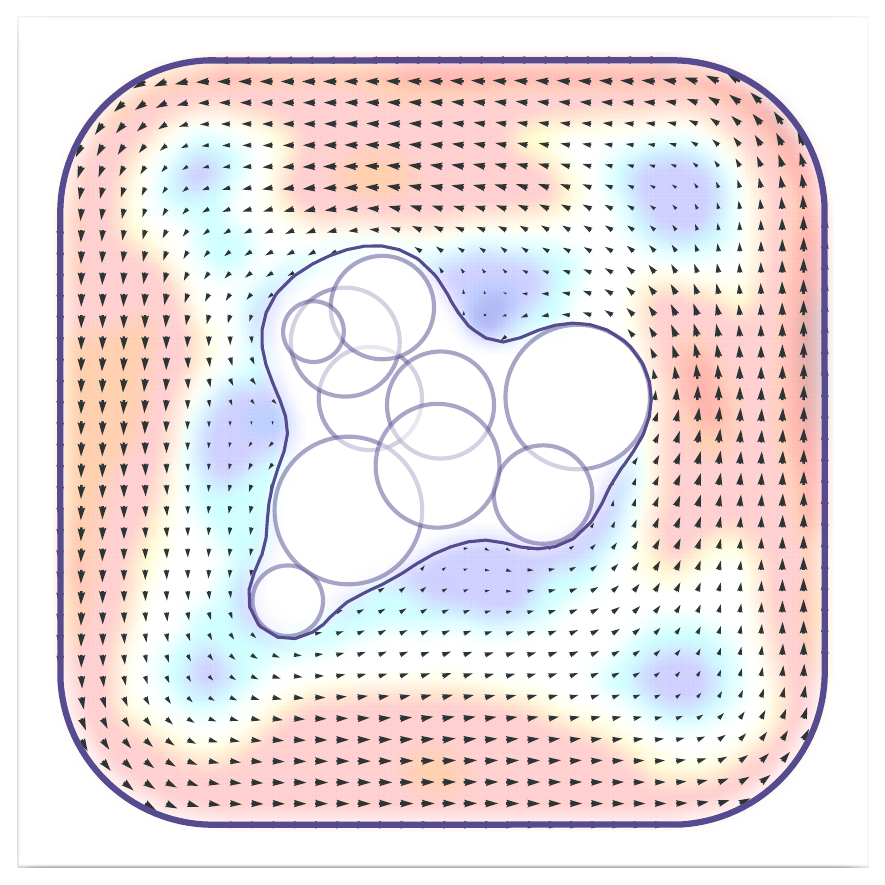}}\hfill
    \parbox{.315\linewidth}{\includegraphics[width=\linewidth,trim={10pt 10pt 10pt 10pt},clip]{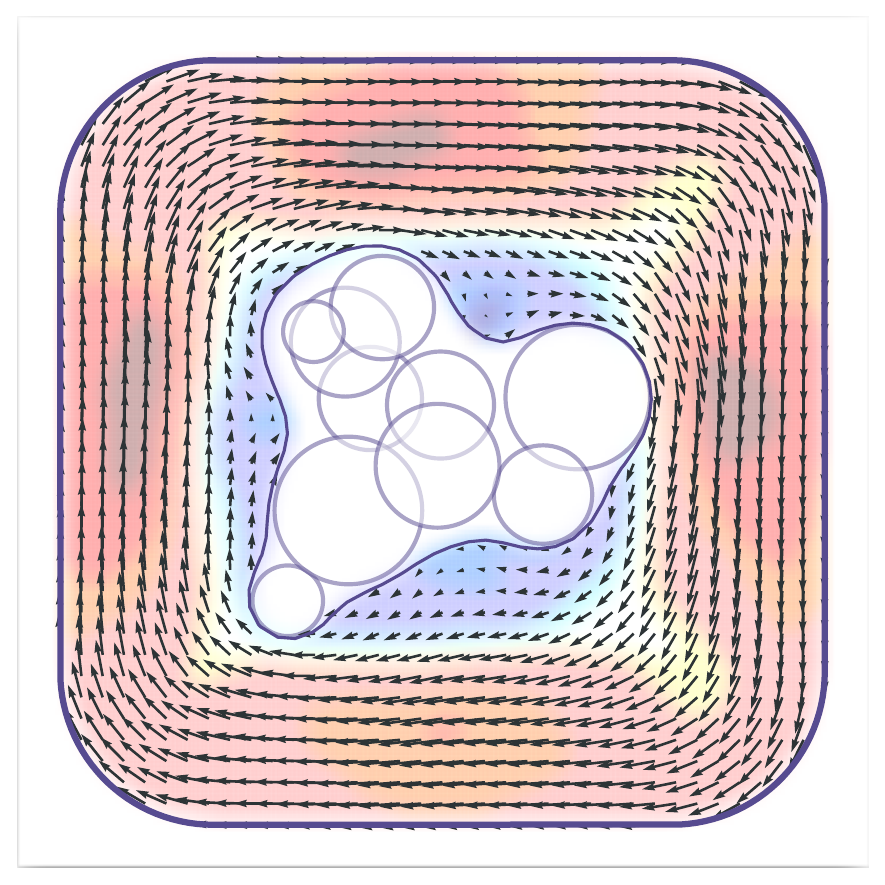}}
    \caption{Bases generated by omitting different parts of our pipeline. Each omission leads to different artifacts.}
    \label{fig:ablation}
\end{figure}

We showcase the importance of three main features of our pipeline (\Cref{fig:ablation}).

\paragraph{Smoothness loss} This is the only loss that is not motivated by fundamental physical law. We included it to bias the MLP towards smooth solutions (which our bases must be). Adding it improves the reliability of the training process: in our experiments, without it, we obtain unreasonable bases more frequently. The last figure in the second row of \Cref{fig:ablation} shows that without the smoothness loss the MLP fails to generate a valid basis.

\paragraph{Rounded Corners} We decided to smooth the corners of the domain to ensure that we can generate a smooth solution. Without it, the training fails to transition around the corners (where there should be a discontinuity) and generate bases that push the flow outside the domain.

\paragraph{Number of Samples} \review{We sample $10^6$ points for each input shape, as physics-based unsupervised losses require a sufficient number of samples to ensure convergence. Reducing the number of samples often leads to poor convergence, with the model violating physical constraints and producing trivial basis functions.}

\subsection{Animations}\label{sec:anim}
We run all our experiments on a GeForce RTX 3090. Generating the initial fluid is interactive, while the simulation runs at 40 frames per second for 250 thousand points. \review{To visualize the flow, we render water displacement using the magnitude of the velocity field and show particle trajectories as they move along the flow.  We refer to the additional materials for all the videos of our animations.}

\begin{figure*}
  \centering\footnotesize
  \includegraphics[width=.11\linewidth]{figures2/edit/1-1.pdf}  \hfill
  \includegraphics[width=.21\linewidth]{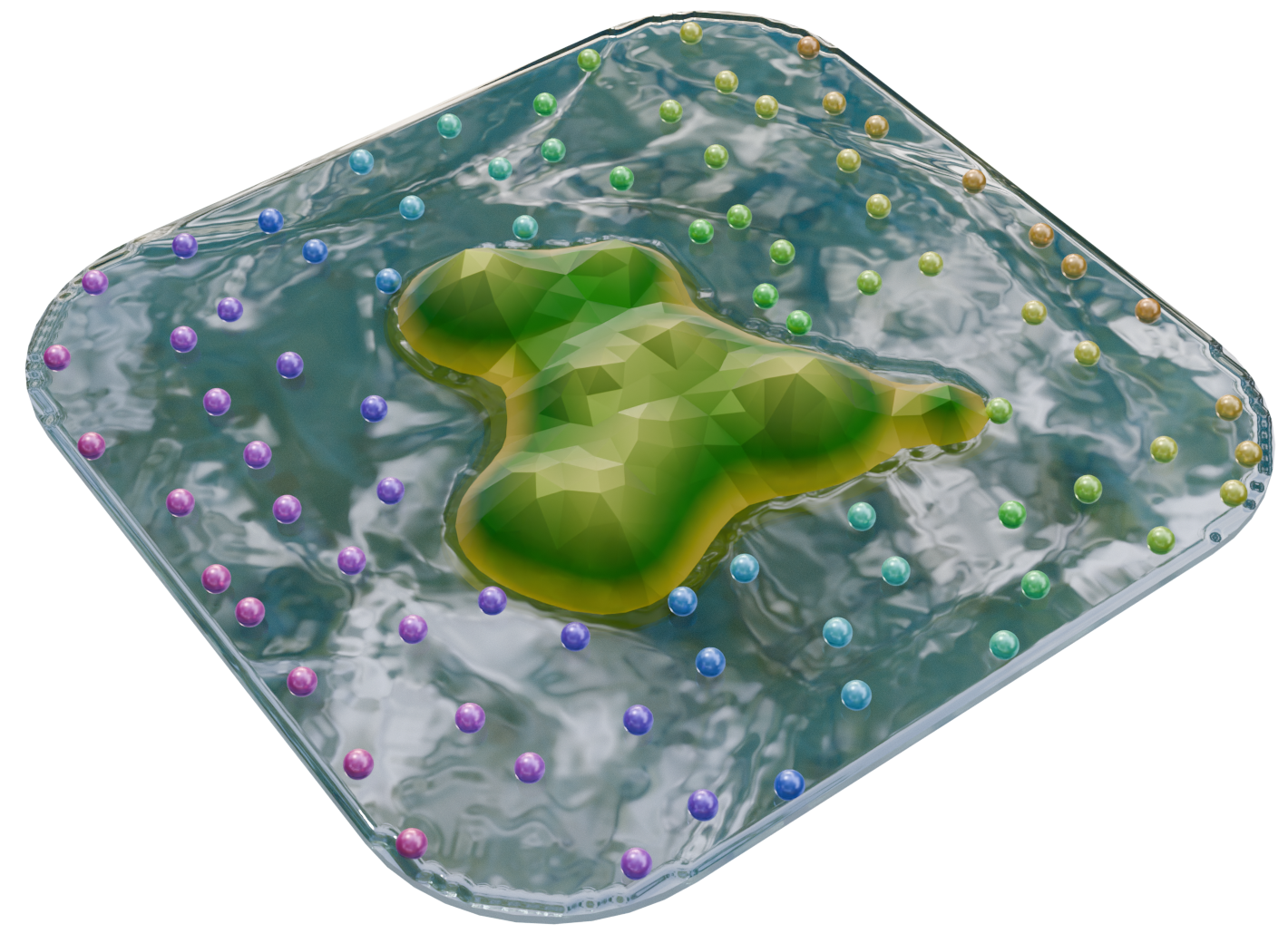}  \hfill
  \includegraphics[width=.21\linewidth]{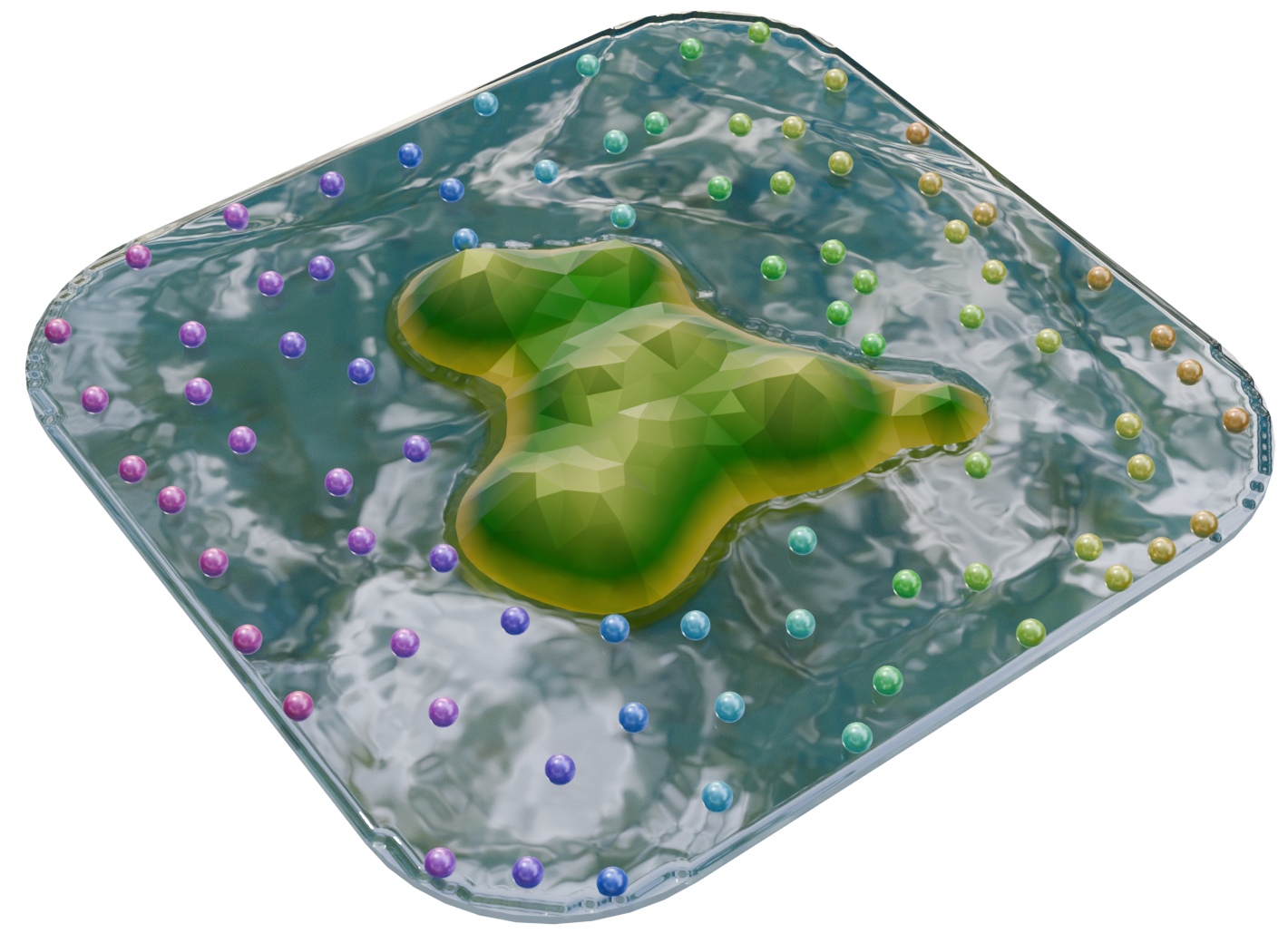}  \hfill
  \includegraphics[width=.21\linewidth]{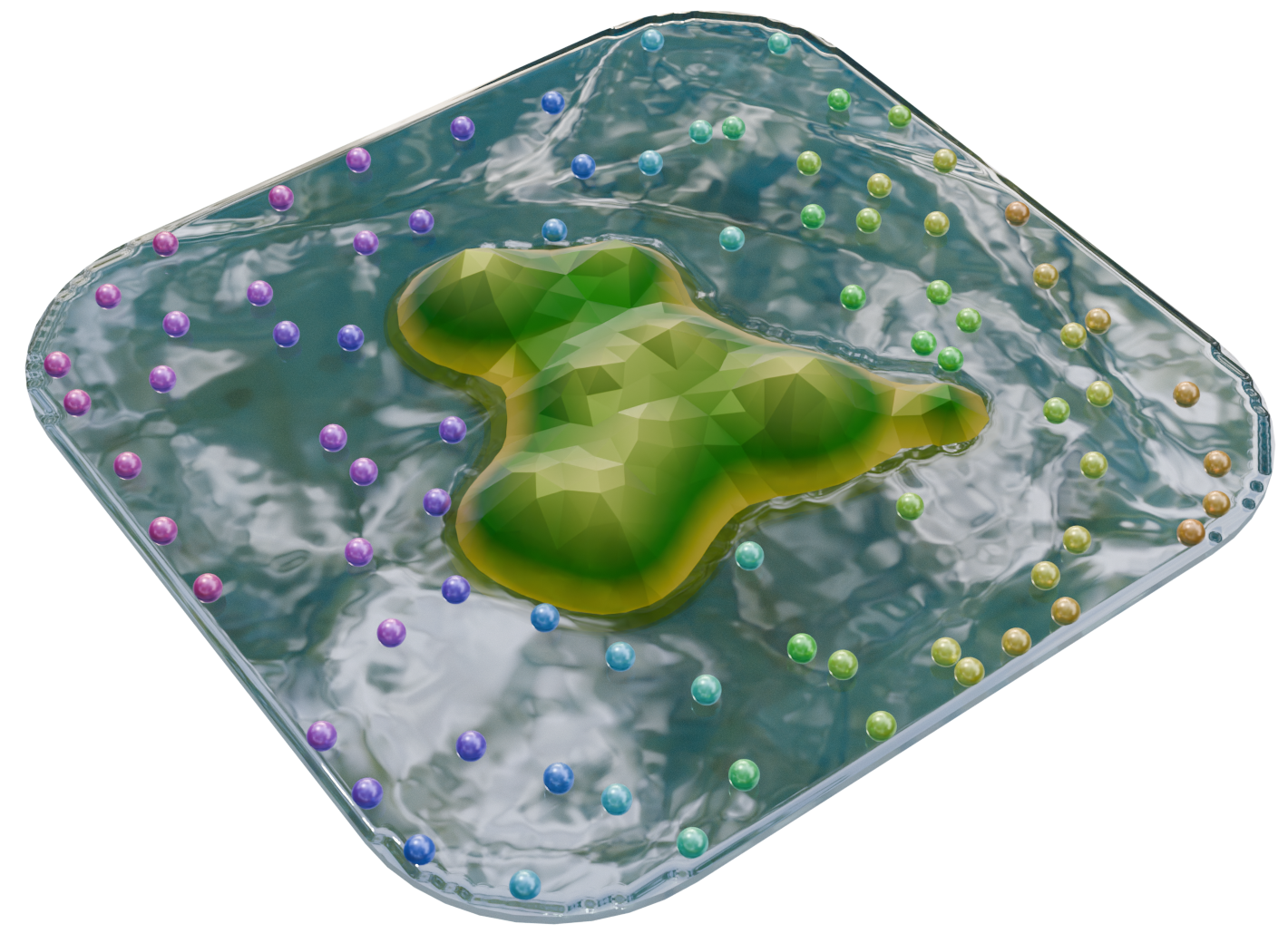}  \hfill
  \includegraphics[width=.21\linewidth]{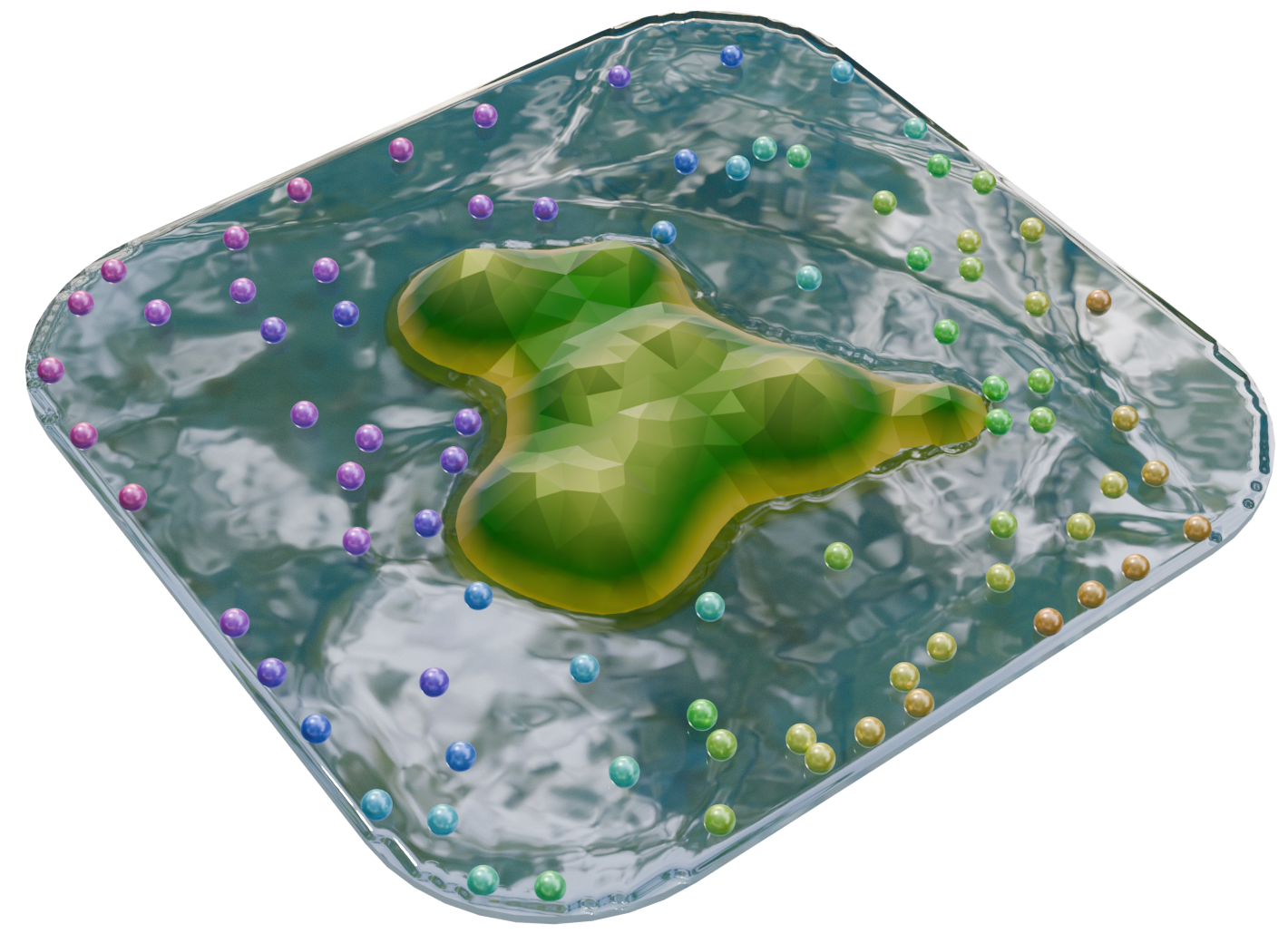}
    \parbox{.11\linewidth}{\centering Input curves}\hfill
  \parbox{.21\linewidth}{\centering $t=0$}\hfill
  \parbox{.21\linewidth}{\centering $t=0.165s$}\hfill
  \parbox{.21\linewidth}{\centering $t=0.33s$}\hfill
  \parbox{.21\linewidth}{\centering $t=0.495$}\\[1em]
\includegraphics[width=.11\linewidth]{figures2/edit/1-2.pdf}  \hfill
  \includegraphics[width=.21\linewidth]{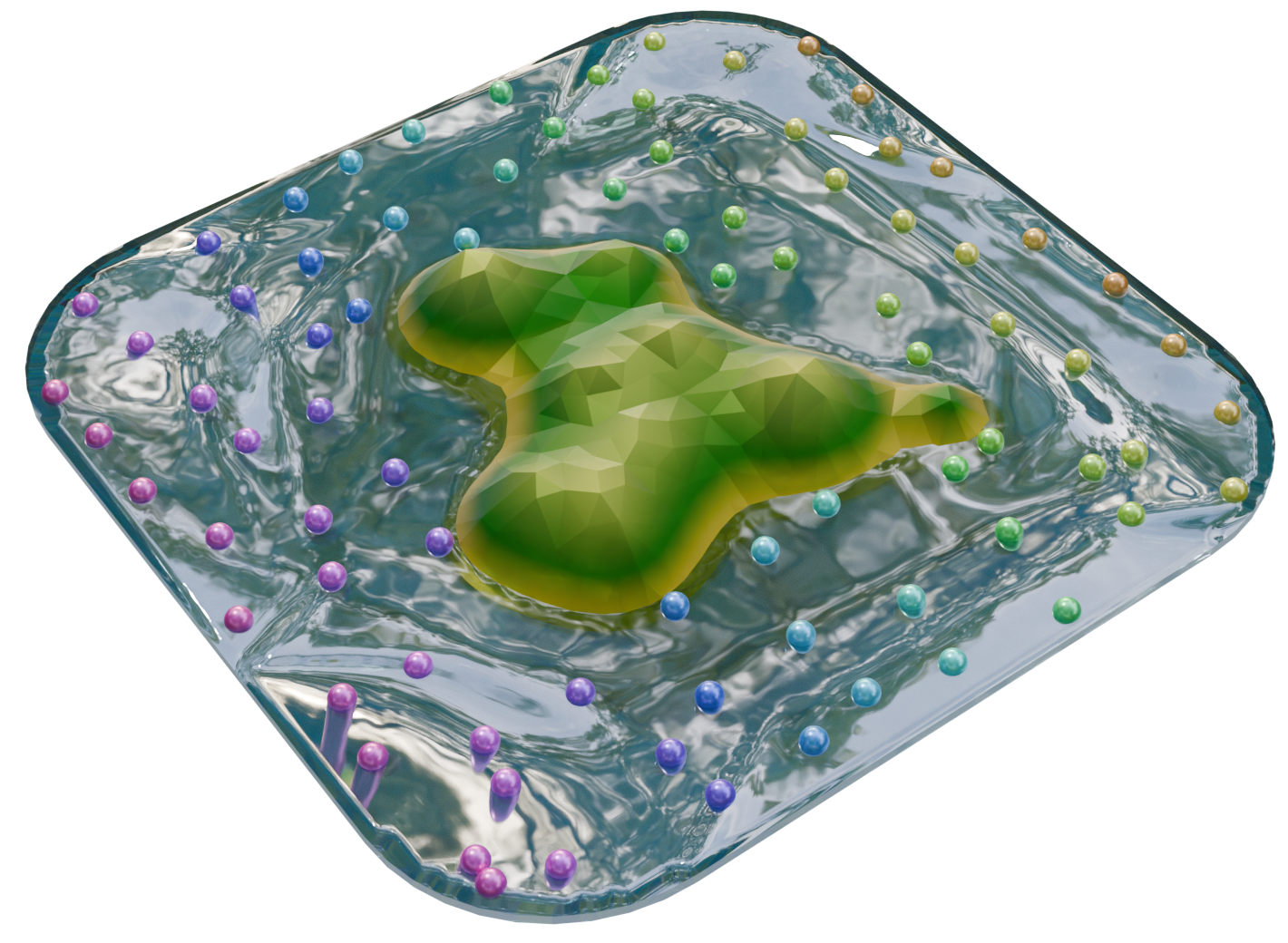}  \hfill
  \includegraphics[width=.21\linewidth]{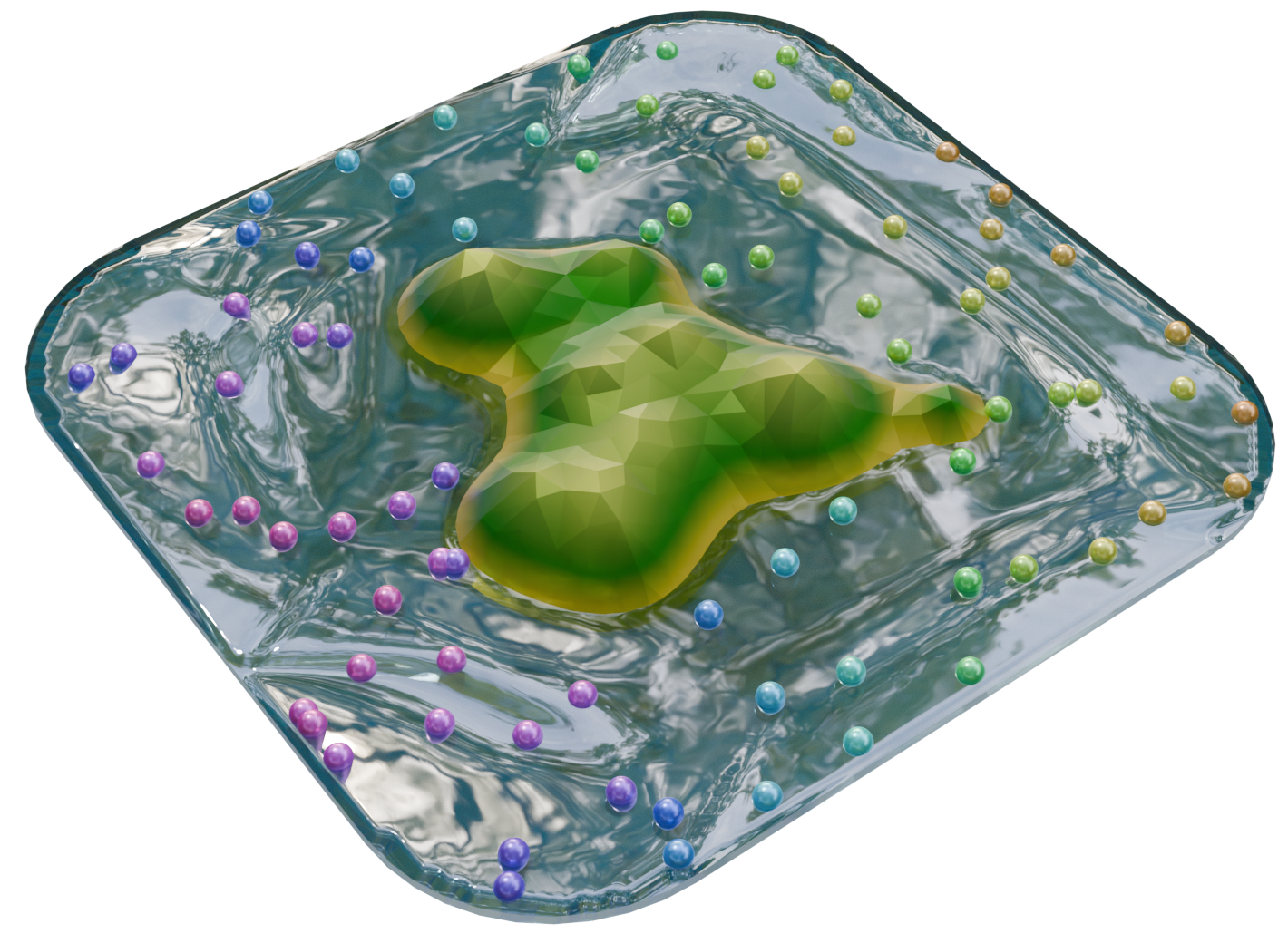}  \hfill
  \includegraphics[width=.21\linewidth]{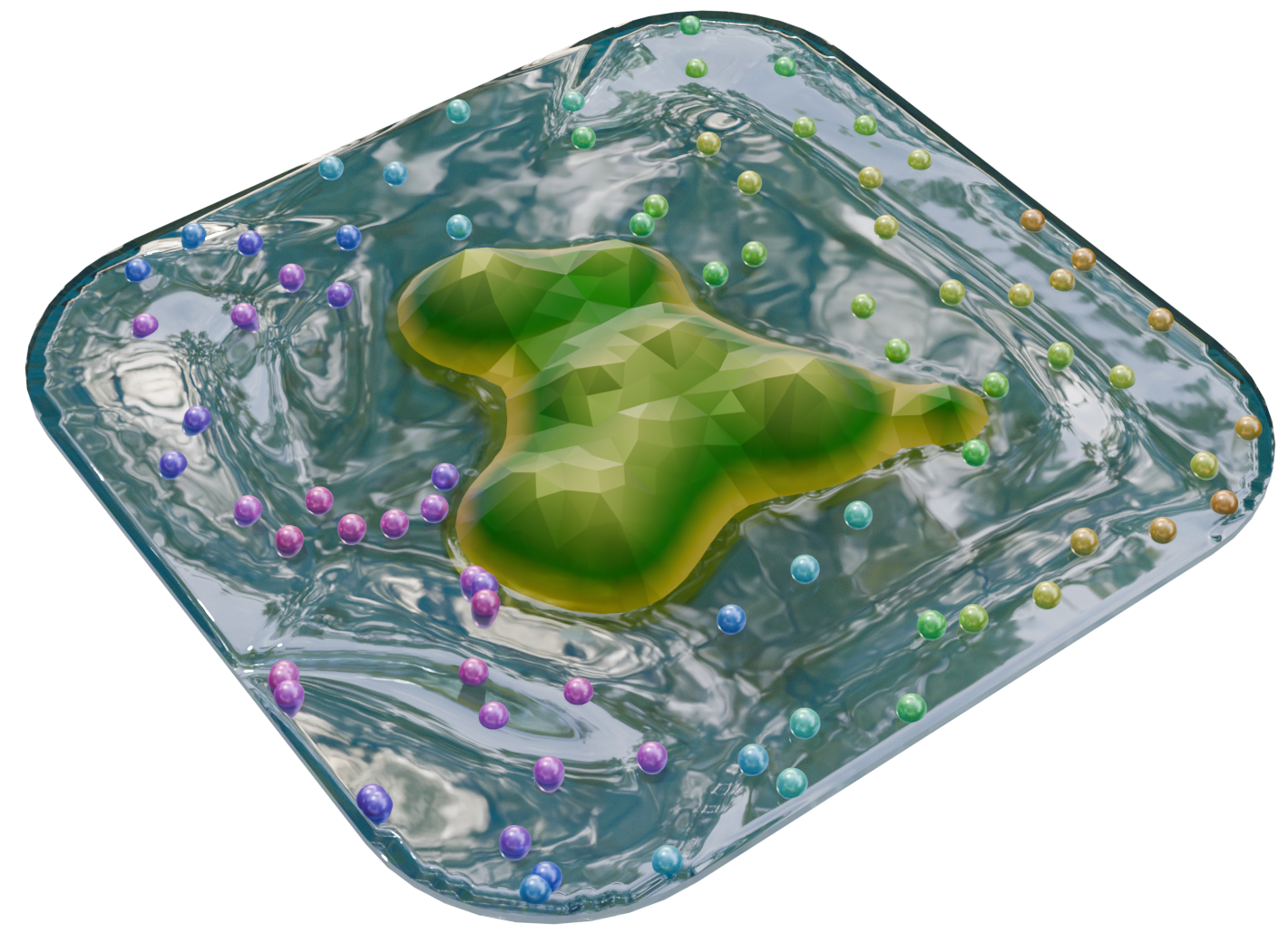}  \hfill
  \includegraphics[width=.21\linewidth]{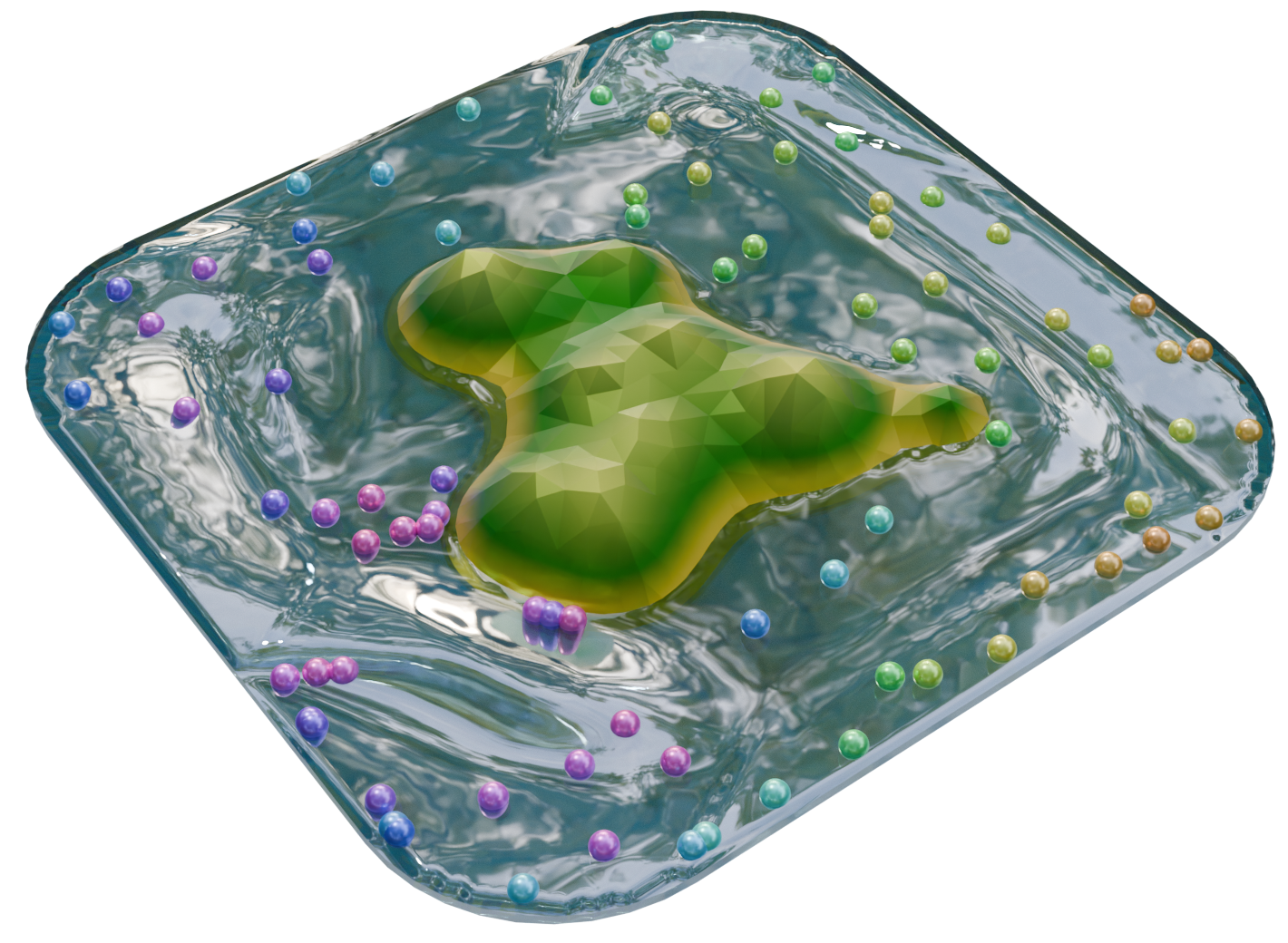}
    \parbox{.11\linewidth}{\centering Input curves}\hfill
  \parbox{.21\linewidth}{\centering $t=0$}\hfill
  \parbox{.21\linewidth}{\centering $t=0.15s$}\hfill
  \parbox{.21\linewidth}{\centering $t=0.25s$}\hfill
  \parbox{.21\linewidth}{\centering $t=0.37$}
  \caption{\review{Example of different fluid animation frames for different input sketches. The flow adapts to the position of the circles and the curves and produces natural animation. The top animation depicts a calm fluid surface, whereas the bottom demonstrates a turbulent scene with multiple vortices.}}
  \label{fig:anim2d}
\end{figure*}

\paragraph{Two-dimensional Editing}
\review{\Cref{fig:anim2d} shows how the input sketch leads to a realistic fluid simulation. If the input flow is tamed, the simulation simulation converges to a calm stationary velocity. By starting with more vortices, the flow remains turbulent even in later frames.}

\paragraph{Moving Boundary}
\review{
Our method naturally accommodates moving boundaries, as demonstrated by various examples involving translation and rotation for a single component and separate components, and even changes in the domain's topology. These animations reveal how interesting fluid's behaviors appears with such dynamics. Our method's ability to handle moving boundaries makes it well-suited for integration into game engines. It can enrich user experience by enabling real-time generation of physically realistic fluid flows when characters interact with water, such as walking or moving through it.

\Cref{fig:teaser} bottom, where a duck moves around a stationary rock, shows an example of moving boundaries with multiple disconnected components \revise{using our time-dependent bases (\Cref{fig:duck} top)}. From another angle (\Cref{fig:duck} bottom) we show that, as the duck pushes into different corners, pre-existing vortices are compressed and eventually dissipated.

\begin{figure*}
  \centering\footnotesize
  \includegraphics[width=.19\linewidth,trim={0.38cm 0.38cm 0.38cm 0.38cm},clip]{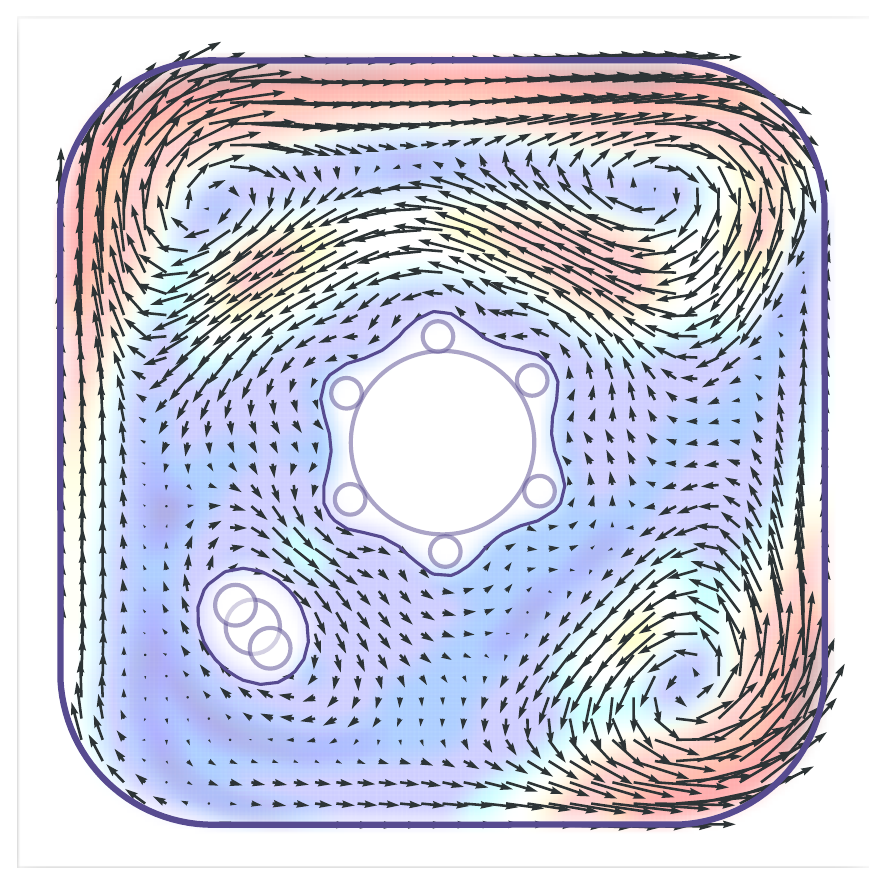}  \hfill
  \includegraphics[width=.19\linewidth,trim={0.38cm 0.38cm 0.38cm 0.38cm},clip]{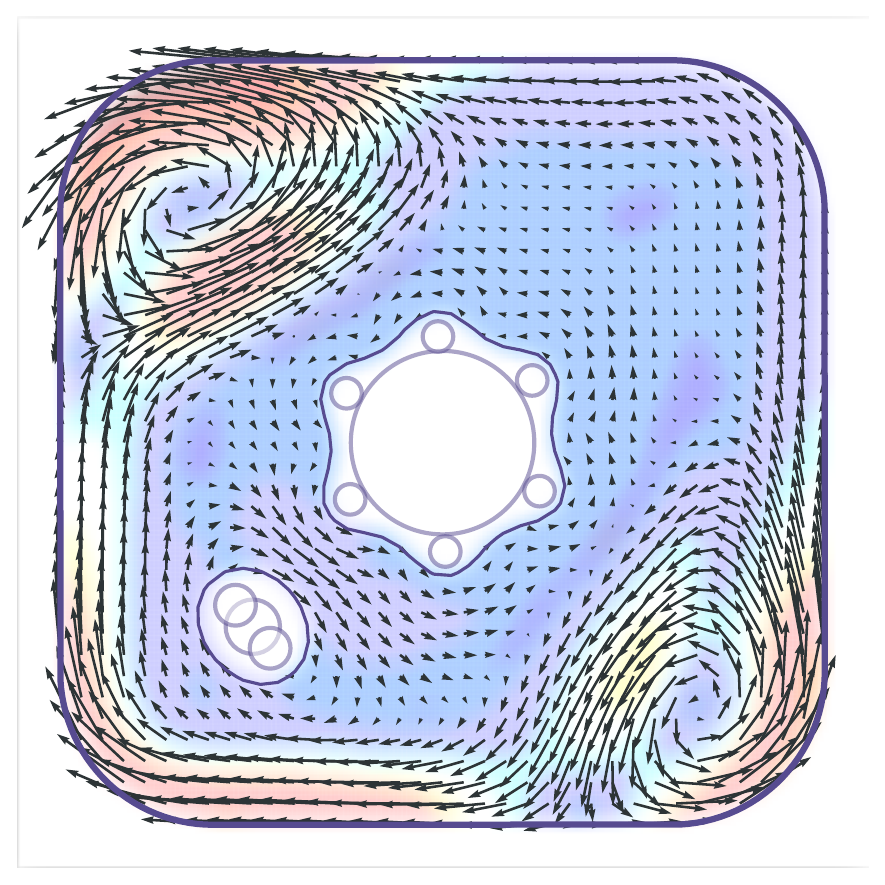}  \hfill
  \includegraphics[width=.19\linewidth,trim={0.38cm 0.38cm 0.38cm 0.38cm},clip]{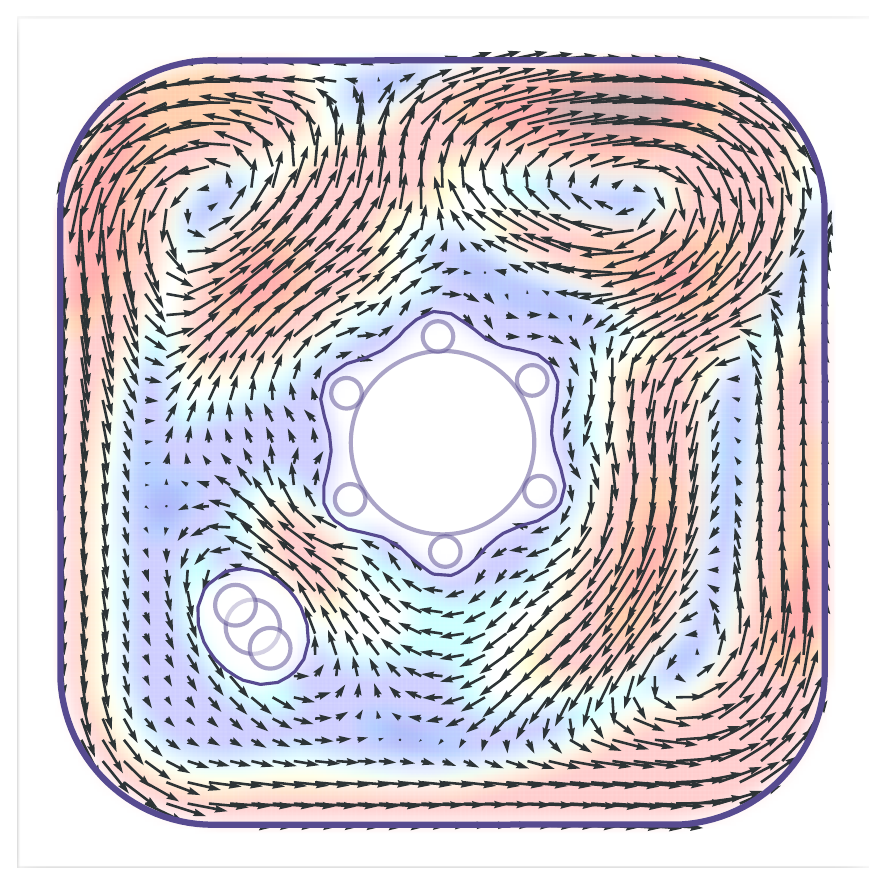}  \hfill
  \includegraphics[width=.19\linewidth,trim={0.38cm 0.38cm 0.38cm 0.38cm},clip]{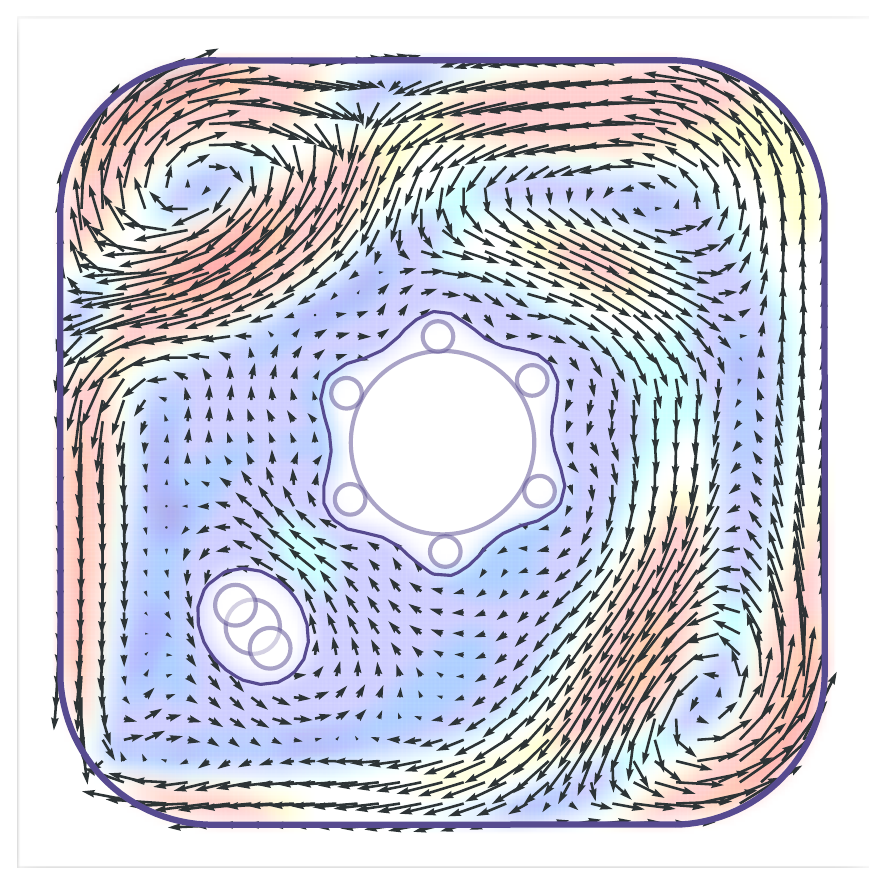}  \hfill
  \includegraphics[width=.19\linewidth,trim={0.38cm 0.38cm 0.38cm 0.38cm},clip]{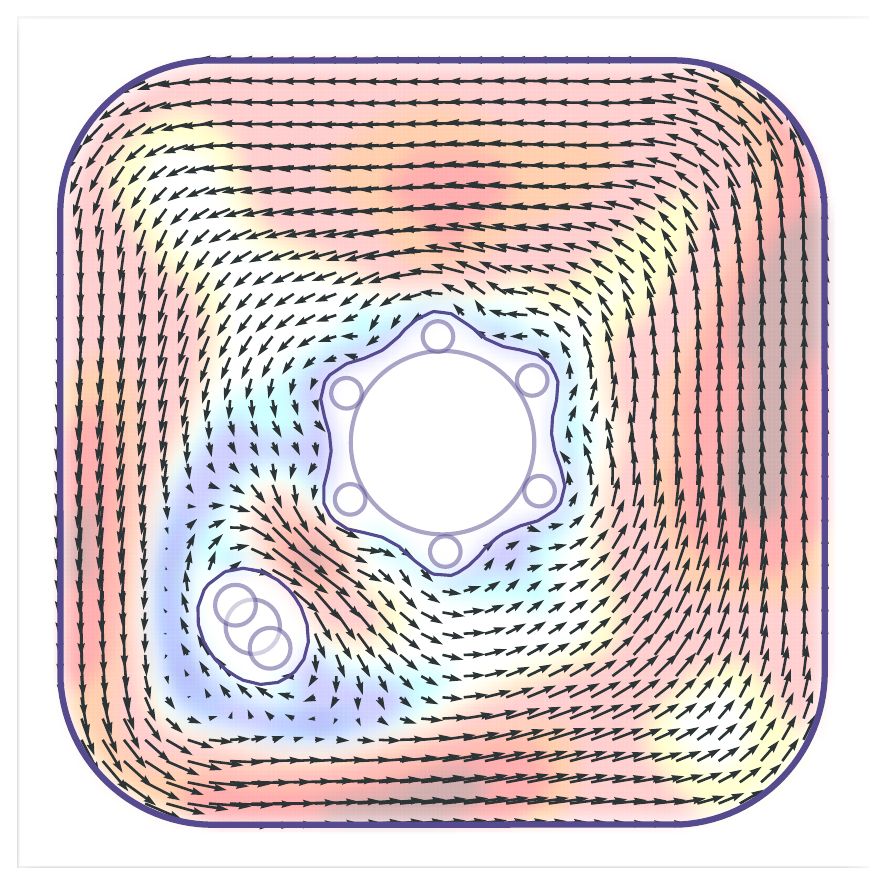}  \hfill
  
  \includegraphics[width=.24\linewidth,clip]{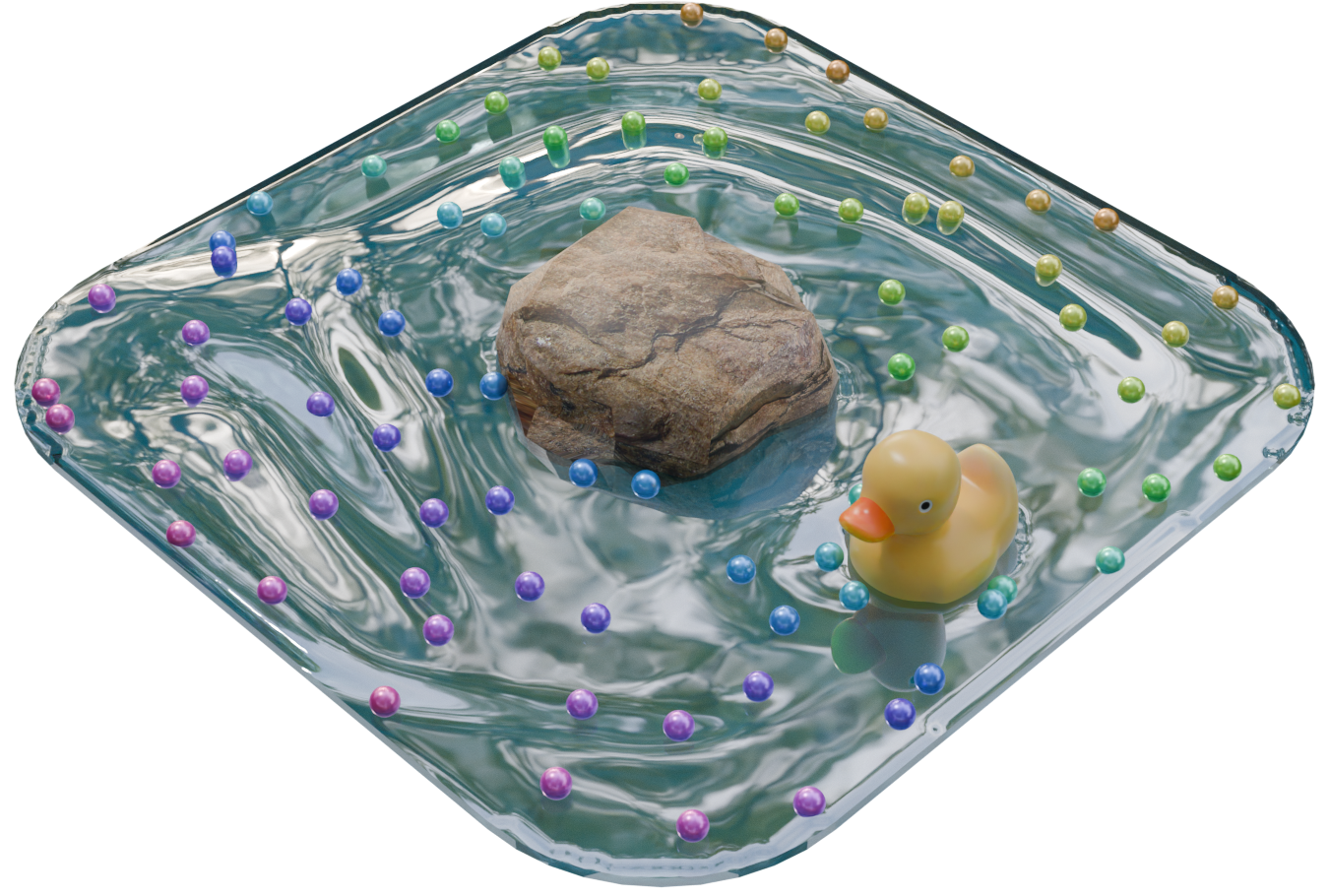}  \hfill
  \includegraphics[width=.24\linewidth,clip]{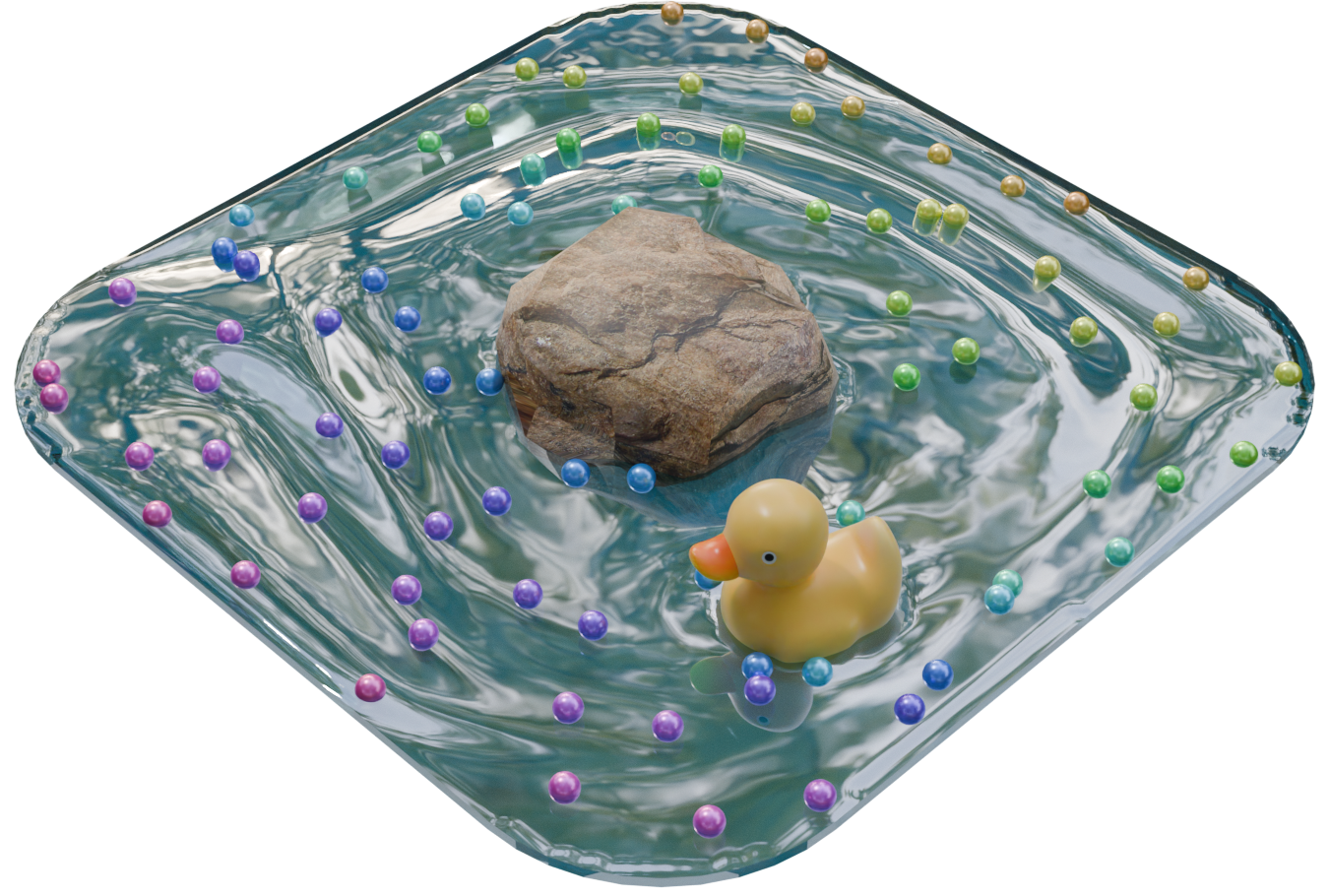}  \hfill
  \includegraphics[width=.24\linewidth,clip]{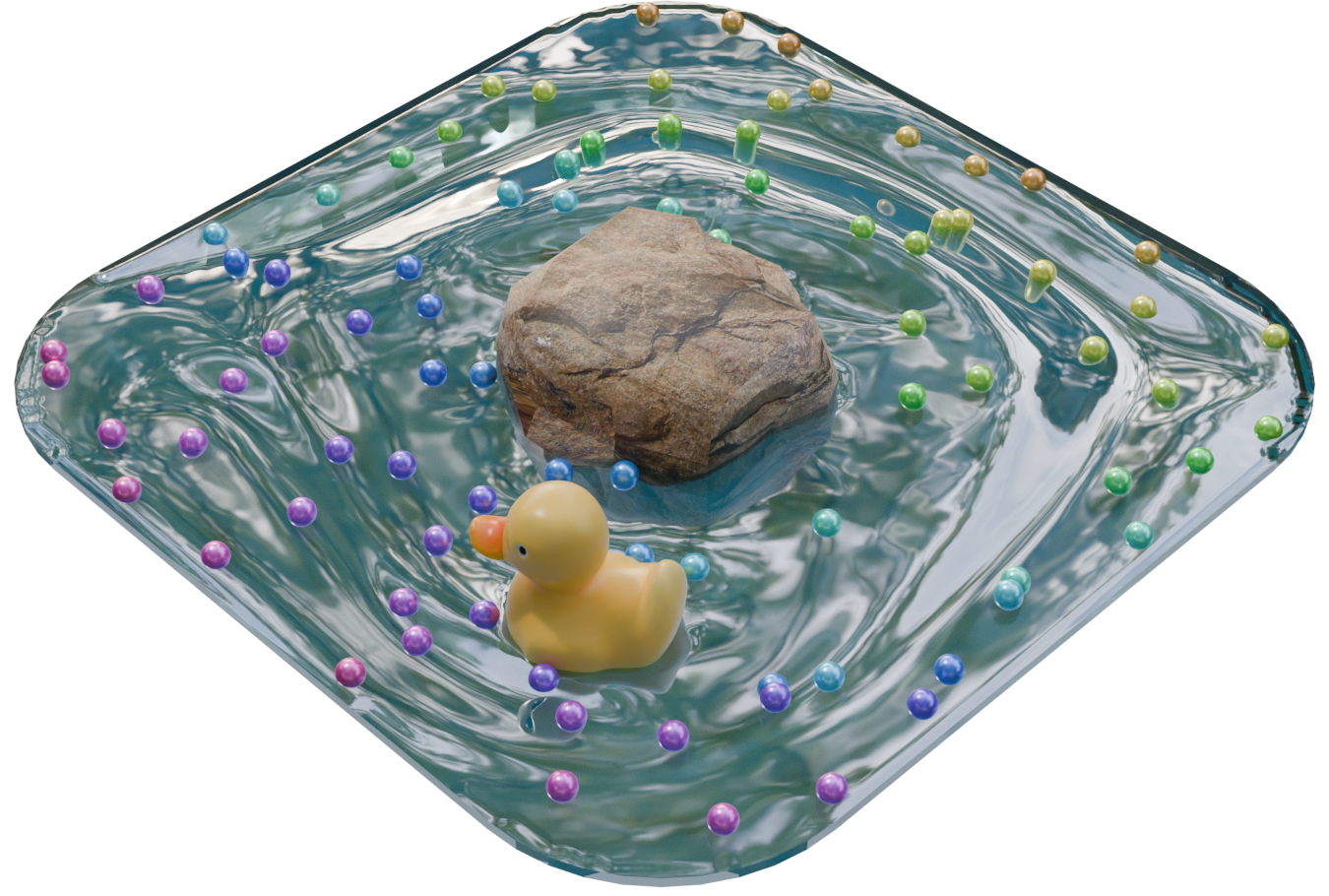}  \hfill
  \includegraphics[width=.24\linewidth,clip]{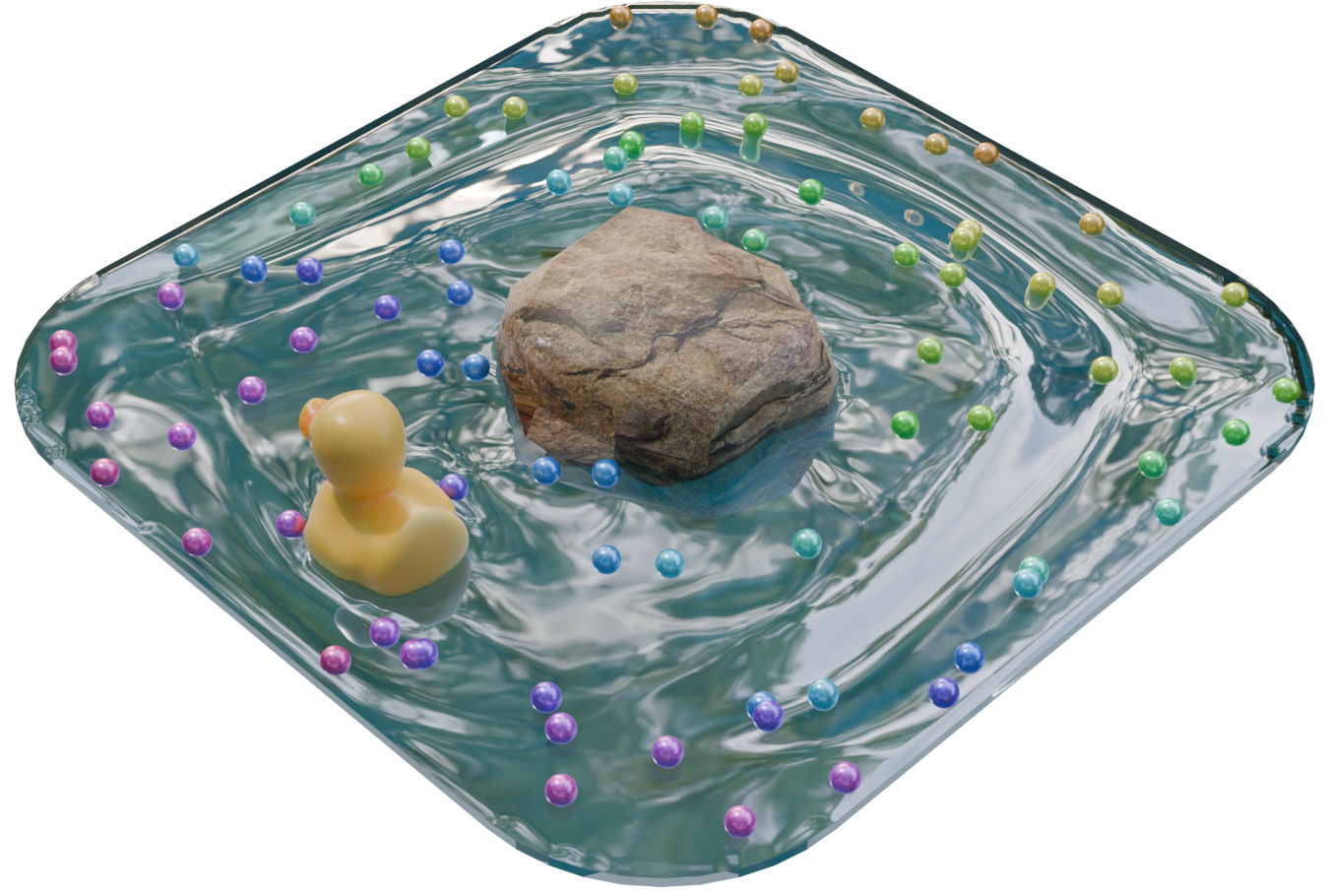}  \par
  
  \parbox{.24\linewidth}{\centering$t=0.025s$}\hfill
  \parbox{.24\linewidth}{\centering$t=0.0375s$}\hfill
  \parbox{.24\linewidth}{\centering$t=0.055s$}\hfill
  \parbox{.24\linewidth}{\centering$t=0.07s$}\hfill
  \caption{\review{A rubber duck moves around a stationary rock (bottom). As it pushes toward the bottom and left corners, the vortices in those regions are compressed and gradually vanish due to the interaction with the duck's movement.} \revise{The top shows the bases for frame 0.}}
  \label{fig:duck}
\end{figure*}

We first illustrate the effect of a spinning fan at different speeds (\Cref{fig:fan}). When the fan spins rapidly, particles are pushed away from the blade region, effectively preventing them from entering. In contrast, with a slower rotation, particles are able to approach and circulate near the blades.

\begin{figure*}
  \centering\footnotesize

Slow fan.\\
  \includegraphics[width=.19\linewidth,clip]{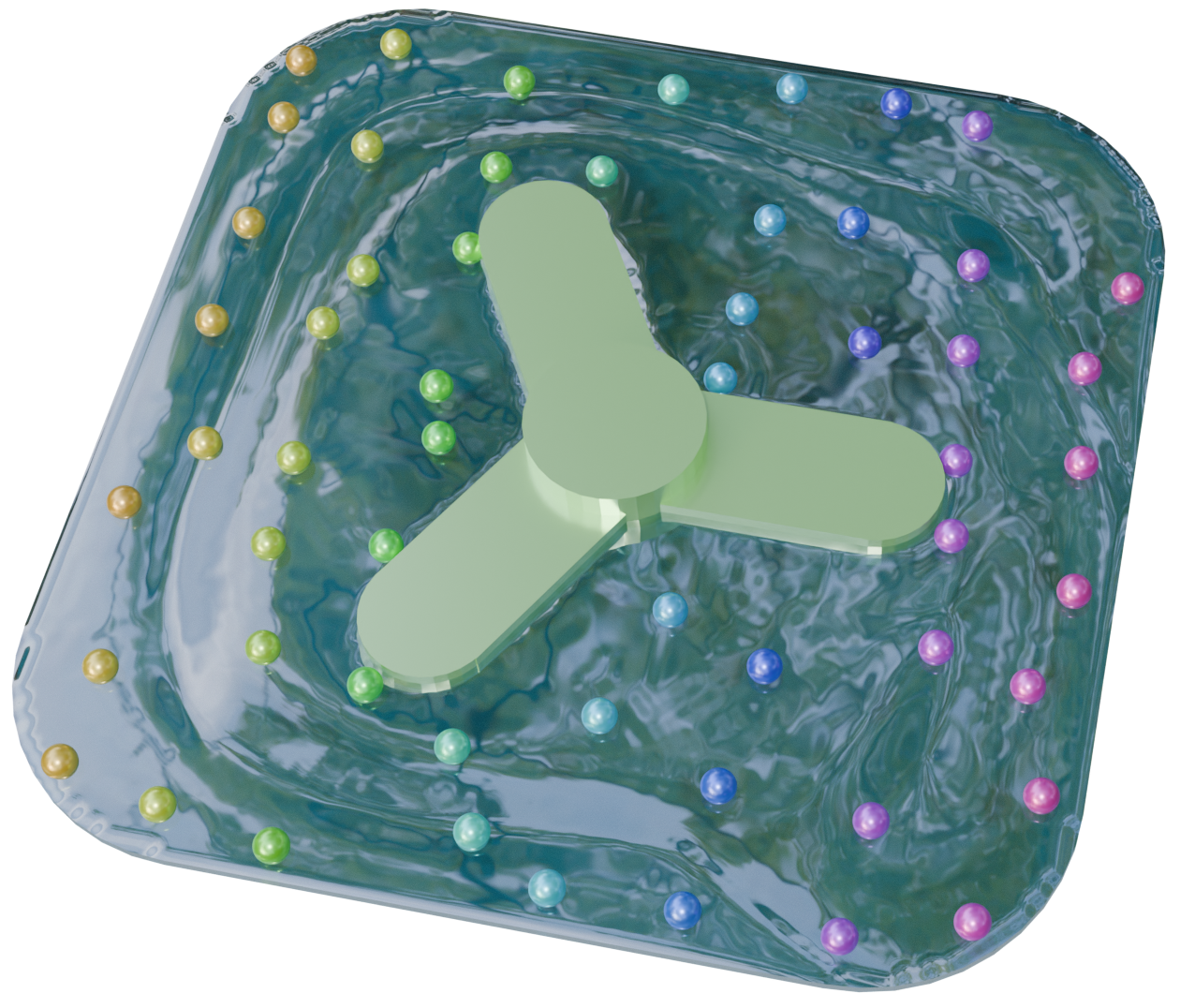}  \hfill
  \includegraphics[width=.19\linewidth,clip]{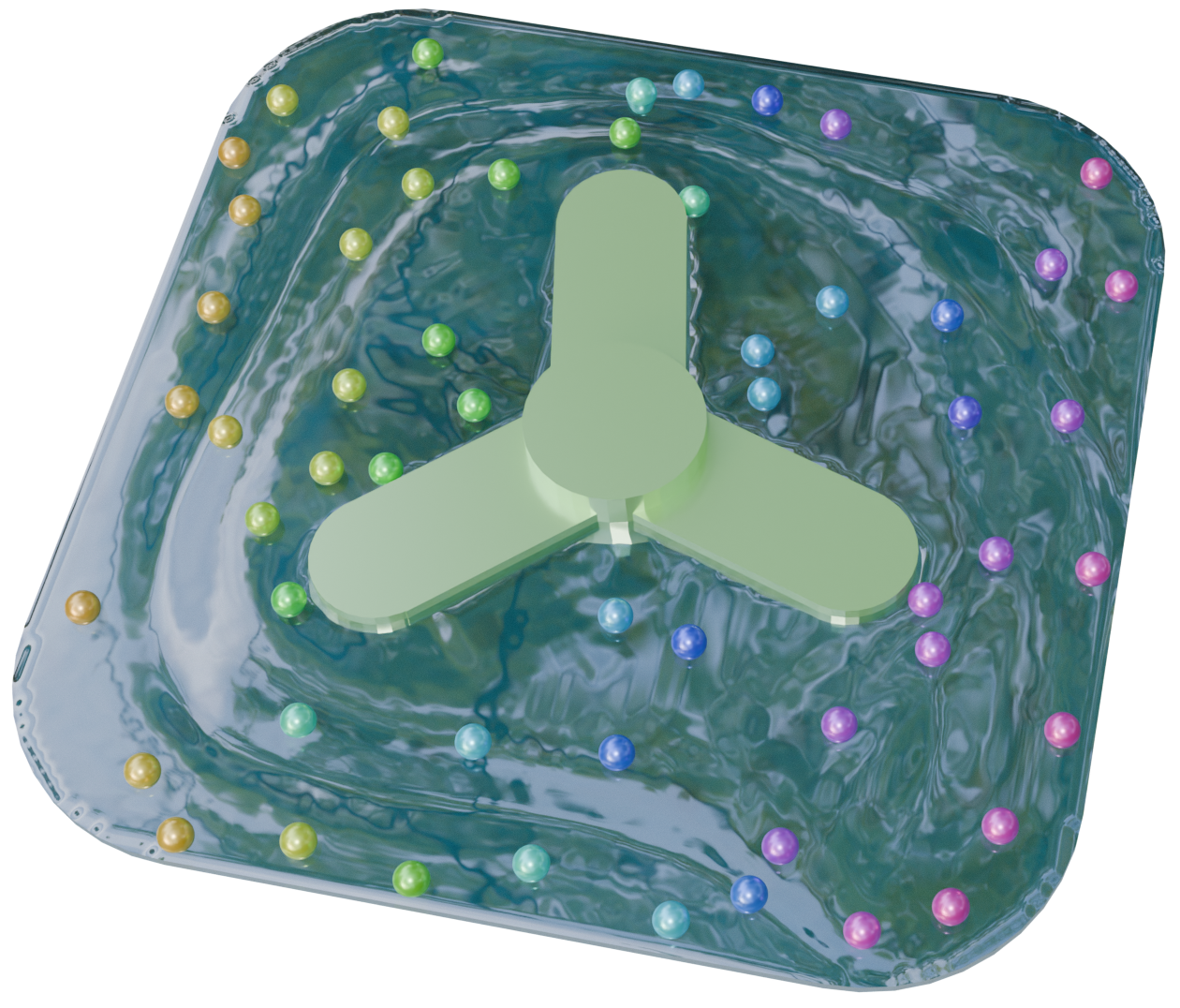}  \hfill
  \includegraphics[width=.19\linewidth,clip]{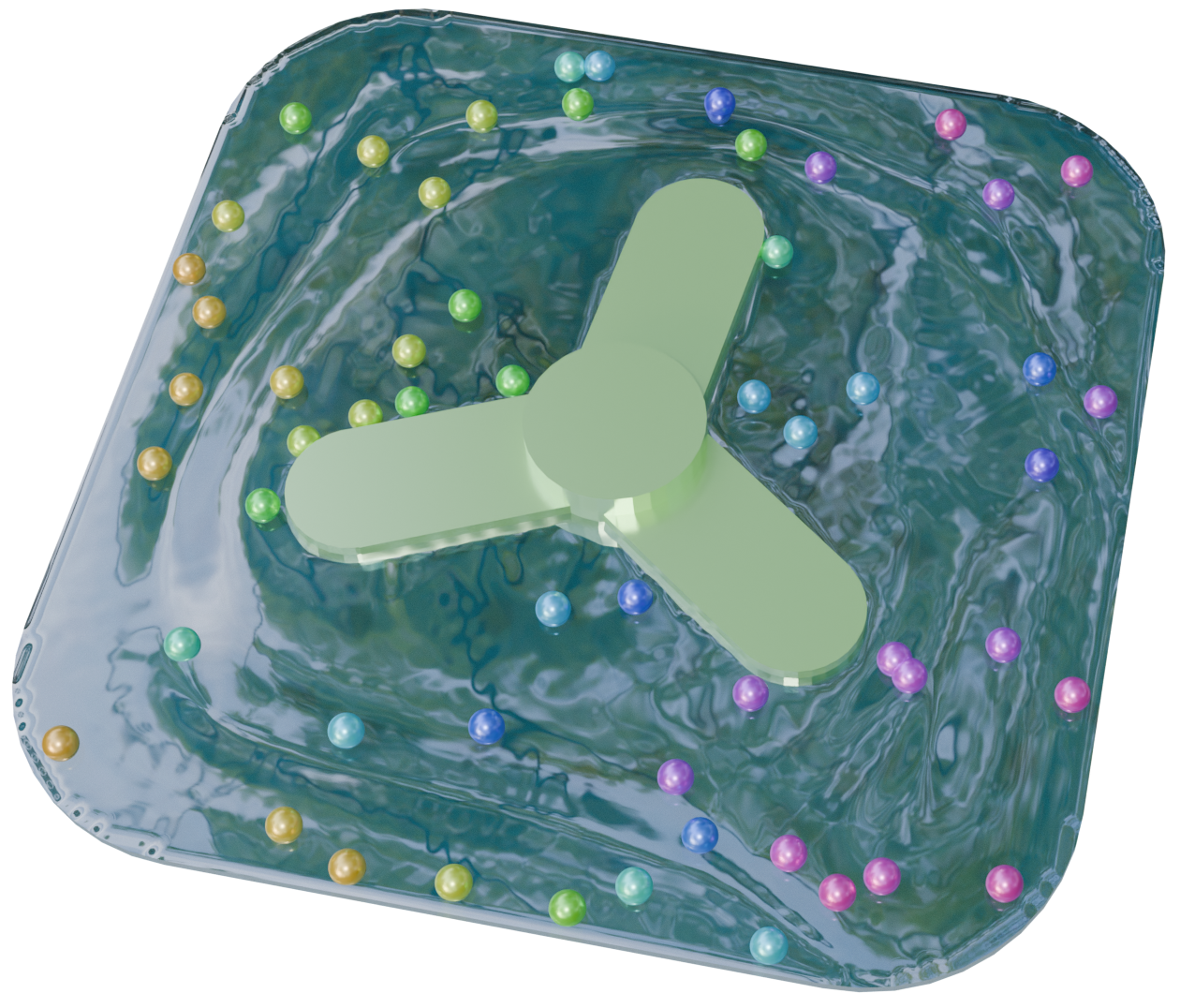}  \hfill
  \includegraphics[width=.19\linewidth,clip]{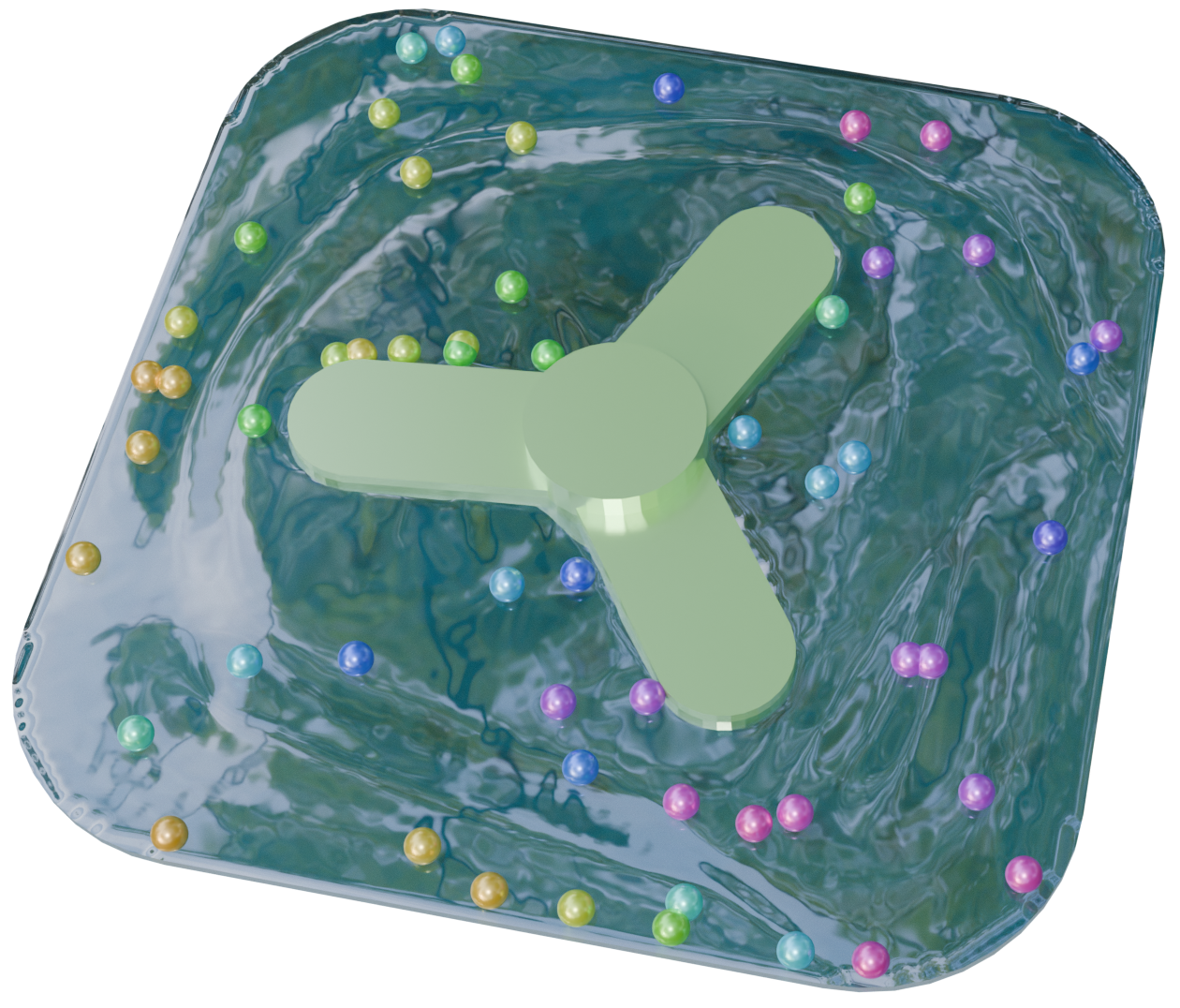}  \hfill
  \includegraphics[width=.19\linewidth,clip]{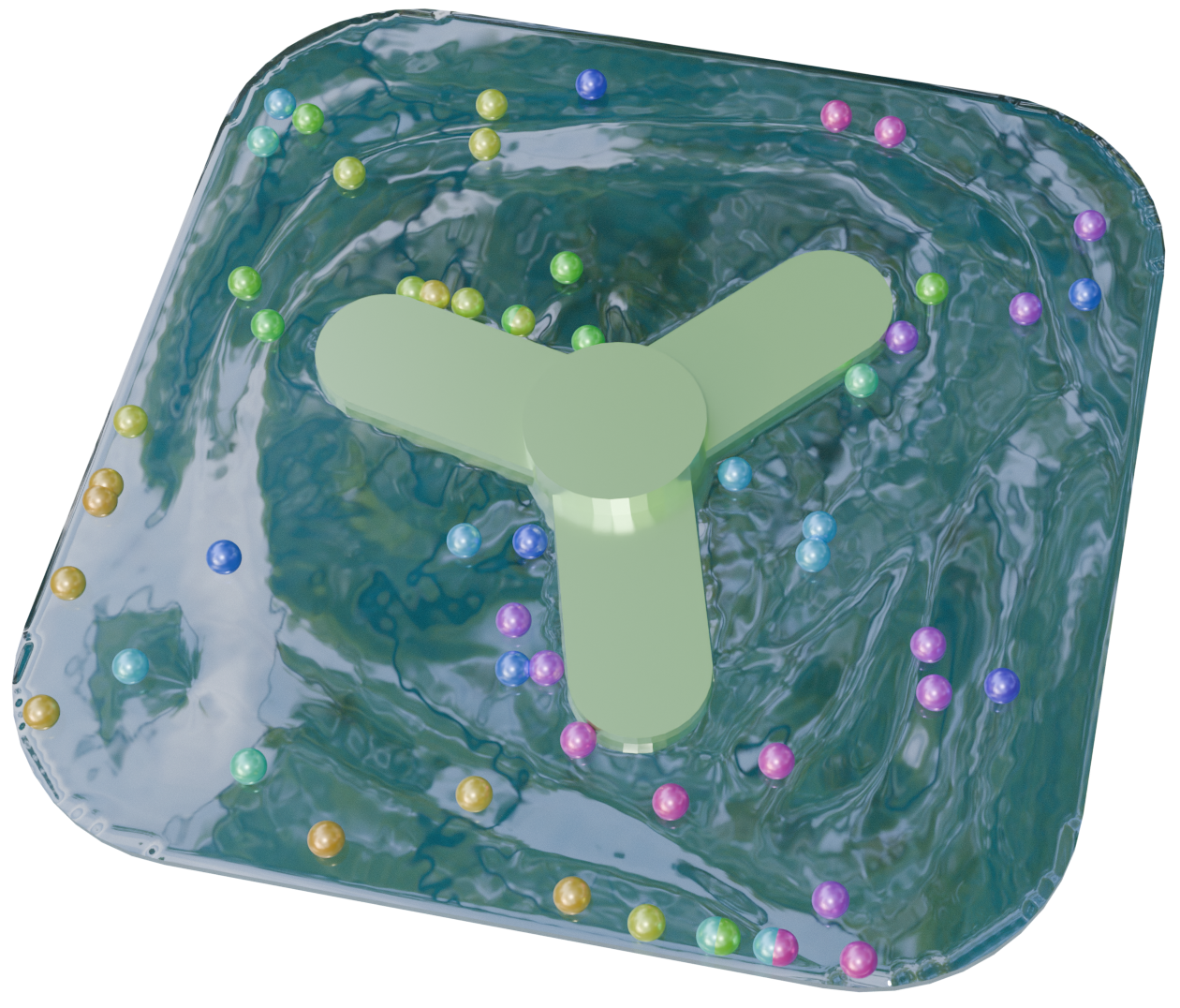}   \\[1em]
  
  Fast fan.\\
  \includegraphics[width=.19\linewidth,clip]{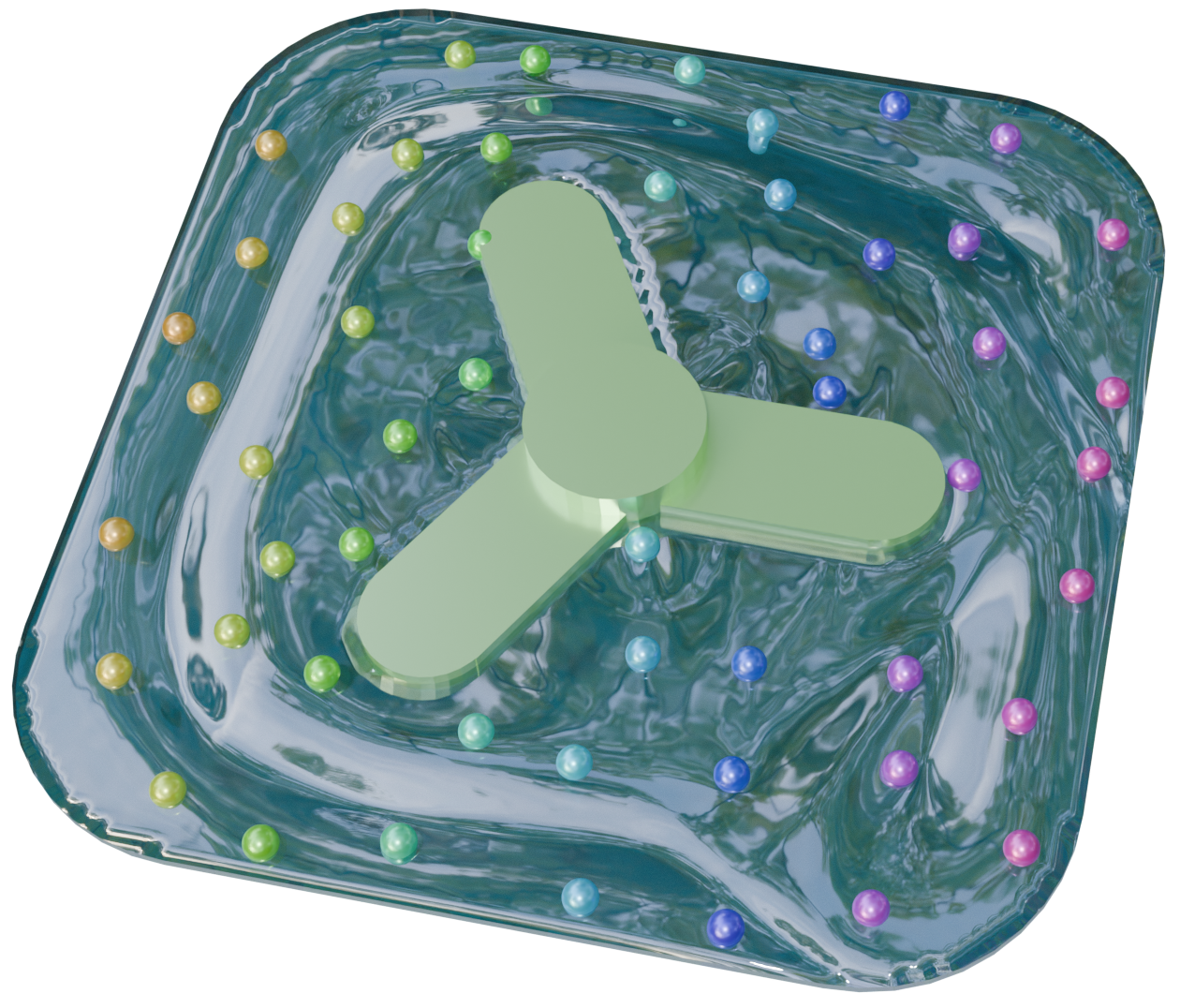}  \hfill
  \includegraphics[width=.19\linewidth,clip]{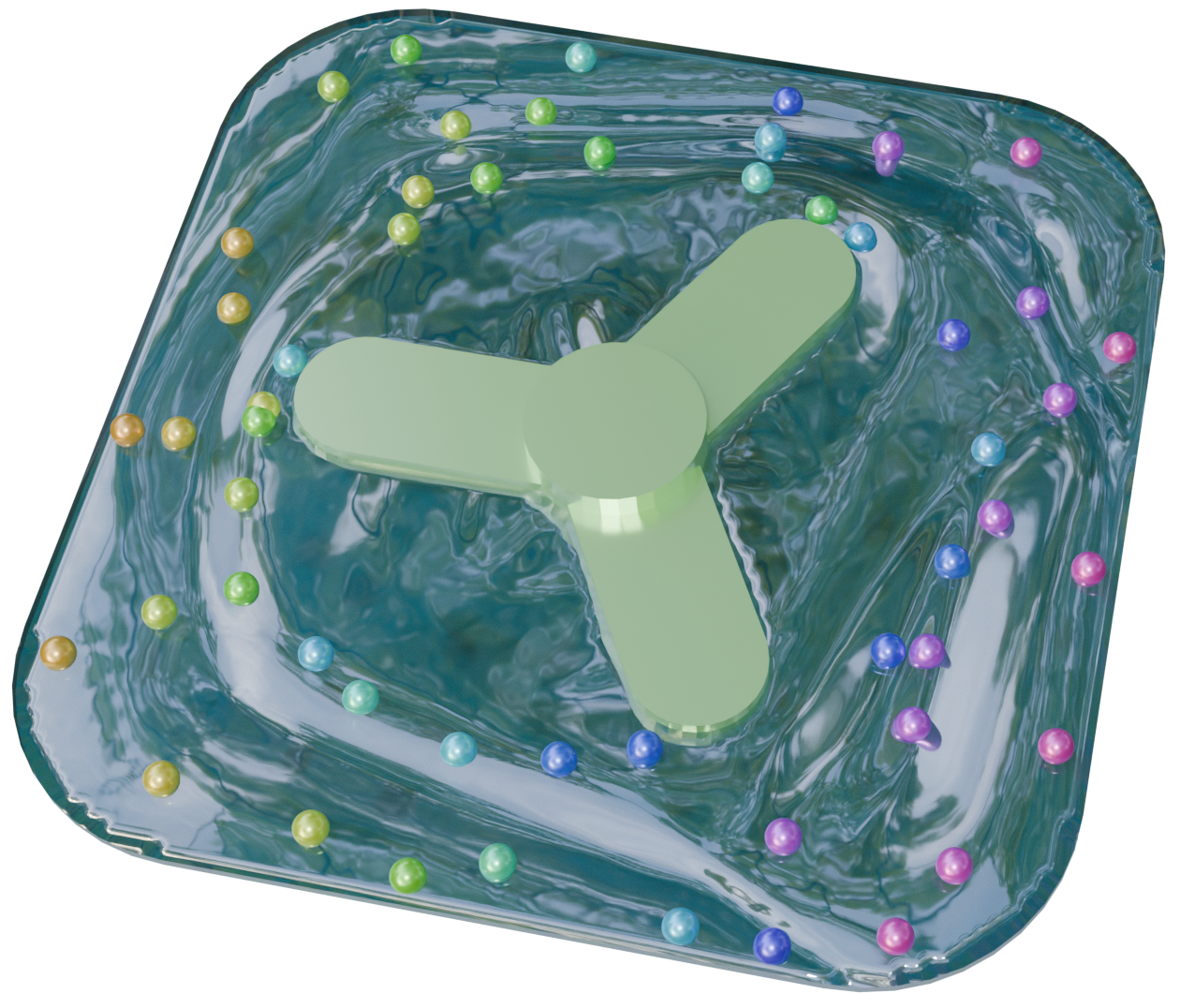}  \hfill
  \includegraphics[width=.19\linewidth,clip]{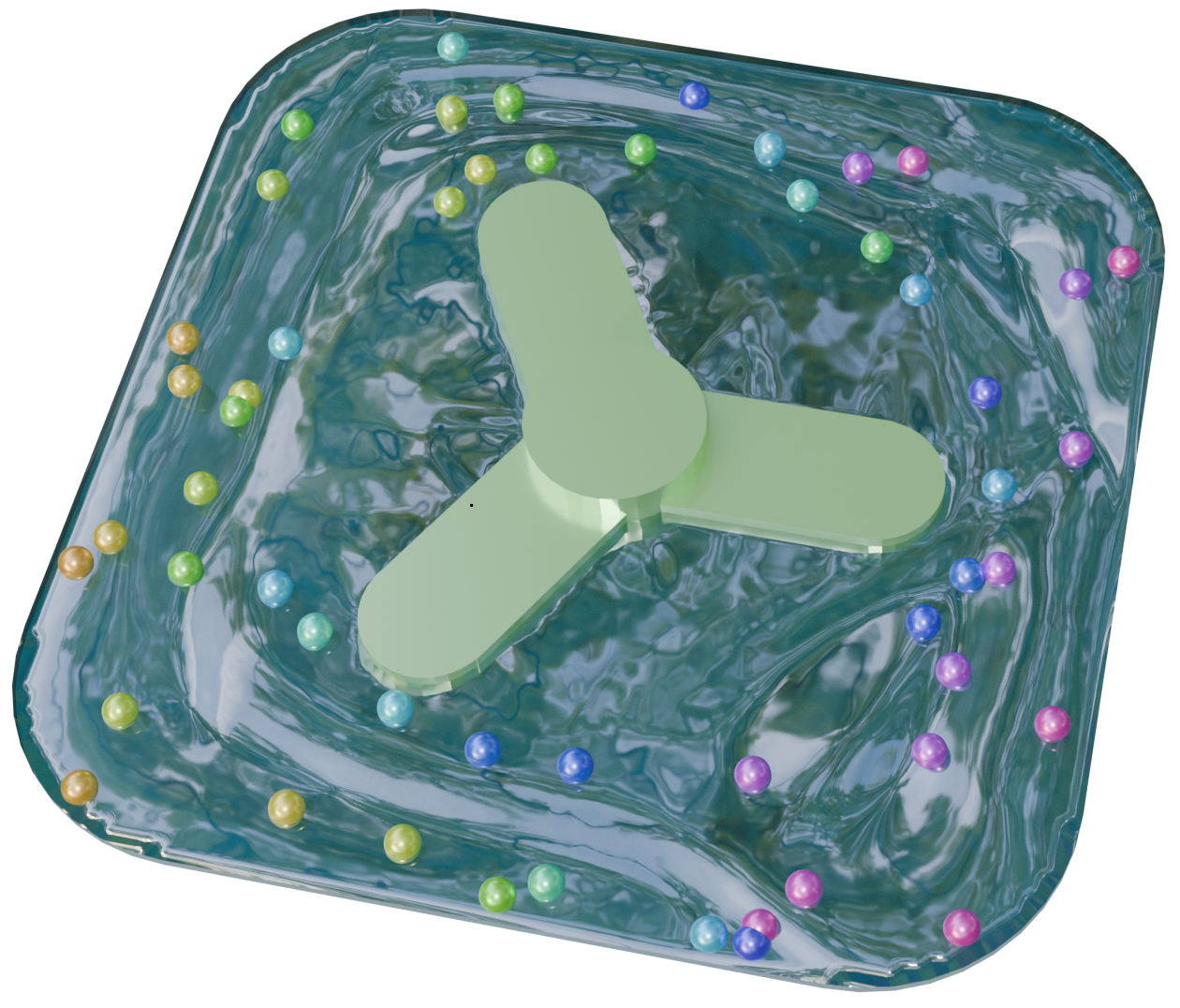}  \hfill
  \includegraphics[width=.19\linewidth,clip]{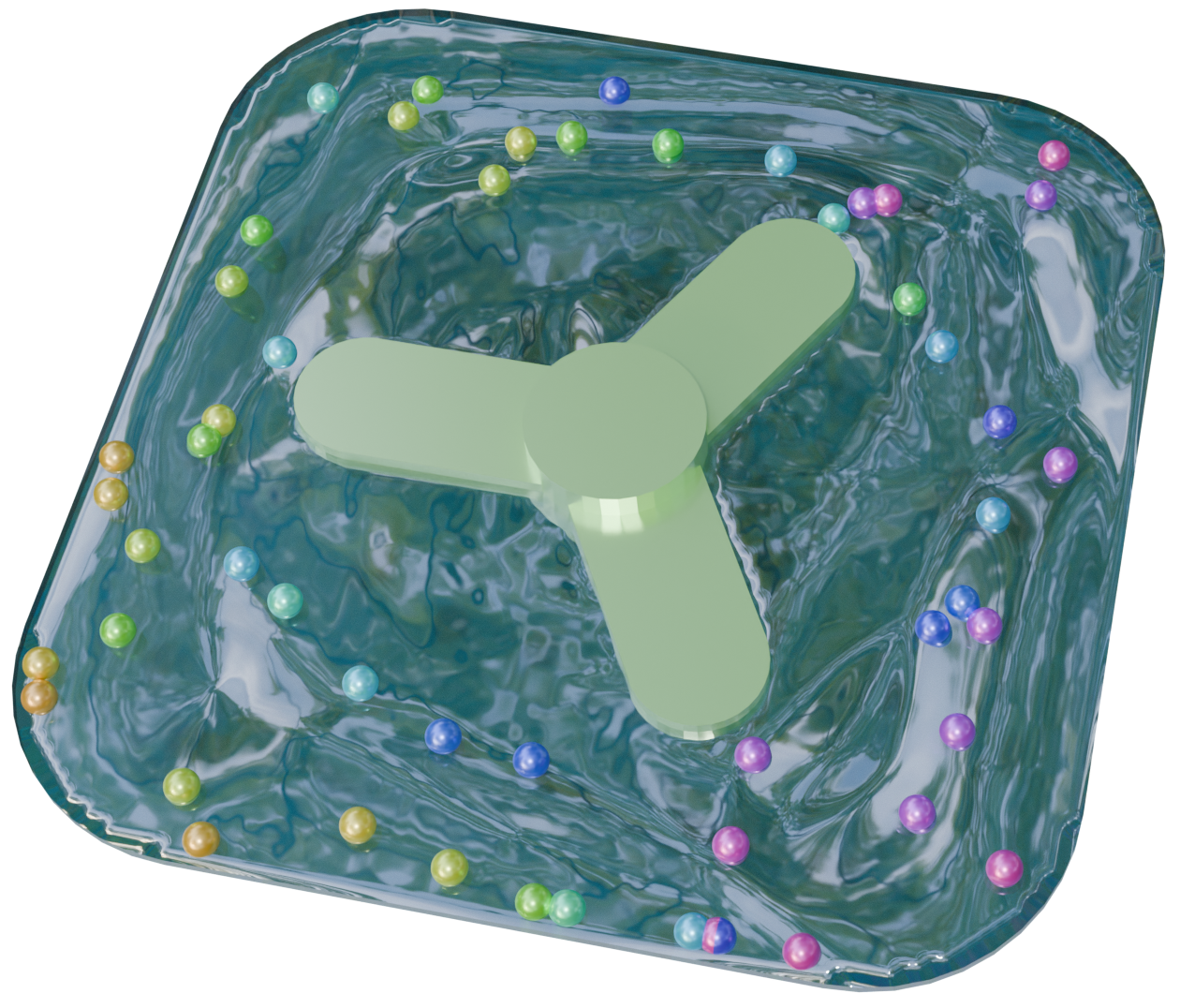}  \hfill
  \includegraphics[width=.19\linewidth,clip]{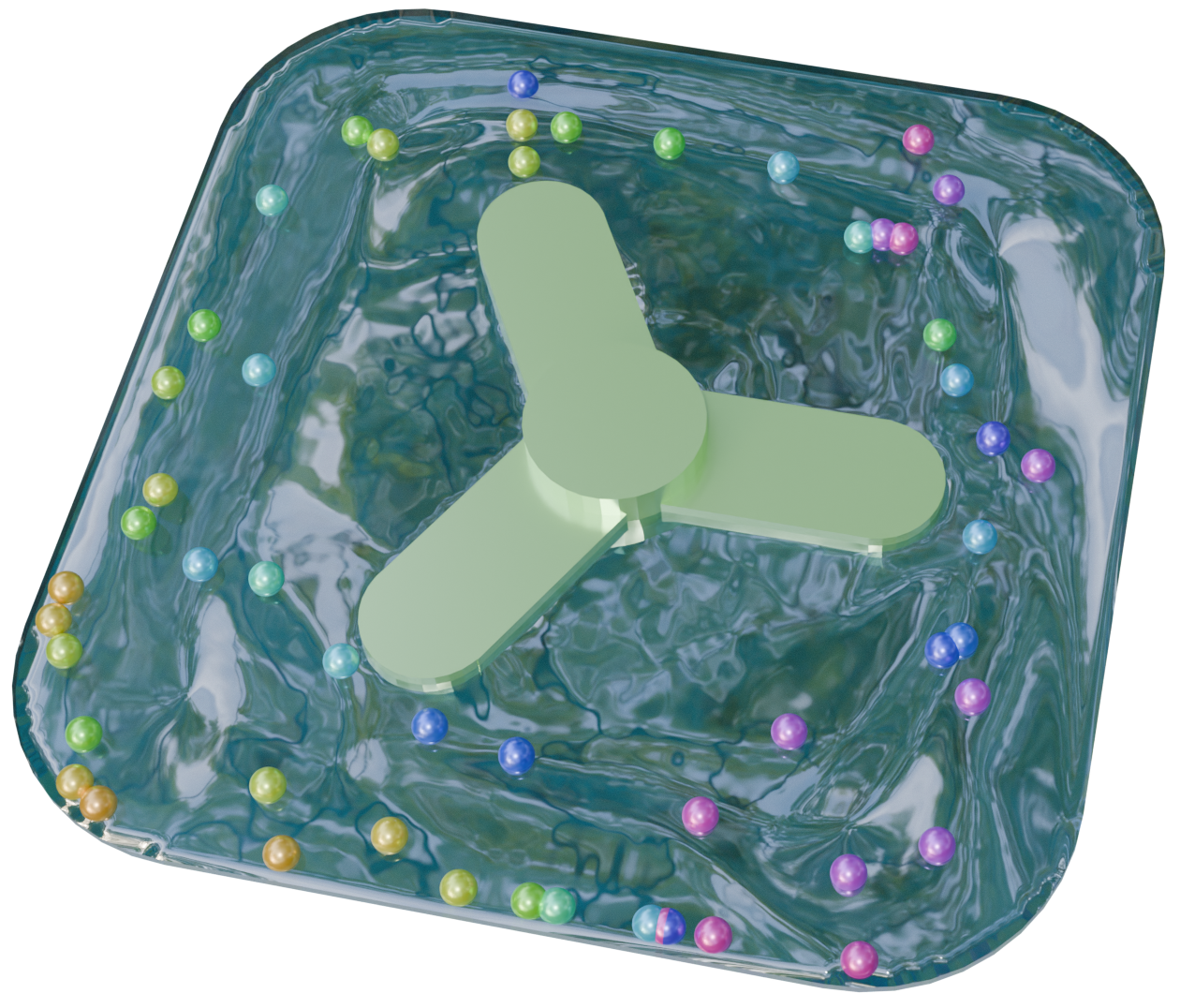}   \par
  
  \parbox{.19\linewidth}{\centering$t=0s$}\hfill
  \parbox{.19\linewidth}{\centering$t=0.25s$}\hfill
  \parbox{.19\linewidth}{\centering$t=0.5s$}\hfill
  \parbox{.19\linewidth}{\centering$t=0.75s$}\hfill
  \parbox{.19\linewidth}{\centering$t=1s$}
  \caption{\review{Two-dimensional fluid animation using our neural kinematic bases. We visualize the magnitude of the velocity as a height-field in the background, while we advect particles to better illustrate movement.}}
  \label{fig:fan}
\end{figure*}

\Cref{fig:gripper} shows a gripper opening and closing, altering the genus of the domain. During the closing motion, both inward and outward flows appears and some particles are carried into the gripper while others are pushed out. Once the gripper is fully closed, the flow circulates around it, forming a loop and particles inside the gripper become trapped. When the gripper reopens, the flow pattern reverses, and the trapped particles are released while new particles are drawn in by the surrounding flow.}

\begin{figure*}
  \centering\footnotesize
  \includegraphics[width=.19\linewidth,trim={80pt 0pt 80pt 0pt}, clip]{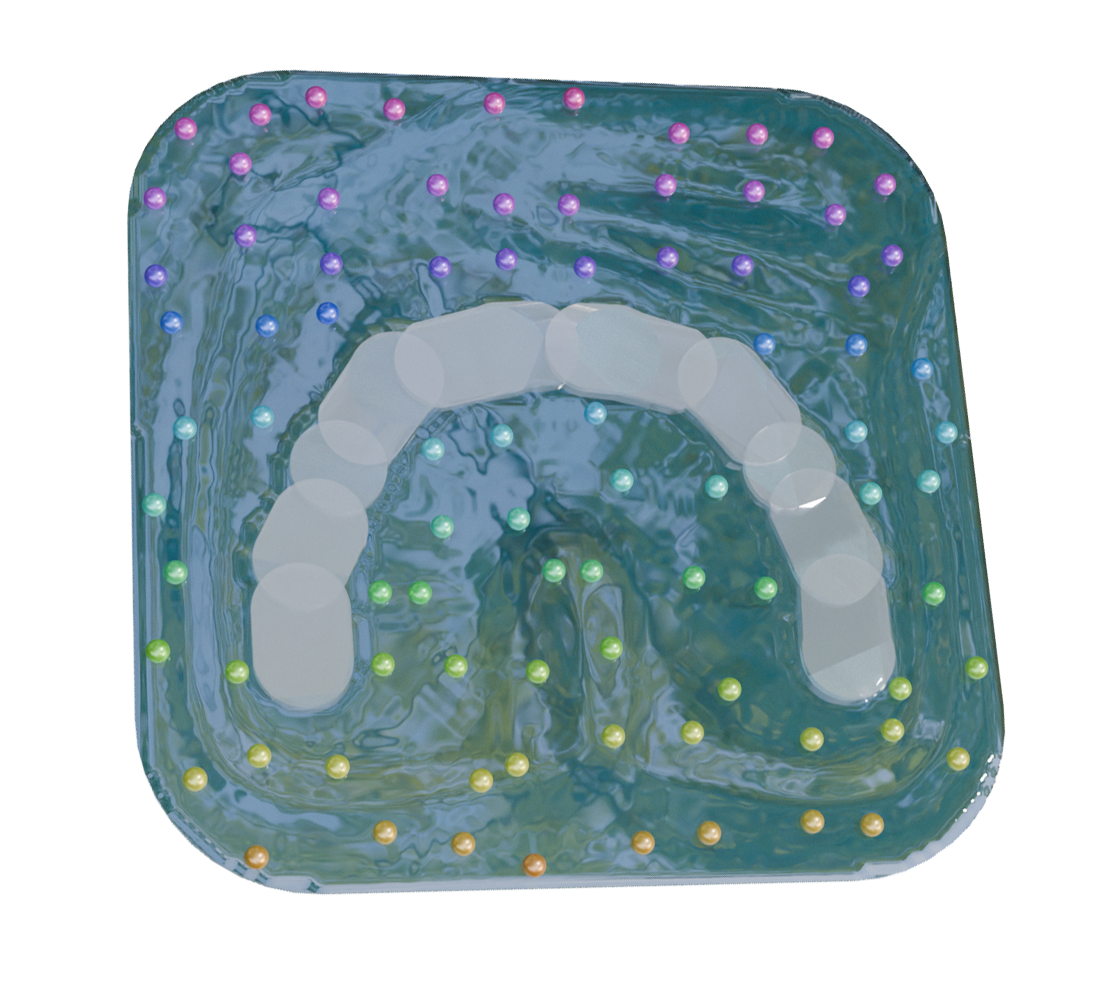}  \hfill
  \includegraphics[width=.19\linewidth,trim={80pt 0pt 80pt 0pt},clip]{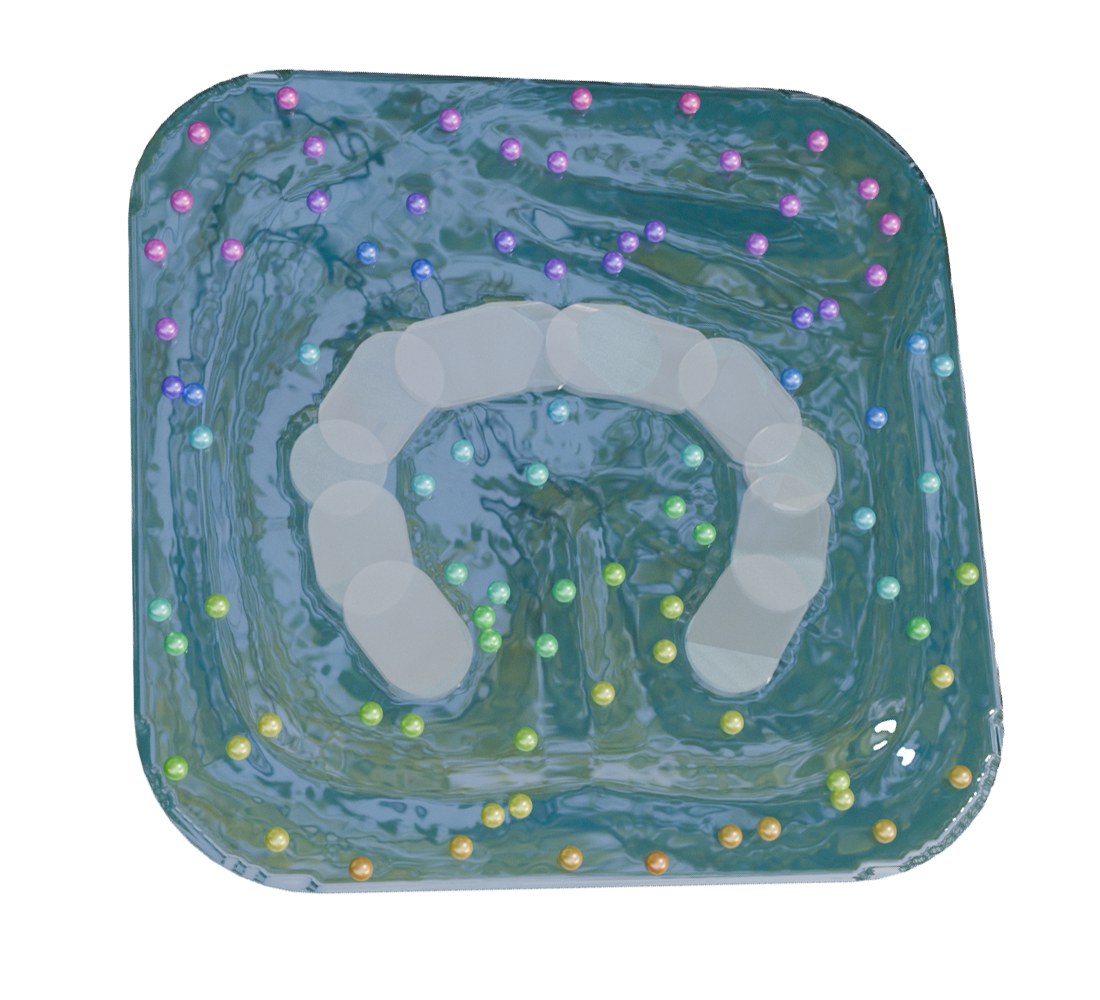}  \hfill
  \includegraphics[width=.19\linewidth,trim={80pt 0pt 80pt 0pt},clip]{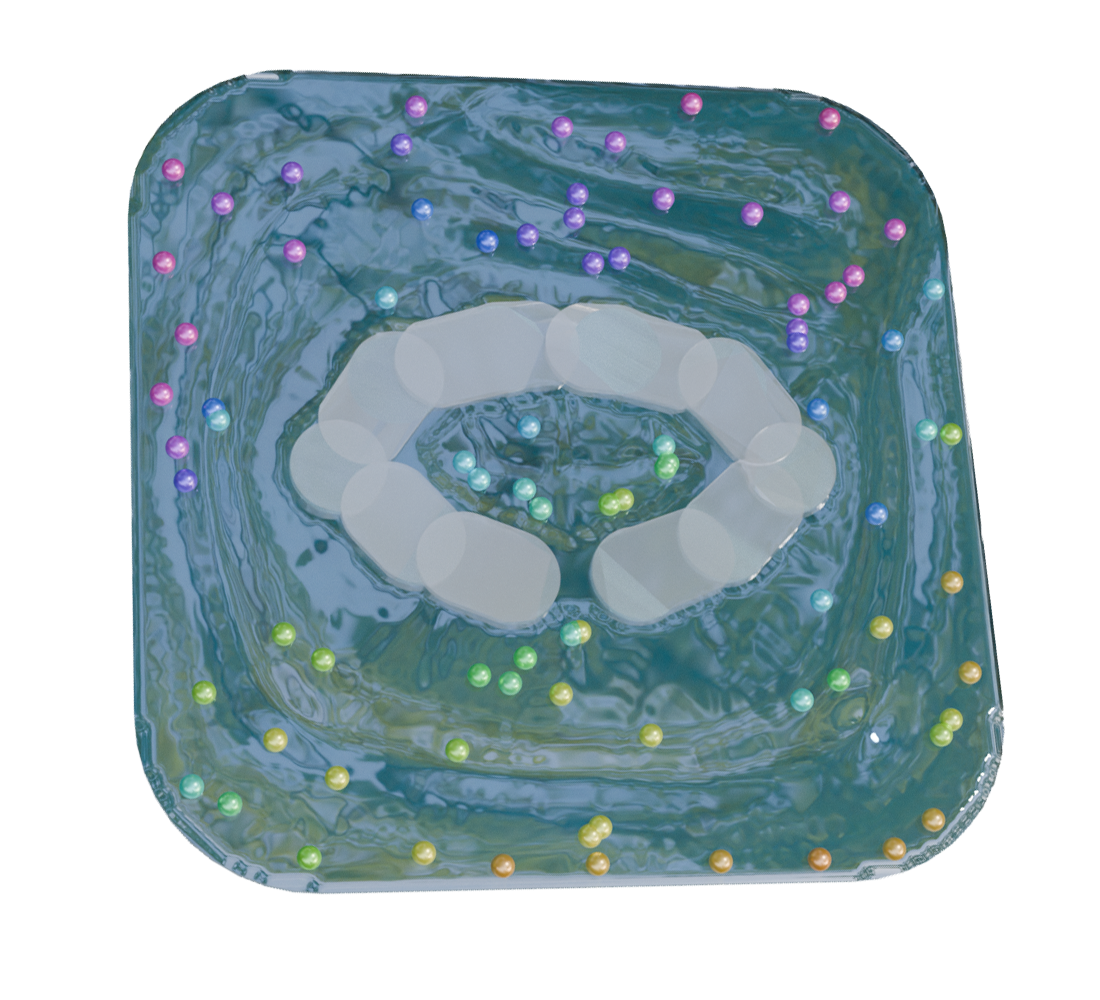}  \hfill
  \includegraphics[width=.19\linewidth,trim={80pt 0pt 80pt 0pt},clip]{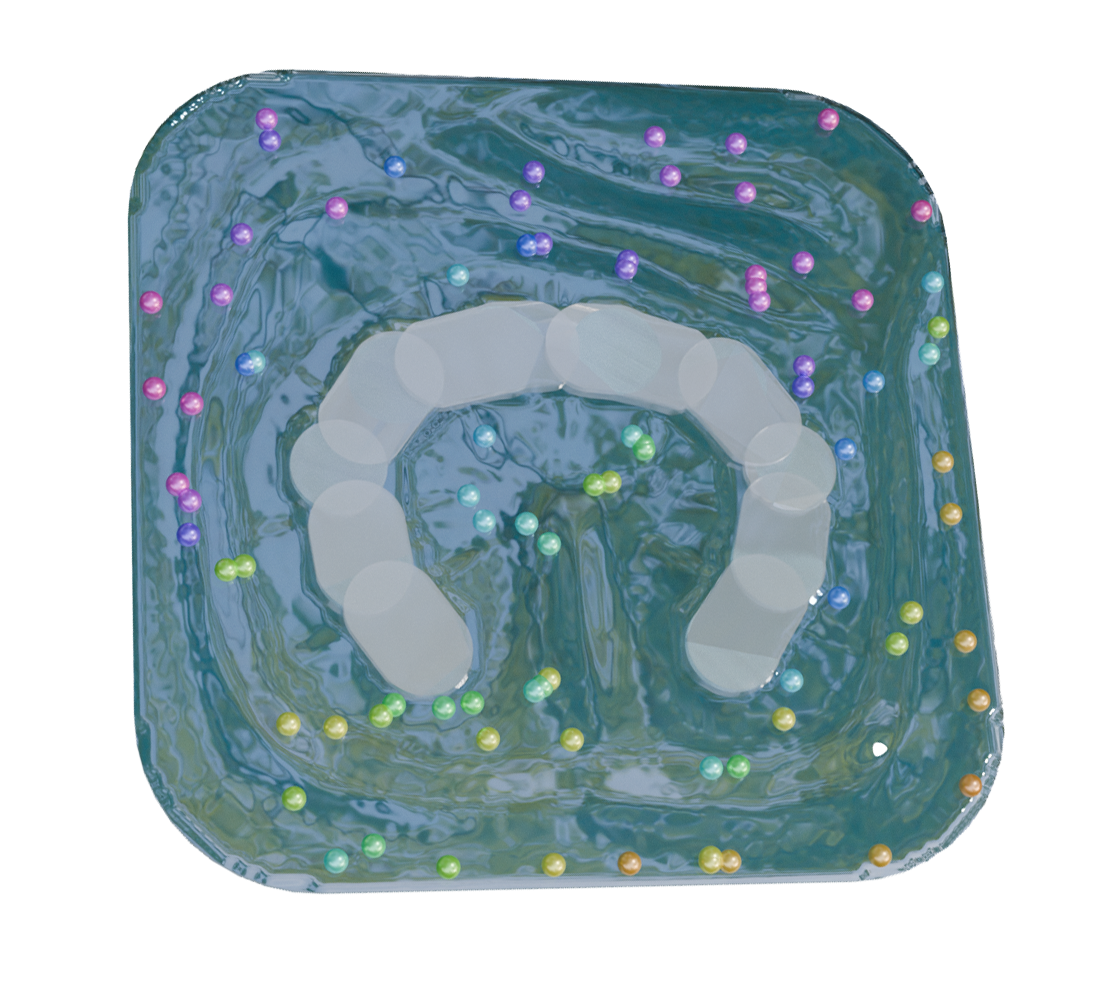}  \hfill
  \includegraphics[width=.19\linewidth,trim={80pt 0pt 80pt 0pt},clip]{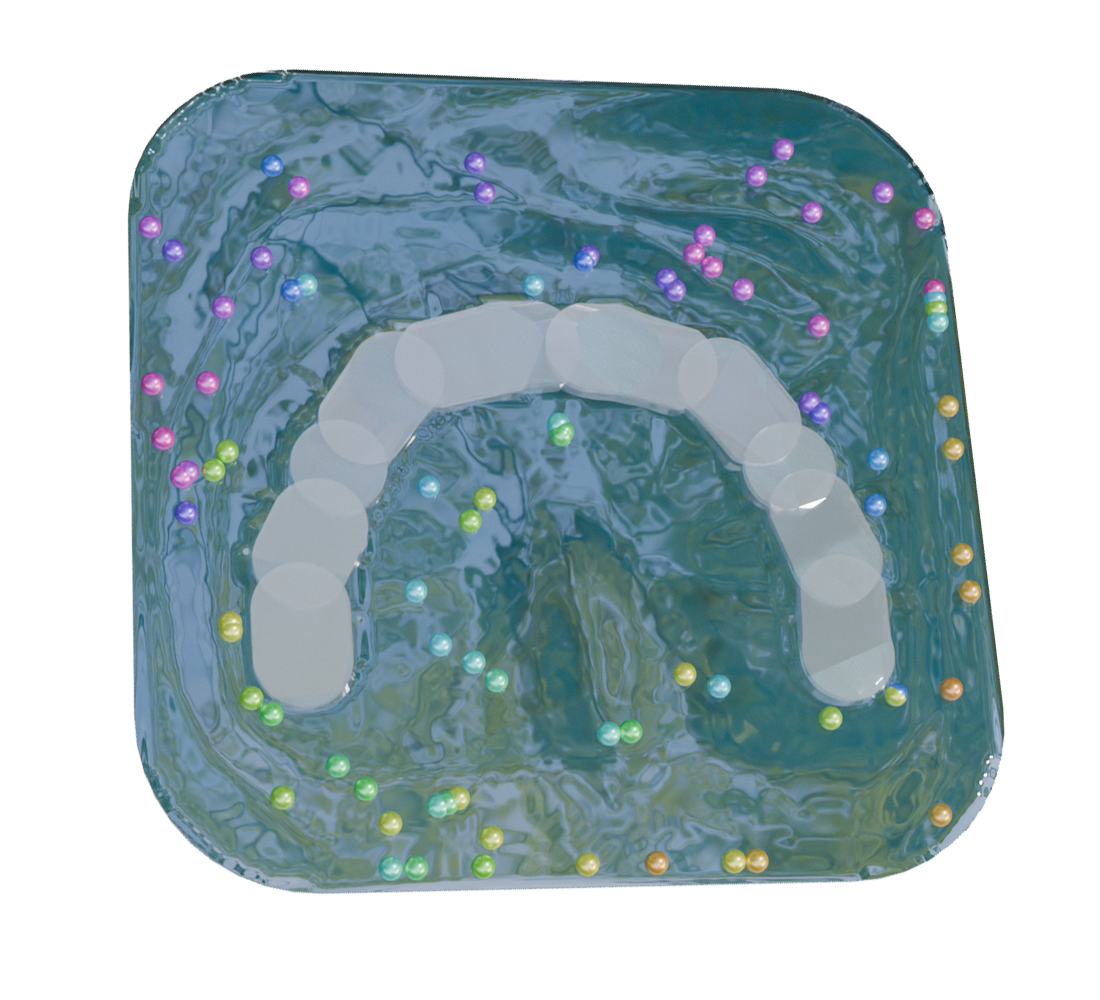}   \par
  \parbox{.19\linewidth}{\centering$t=0s$}\hfill
  \parbox{.19\linewidth}{\centering$t=0.25s$}\hfill
  \parbox{.19\linewidth}{\centering$t=0.5s$}\hfill
  \parbox{.19\linewidth}{\centering$t=0.75s$}\hfill
  \parbox{.19\linewidth}{\centering$t=1s$}
  \caption{\review{Example of fluid animation where the domain changes genus. When the cavity appear, the flow becomes ``trapped.'' It is the released as it opens. }}
  \label{fig:gripper}
\end{figure*}

\begin{figure*}
  \centering\footnotesize
  \includegraphics[width=.19\linewidth,]{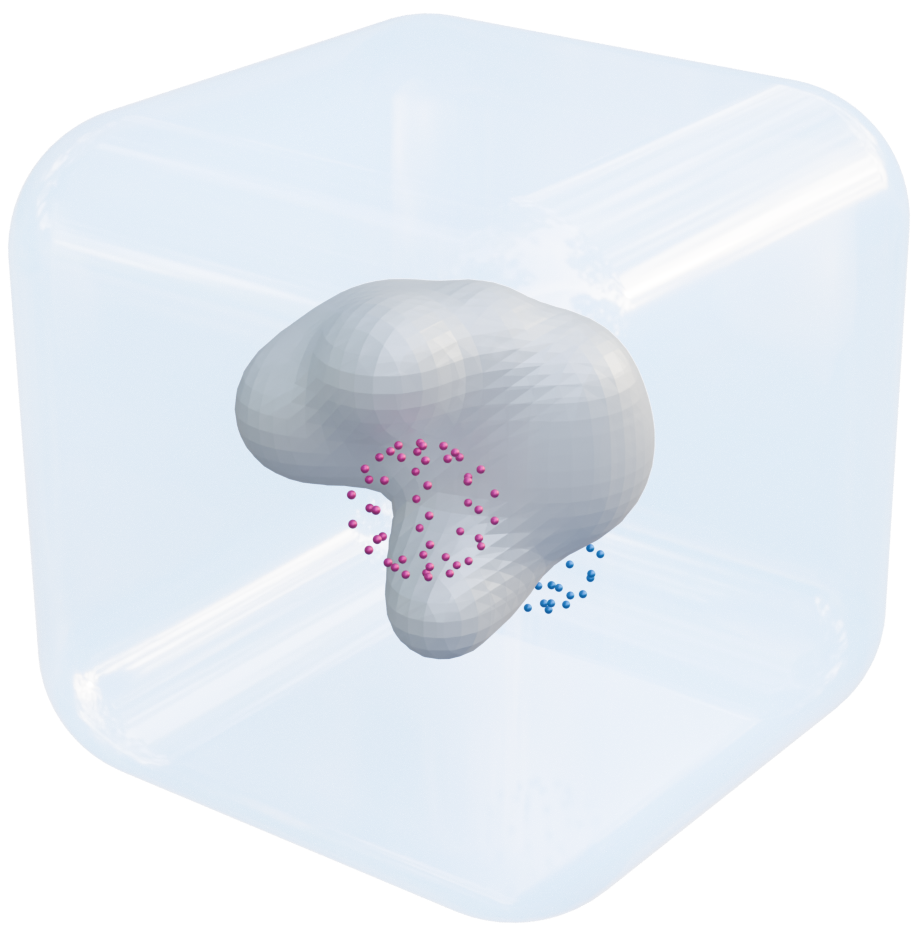}  \hfill
  \includegraphics[width=.19\linewidth]{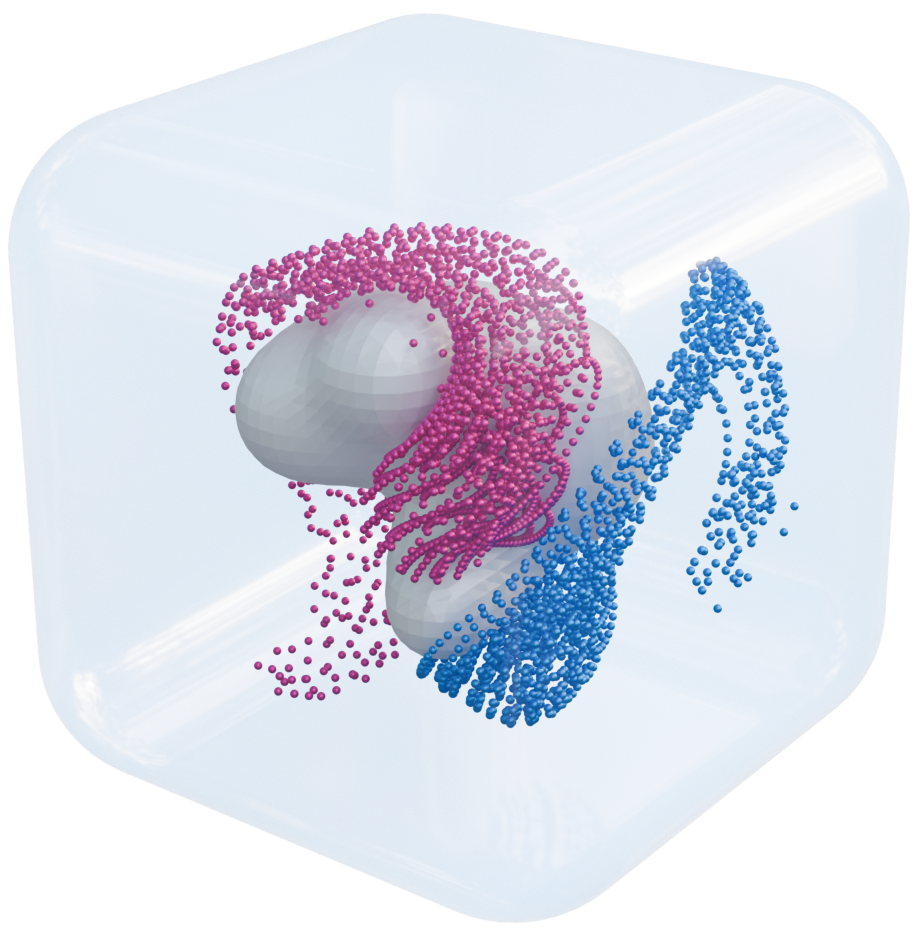}  \hfill
  \includegraphics[width=.19\linewidth]{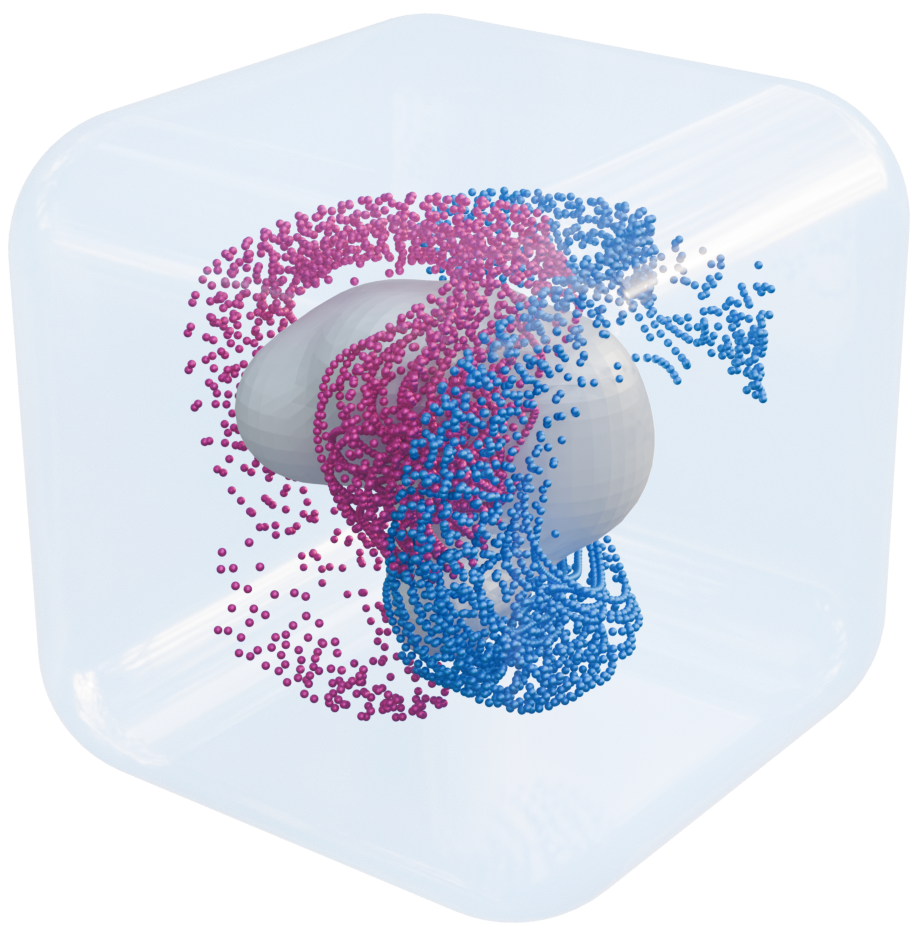}  \hfill
  \includegraphics[width=.19\linewidth]{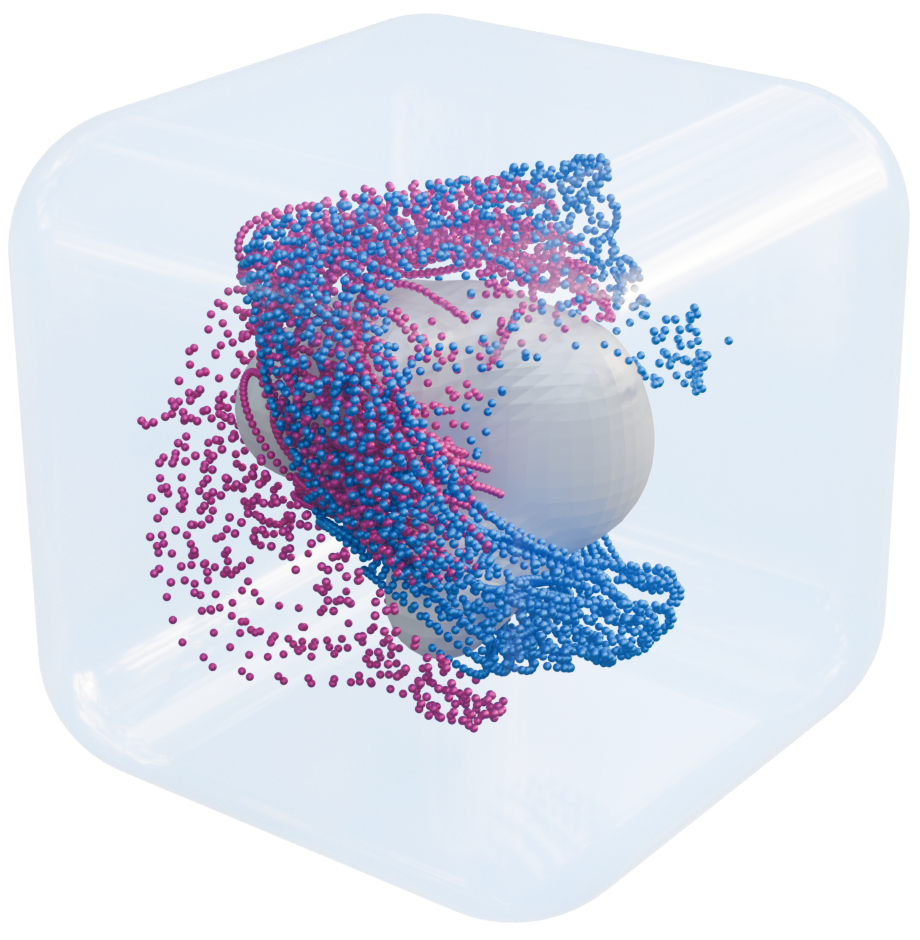}  \hfill
  \includegraphics[width=.19\linewidth]{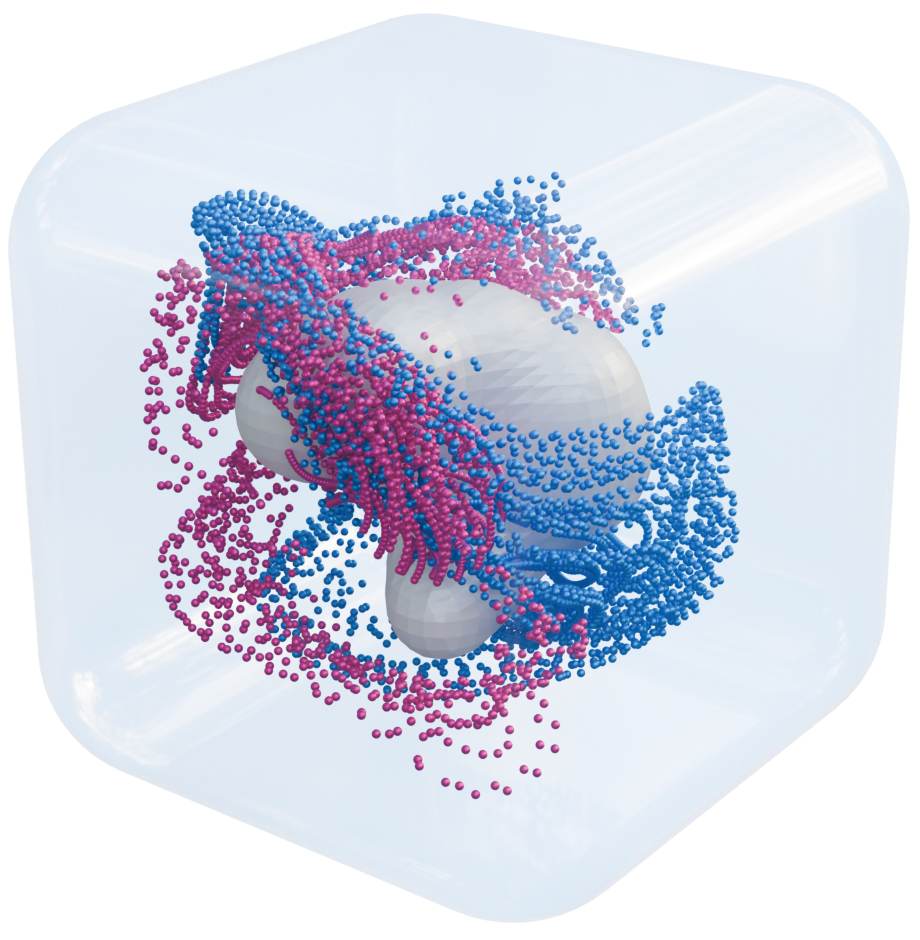}   \par

  \parbox{.19\linewidth}{\centering$t=0s$}\hfill
  \parbox{.19\linewidth}{\centering$t=4.15s$}\hfill
  \parbox{.19\linewidth}{\centering$t=5.75$}\hfill
  \parbox{.19\linewidth}{\centering$t=7.5s$}\hfill
  \parbox{.19\linewidth}{\centering$t=9.65s$}
  \caption{\review{Three-dimensional fluid animation using our neural kinematic bases of a fixed obstacle. \revise{We show particle traces from two spherical sources advected by the velocity field.}}}
  \label{fig:3d}
\end{figure*}

\paragraph{Three-dimensional} 
\review{
All our methods can be generalized to three dimensions. The domain is now represented by ten spheres and their radii (40 parameters instead of 30); the MLP network accepts 3D points and produces a volumetric vector field (\Cref{fig:teaser} top shows some bases for a fan). \Cref{fig:teaser} middle shows an animation of a moving 3D fan where the flow moves around the blade, while \Cref{fig:3d} shows an animation with static boundaries. \revise{Both figures show the particle traces from spherical sources (in different colors) advected by the velocity field.}}

%% file: 06-conclustions.tex
\section{Conclusions}

We introduce a novel neural kinematic base represented by an MLP that captures fundamental physical properties and generalizes to different domains. We show how using our bases and a standard advection integrator, we can generate two- and three-dimensional animations of \review{moving obstacles}.

To keep the training times reasonable, we decided to use only ten bases, but our method should be able to generate more bases. Additionally, we could represent the domain as a collection of capsules, thus allowing for a more detailed domain geometry.

Our method enforces Dirichlet boundary conditions, aligning the bases with the domain boundaries. Extending it to support other types, such as Neumann conditions, would require additional mechanisms to specify and blend boundary constraints. Our bases are inherently smooth, which produces smooth fluid animations and boundary geometries but limits applicability to non-smooth boundaries. Moreover, because Dirichlet conditions prevent prescribing velocity normal to obstacles, obstacles cannot actively push the fluid, resulting in quasi-static behavior. We believe that these phenomena are interesting and we leave their investigation as future work.

Finally, since our bases are encoded with an MLP, they are fully differentiable. Inverse design, for instance, could be an interesting avenue for future work. For instance, it could involve optimizing the position of the circle to match a given animation.

%% file: 08-ack.tex
\begin{acks}
The Bellairs Workshop on Computer Animation was instrumental in the conception of the research presented in this paper.  
This work was partially supported by the NSERC grants DGECR-2021-00461, RG-PIN 2021-03707, RGPIN-2024-04523, and RGPIN-2024-04605, FRQNT 365040, as well as a gift from Adobe Research.  
This work was also supported in part through the Digital Research Alliance of Canada resources, services, and staff expertise.
\end{acks}

%% file: 09-append.tex
\clearpage
\section{Comparisons}\label{sec:append}

\begin{figure}
  \centering\footnotesize
  \parbox{.05\linewidth}{~}\hfill
    \parbox{.46\linewidth}{\centering \citet{Cui2018}}\hfill
  \parbox{.46\linewidth}{\centering Ours}\par
    \parbox{.05\linewidth}{\rotatebox{90}{\centering frame 0}}\hfill
  \parbox{.46\linewidth}{\includegraphics[width=\linewidth]{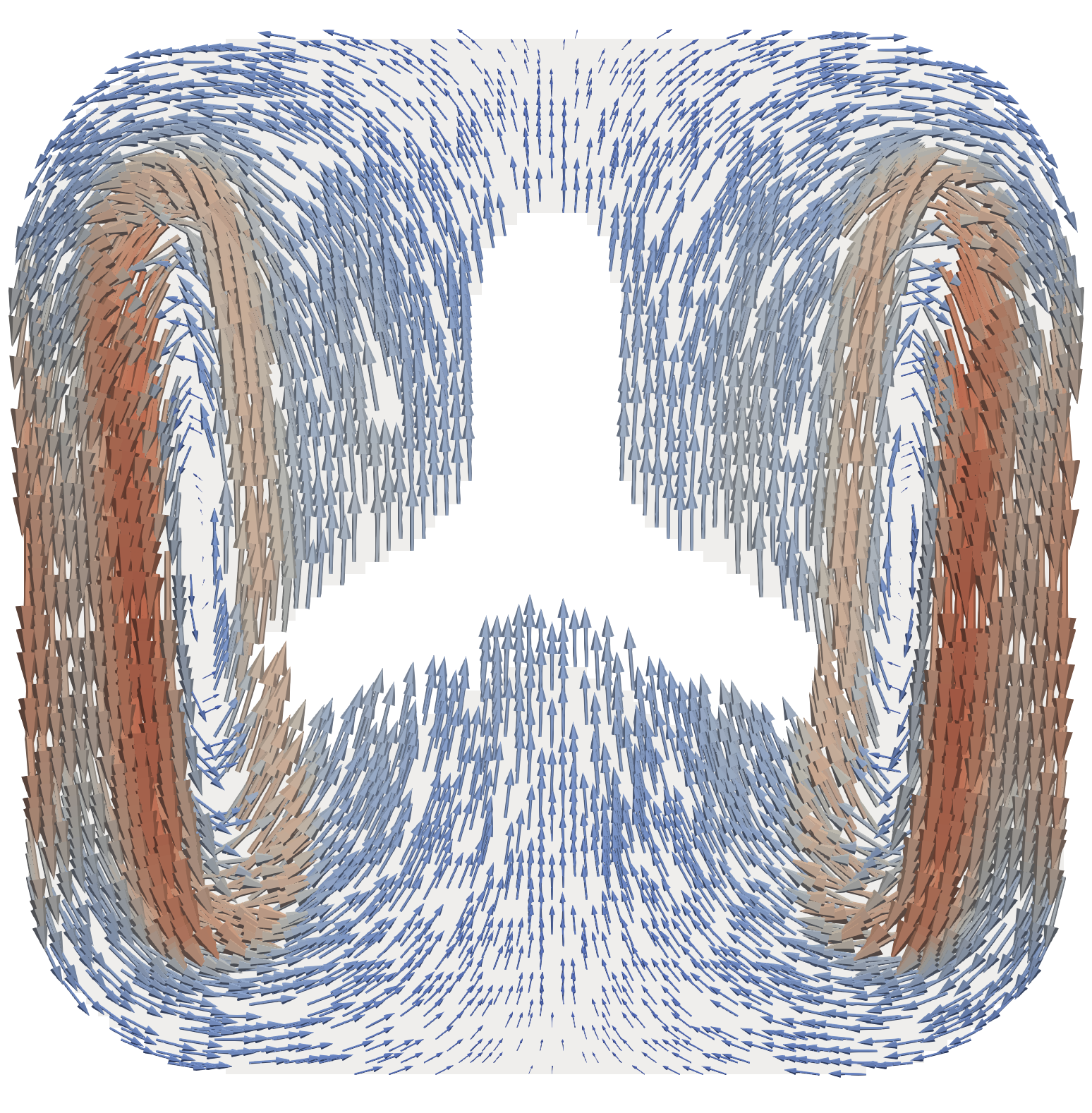}}  \hfill
  \parbox{.46\linewidth}{\includegraphics[width=\linewidth,trim={20pt 20pt 20pt 20pt},clip]{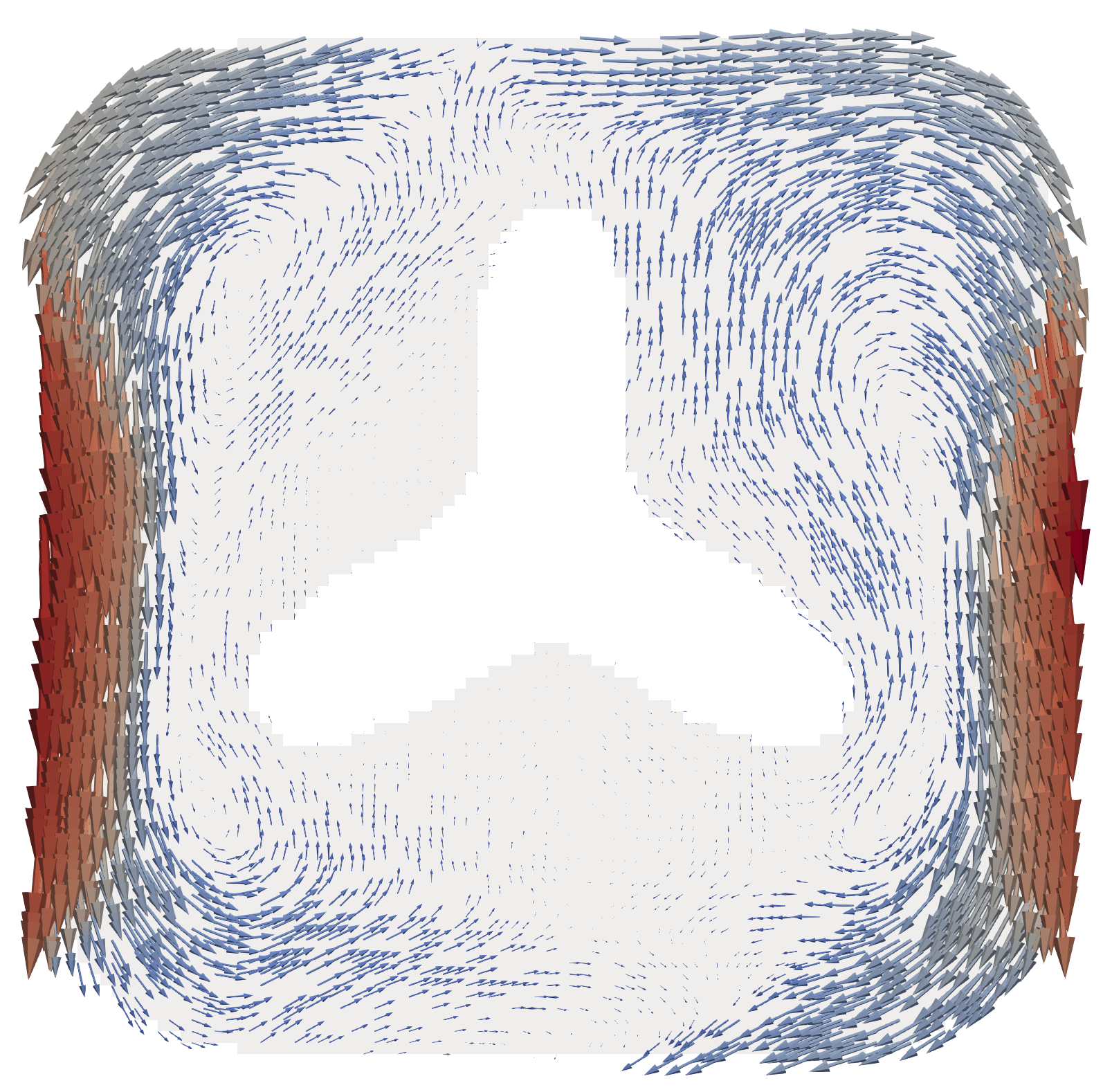}}\par
  \parbox{.05\linewidth}{\rotatebox{90}{\centering frame 100}}\hfill
  \parbox{.46\linewidth}{\includegraphics[width=\linewidth]{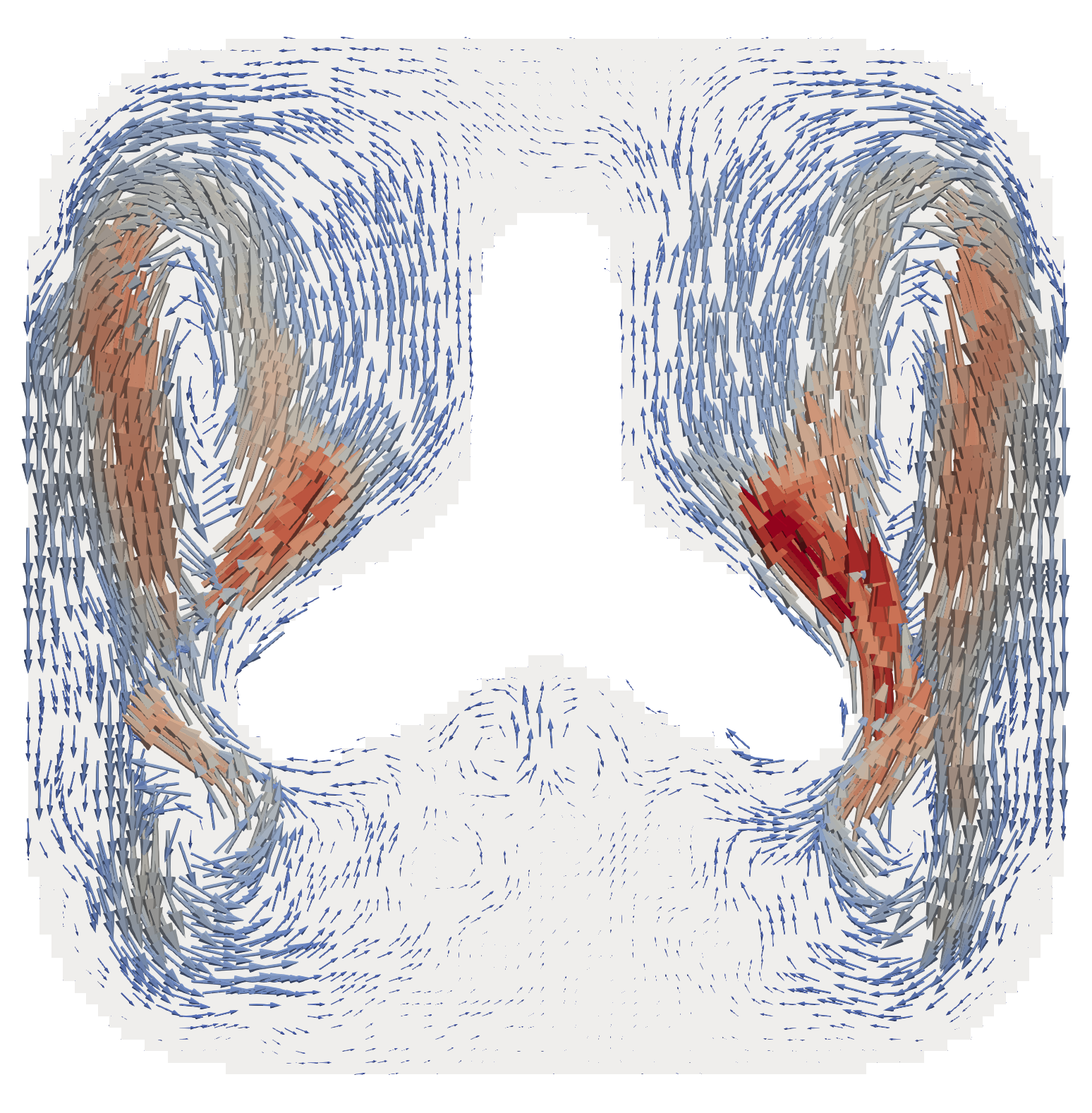}}  \hfill
  \parbox{.46\linewidth}{\includegraphics[width=\linewidth, trim={20pt 20pt 20pt 20pt},clip]{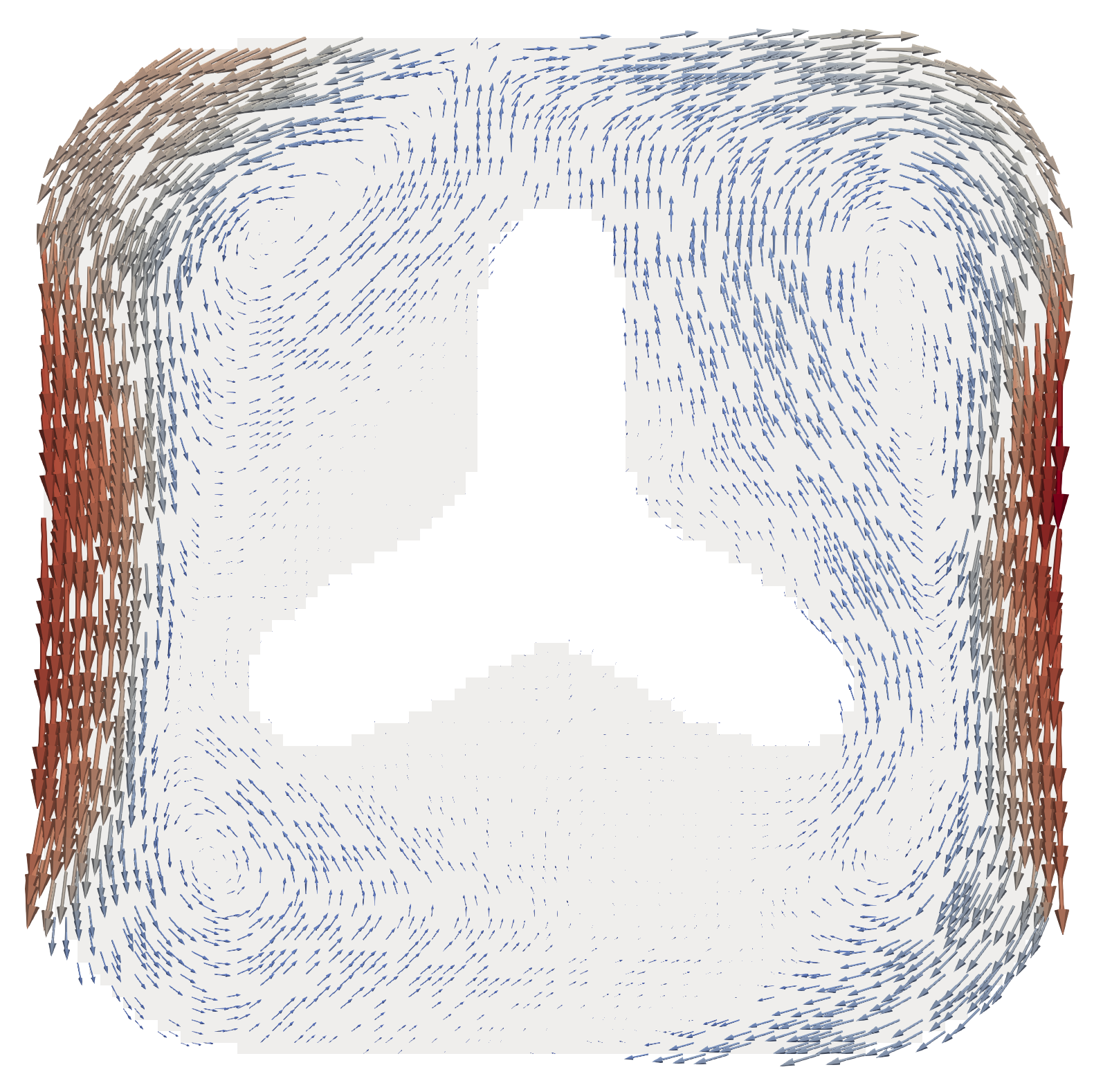}}
    \caption{Comparison with a fixed fan as obstacle with velocity field initialized from a large smoke.}
  \label{fig:comparel}
\end{figure}

\begin{figure}
  \centering\footnotesize
  \parbox{.05\linewidth}{~}\hfill
    \parbox{.46\linewidth}{\centering \citet{Cui2018}}\hfill
  \parbox{.46\linewidth}{\centering Ours}\par
    \parbox{.05\linewidth}{\rotatebox{90}{\centering frame 0}}\hfill
  \parbox{.46\linewidth}{\includegraphics[width=\linewidth]{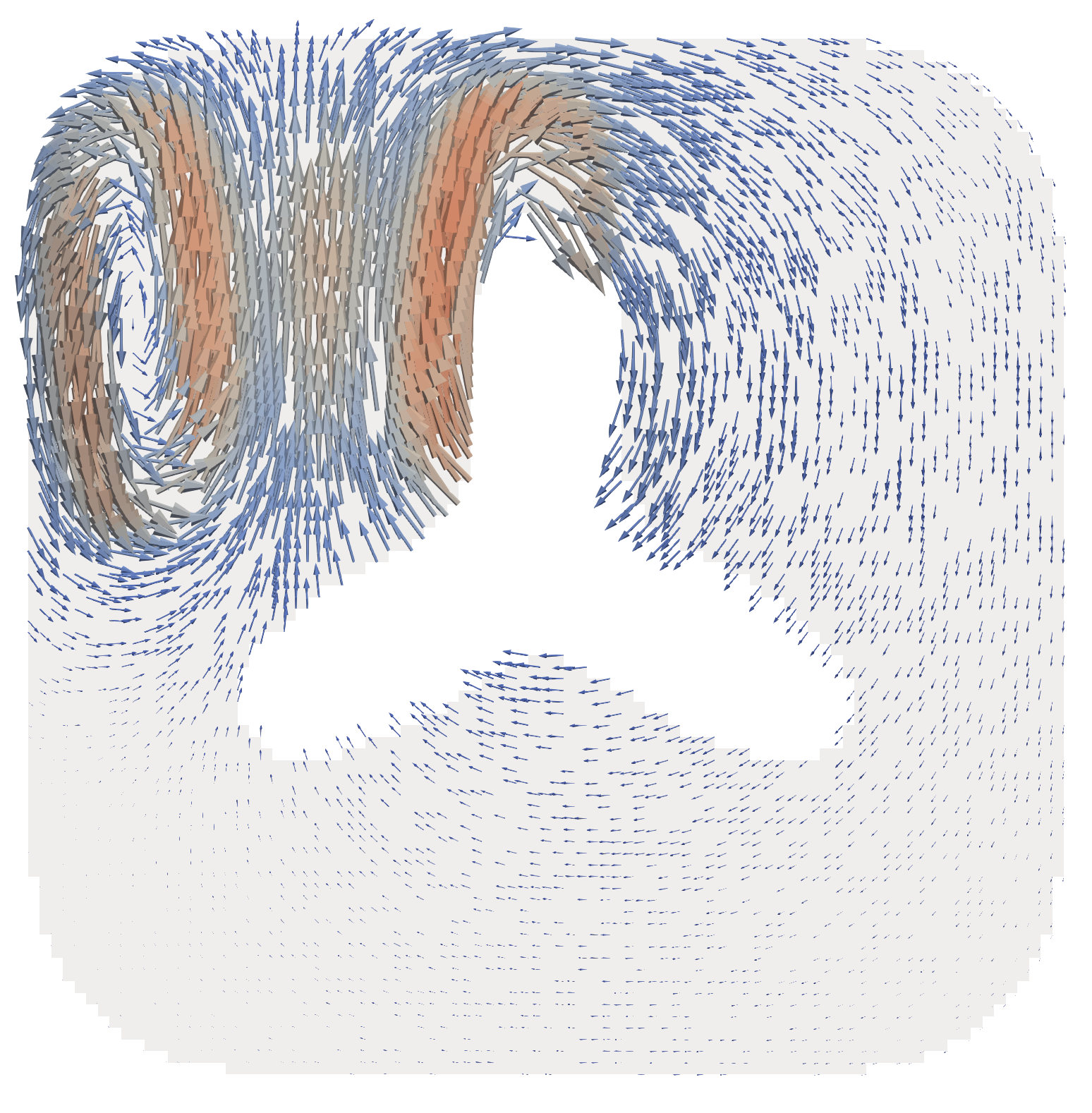}}  \hfill
  \parbox{.46\linewidth}{\includegraphics[width=\linewidth,trim={20pt 20pt 20pt 20pt},clip]{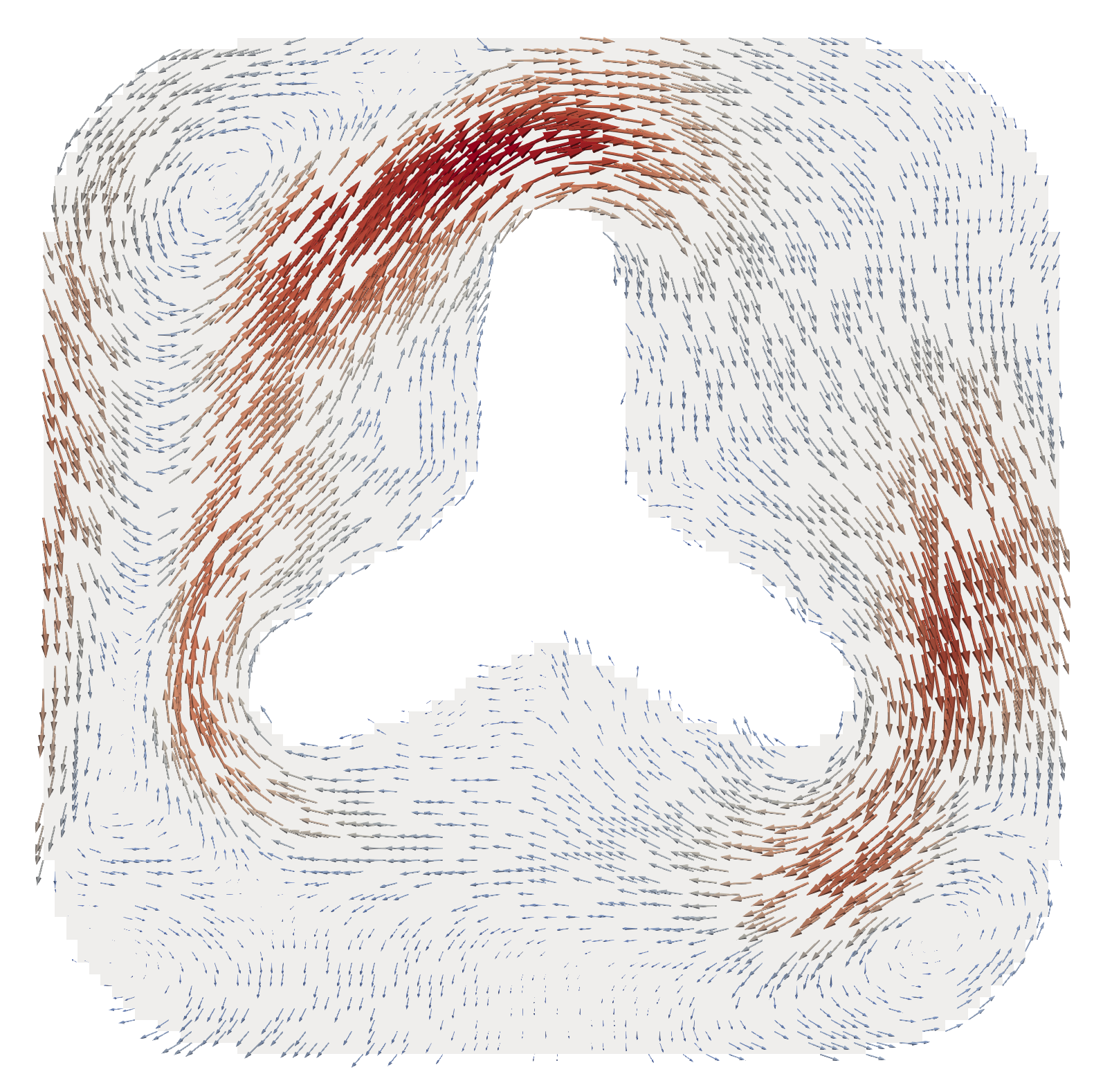}}\par
  \parbox{.05\linewidth}{\rotatebox{90}{\centering frame 100}}\hfill
  \parbox{.46\linewidth}{\includegraphics[width=\linewidth]{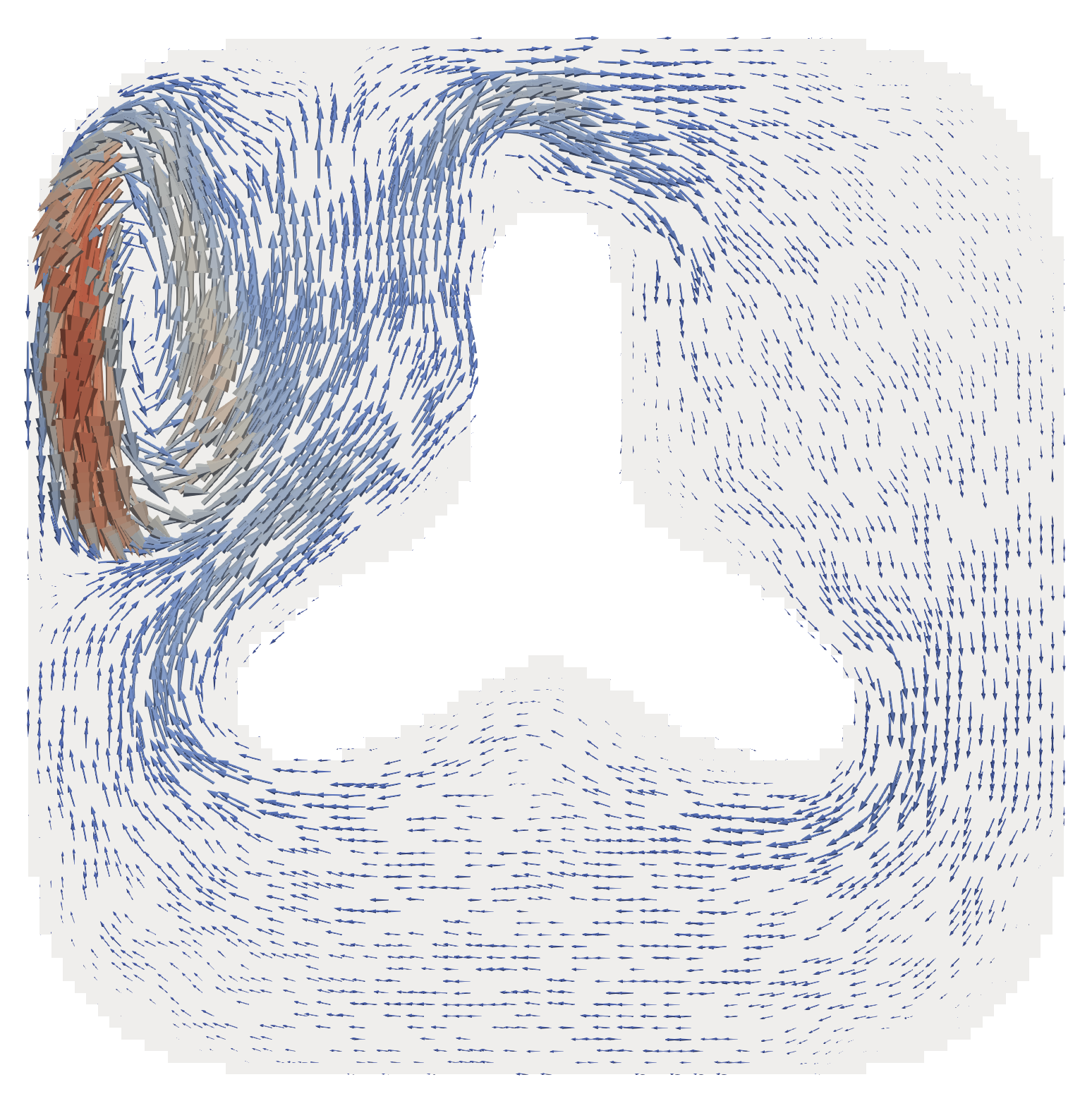}}  \hfill
  \parbox{.46\linewidth}{\includegraphics[width=\linewidth, trim={20pt 20pt 20pt 20pt},clip]{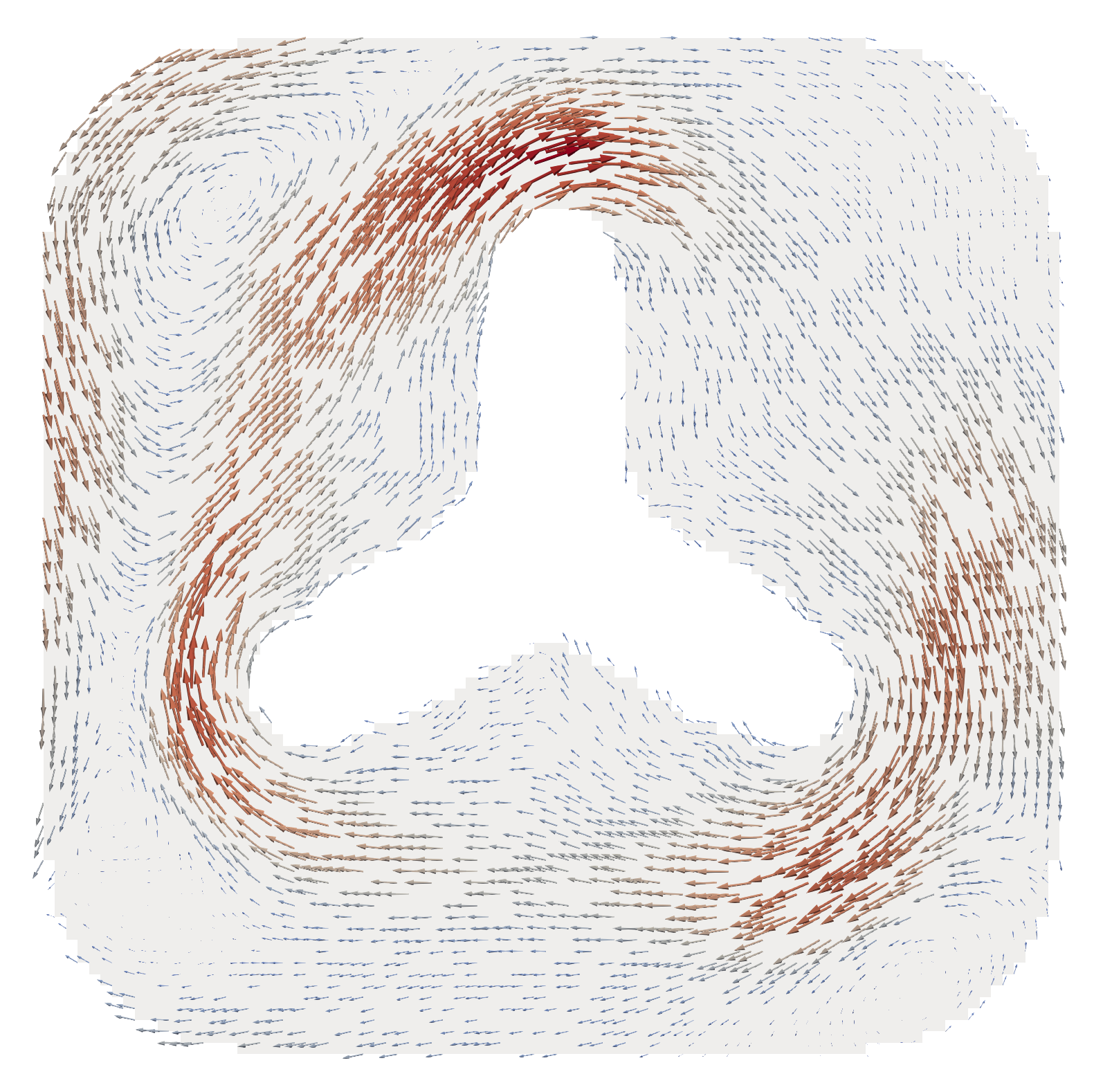}}
    \caption{Comparison with a fixed fan as obstacle with velocity field initialized from a small smoke.}
  \label{fig:compares}
\end{figure}

We evaluate two 2D scenes with different smoke scales (figures~\ref{fig:comparel} and~\ref{fig:compares}), fitting the velocity field from \citet{Cui2018} to our bases. Since the two methods differ in their representational capacity, our approach performs less favorably when fitting fields with many zero values (small-scale case). However, it demonstrates better alignment with obstacle boundary conditions as our bases account for obstacle boundaries during generation rather than relying on post-processing.